\documentclass[zpreprint,zbstnp]{zeus_paper}

\usepackage[english]{babel}

\chardef\usc=95
\chardef\til=126
\catcode`\@=11 % @ signs are now treated as letters
\DeclareRobustCommand\xdotspace{\futurelet\@let@token\@xdotspace}
\def\@xdotspace{%
  \ifx\@let@token.\else
  \ifx\@let@token\bgroup.\else
  \ifx\@let@token\egroup.\else
  \ifx\@let@token\/.\else
  \ifx\@let@token\ .\else
  \ifx\@let@token~.\else
  \ifx\@let@token!.\else
  \ifx\@let@token,.\else
  \ifx\@let@token:.\else
  \ifx\@let@token;.\else
  \ifx\@let@token?.\else
  \ifx\@let@token/.\else
  \ifx\@let@token'.\else
  \ifx\@let@token).\else
  \ifx\@let@token-.\else
  \ifx\@let@token\@xobeysp.\else
  \ifx\@let@token\space.\else
  \ifx\@let@token\@sptoken.\else
   .\space
   \fi\fi\fi\fi\fi\fi\fi\fi\fi\fi\fi\fi\fi\fi\fi\fi\fi\fi}
\catcode`\@=12 % @ signs are no longer letters

\newcommand{\stru}[2]{%
   \relax\ifmmode\hbox{\vrule height#1 depth#2 width0pt}%
   \else\vrule height#1 depth#2 width0pt\fi}
\newcommand{\Ronum}[1]{\uppercase\expandafter{\romannumeral#1}}
\newcommand{\ronum}[1]{\expandafter{\romannumeral#1}}
\DeclareRobustCommand{\LaTeXZ}{%
  \LaTeX\kern-.05em4\kern-.1em
  {\raisebox{-0.2ex}{$\scriptstyle\text{ZEUS}$}}\xspace}
\DeclareMathAlphabet{\mathbf}{OT1}{cmr}{bx}{sl}
\newcommand{\eVdist}{\kern-0.06667em}

\newcommand{\slashfrac}[2]{%
  \raisebox{0.5ex}{\ensuremath #1}\kern-0.12em/\kern-0.08em
  \raisebox{-.8ex}{\ensuremath #2}}
\newcommand{\sqr}[3]{%
    {\vcenter{\hrule height.#3ex\hbox{\vrule width.#2ex height#1ex
     \kern#1ex\vrule width.#3ex}\hrule height.#2ex}}}

\catcode`\@=11 % @ signs are now treated as letters
\newcommand{\parenbar}{\mathpalette\p@renb@r}
\def\p@renb@r#1#2{\vbox{%
  \ifx#1\scriptscriptstyle \dimen@.7em\dimen@ii.2em\else
  \ifx#1\scriptstyle \dimen@.8em\dimen@ii.25em\else
  \dimen@1em\dimen@ii.4em\fi\fi \offinterlineskip
  \ialign{\hfill##\hfill\cr
    \vbox{\hrule width\dimen@ii}\cr
    \noalign{\vskip-.3ex}%
    \hbox to\dimen@{$\mathchar300\hfil\mathchar301$}\cr
    \noalign{\vskip-.3ex}%
    $#1#2$\cr}}}
\catcode`\@=12 % @ signs are no longer letters

\newcommand{\IP}{{\rm I$\kern-0.01667em$P}\xspace}

\mathchardef\qsm=63
\mathchardef\pls=43
\mathchardef\mns=512
\mathchardef\plm=518
\mathchardef\eql=61
\mathchardef\smallleft=300
\mathchardef\smallright=301
\mathchardef\les=316
\mathchardef\gre=318
\mathchardef\leq=532
\mathchardef\grq=533
\catcode`\@=11 % @ signs are now treated as letters
\newcounter{pict@width}
\newcounter{pict@height}
\newlength{\pict@scale}
\newcommand{\psfigadd}[4]{%
\setcounter{pict@width}{1*\ratio{#2+\pict@scale/2}{\pict@scale}}
\setcounter{pict@height}{1*\ratio{#3+\pict@scale/2}{\pict@scale}}
\setlength{\unitlength}{\pict@scale}
\hbox to #2{\hspace{-\fill}\begin{picture}(\thepict@width,\thepict@height)
\put(0,0){\psfig{figure=#1,width=#2,height=#3,clip=}}
\SetScale{0.283466457}
\SetWidth{1.763889}
{#4}
\end{picture}}
}
\newcounter{pict@widthfst}
\newcounter{pict@widthscd}
\newcounter{pict@widthtot}
\newcommand{\psfigaddtwo}[7]{%
\setcounter{pict@widthfst}{1*\ratio{#2+\pict@scale/2}{\pict@scale}}
\setcounter{pict@widthscd}{1*\ratio{#2+#4+\pict@scale/2}{\pict@scale}}
\setcounter{pict@widthtot}{1*\ratio{#2+#4+#6+\pict@scale/2}{\pict@scale}}
\setcounter{pict@height}{1*\ratio{#3+\pict@scale/2}{\pict@scale}}
\setlength{\unitlength}{\pict@scale}
\hbox{\hspace{-\fill}\begin{picture}(\thepict@widthtot,\thepict@height)
\put(0,0){\psfig{figure=#1,width=#2,height=#3,clip=}}
\put(\thepict@widthscd,0){\psfig{figure=#5,width=#6,height=#3,clip=}}
\SetScale{0.283466457}
\SetWidth{1.763889}
{#7}
\end{picture}}
}
\newcommand{\psfigror}[4]{%
\setcounter{pict@width}{1*\ratio{#2+\pict@scale/2}{\pict@scale}}
\setcounter{pict@height}{1*\ratio{#3+\pict@scale/2}{\pict@scale}}
\setlength{\unitlength}{\pict@scale}
\hbox{\begin{picture}(\thepict@width,\thepict@height)
\put(0,\thepict@height){\psfig{figure=#1,width=#3,height=#2,clip=,angle=270}}
\SetScale{0.283466457}
\SetWidth{1.763889}
{#4}
\end{picture}}
}
\newcommand{\psfigrol}[4]{%
\setcounter{pict@width}{1*\ratio{#2+\pict@scale/2}{\pict@scale}}
\setcounter{pict@height}{1*\ratio{#3+\pict@scale/2}{\pict@scale}}
\setlength{\unitlength}{\pict@scale}
\hbox{\begin{picture}(\thepict@width,\thepict@height)
\put(0,0){\psfig{figure=#1,width=#3,height=#2,clip=,angle=90}}
\SetScale{0.283466457}
\SetWidth{1.763889}
{#4}
\end{picture}}
}
\catcode`\@=12 % @ signs are no longer letters
\newlength\listtextwidth

\catcode`\@=11 % @ signs are now treated as letters
\newlength{\@tabfninsert}
\newlength{\@tabfnwidth}
\newcommand{\tabfootnote}[2]{%
  \setlength{\@tabfninsert}{0.8em}
  \setlength{\@tabfnwidth}{\textwidth}
  \addtolength{\@tabfnwidth}{-\@tabfninsert}
  \addtolength{\@tabfnwidth}{-0.4em}
  \noindent\makebox[\@tabfninsert][r]{\footnotesize$^{#1}$\hfil}\hfill%
  \parbox[t]{\@tabfnwidth}{\footnotesize #2\hfill}}
\catcode`\@=12 % @ signs are no longer letters

\begin{document}

\prepnum{{DESY--07--102}}

\title{
Three- and four-jet final states in photoproduction at HERA
}

\author{ZEUS Collaboration}
\date{July 2007}

\abstract{
Three- and four-jet final states have been measured in photoproduction at HERA using the ZEUS detector with an integrated luminosity of $121~{\rm pb^{-1}}$.  The results are presented for jets with transverse energy $E_T^{\rm jet}>6$~GeV and pseudorapidity $|\eta^{\rm jet}|<2.4$, in the kinematic region given by the virtuality of the photon $Q^2<1~\rm{GeV^2}$ and the inelasticity $0.2\le y\le0.85$ and in two mass regions defined as $25\le M_{nj}<50$~GeV and $M_{nj}\ge50$~GeV, where $M_{nj}$ is the invariant mass of the $n$-jet system.  The four-jet photoproduction cross section has been measured for the first time and represents the highest-order process studied at HERA.  Both the three- and four-jet cross sections have been compared with leading-logarithmic parton-shower Monte Carlo models, with and without multi-parton interactions.  The three-jet cross sections have been compared to an $\mathcal{O}(\alpha\alpha_s^2)$ perturbative QCD calculation.
}

\makezeustitle

\def\3{\ss}

\pagenumbering{Roman}

\begin{center}
{                      \Large  The ZEUS Collaboration              }
\end{center}
  S.~Chekanov$^{   1}$,
  M.~Derrick,
  S.~Magill,
  B.~Musgrave,
  D.~Nicholass$^{   2}$,
  \mbox{J.~Repond},
  R.~Yoshida\\
 {\it Argonne National Laboratory, Argonne, Illinois 60439-4815}, USA~$^{n}$
\par \filbreak
  M.C.K.~Mattingly \\
 {\it Andrews University, Berrien Springs, Michigan 49104-0380}, USA
\par \filbreak
  M.~Jechow, N.~Pavel~$^{\dagger}$, A.G.~Yag\"ues Molina \\
  {\it Institut f\"ur Physik der Humboldt-Universit\"at zu Berlin,
           Berlin, Germany}
% ...
\newpage
$^{\    1}$ supported by DESY, Germany \\
$^{\    2}$ also affiliated with University College London, UK \\
% ...
$^{\dagger}$ deceased \\
%
%\par         % if index listing & table fit to 1 page, put gap here
%\newpage   % alternatively: go to newpage, if page is too small
                                                         %
% \institute_references_start    % do not touch or move this line !
                                                           %
\begin{tabular}[h]{rp{14cm}}
$^{a}$ &  supported by the Natural Sciences and Engineering Research Council of Canada (NSERC) \\
$^{b}$ &  supported by the German Federal Ministry for Education and Research (BMBF), under
          contract numbers HZ1GUA 2, HZ1GUB 0, HZ1PDA 5, HZ1VFA 5\\
$^{c}$ &  supported in part by the MINERVA Gesellschaft f\"ur Forschung GmbH, the Israel Science
          Foundation (grant no. 293/02-11.2) and the U.S.-Israel Binational Science Foundation \\
$^{d}$ &  supported by the German-Israeli Foundation and the Israel Science Foundation\\
$^{e}$ &  supported by the Italian National Institute for Nuclear Physics (INFN) \\
$^{f}$ &  supported by the Japanese Ministry of Education, Culture, Sports, Science and Technology
          (MEXT) and its grants for Scientific Research\\
$^{g}$ &  supported by the Korean Ministry of Education and Korea Science and Engineering
          Foundation\\
$^{h}$ &  supported by the Netherlands Foundation for Research on Matter (FOM)\\
$^{i}$ &  supported by the Polish State Committee for Scientific Research, grant no.
          620/E-77/SPB/DESY/P-03/DZ 117/2003-2005 and grant no. 1P03B07427/2004-2006\\
$^{j}$ &  partially supported by the German Federal Ministry for Education and Research (BMBF)\\
$^{k}$ &  supported by RF Presidential grant N 8122.2006.2 for the leading
          scientific schools and by the Russian Ministry of Education and Science through its grant
          Research on High Energy Physics\\
$^{l}$ &  supported by the Spanish Ministry of Education and Science through funds provided by
          CICYT\\
$^{m}$ &  supported by the Particle Physics and Astronomy Research Council, UK\\
$^{n}$ &  supported by the US Department of Energy\\
$^{o}$ &  supported by the US National Science Foundation. Any opinion,
findings and conclusions or recommendations expressed in this material
are those of the authors and do not necessarily reflect the views of the
National Science Foundation.\\
$^{p}$ &  supported by the Polish Ministry of Science and Higher Education
as a scientific project (2006-2008)\\
$^{q}$ &  supported by FNRS and its associated funds (IISN and FRIA) and by an Inter-University
          Attraction Poles Programme subsidised by the Belgian Federal Science Policy Office\\
$^{r}$ &  supported by the Malaysian Ministry of Science, Technology and
Innovation/Akademi Sains Malaysia grant SAGA 66-02-03-0048\\
\end{tabular}
                                                           %
% \institute_references_end     % do not touch or move this line !
                                                           %
\newpage

\pagenumbering{arabic}
\pagestyle{plain}

\section{Introduction}

In photoproduction at HERA, a quasi-real photon, emitted by the incoming electron\footnote{From now on the word ``electron'' is used as a generic term for electrons and positrons, unless stated otherwise.}, interacts with the proton. Hard photoproduction~\cite{pl:b79:83,np:b166:413,pr:d21:54,zfp:c6:241,proc:hera:1987:331,prl:61:275,prl:61:682,pr:d39:169,zfp:c42:657,pr:d40:2844} may be categorised at leading order (LO) as being either direct, if the photon interacts as a point-like particle, or resolved, if it fluctuates into a partonic system, and subsequently transfers only a fraction of its momentum to the hard interaction.

Photoproduction collisions at HERA can lead to final states with multiple jets.  Multi-jet events are of particular interest as they are produced by processes that are manifestly beyond LO in the strong coupling constant, $\alpha_s$.  Presently, predictions from perturbative quantum chromodynamics (pQCD) pertaining to multi-jet final states in photoproduction are only available up to $\mathcal{O}(\alpha\alpha_s^{2})$, where $\alpha$ is the fine structure constant.

\begin{figure}
\begin{center}
\includegraphics*[angle=0,scale=0.45]{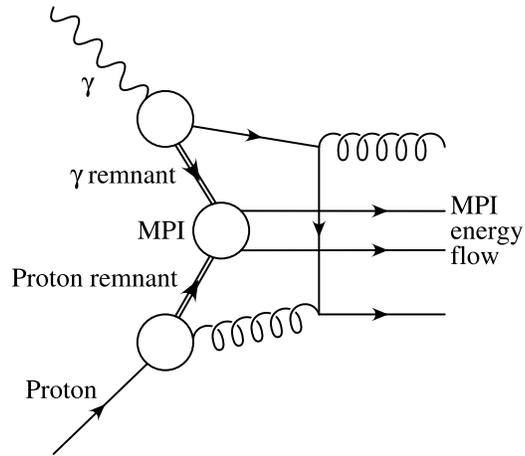}
\caption{
A schematic representation of an event exhibiting a secondary scatter (MPI).
\label{fig:MPI}}
\end{center}
\end{figure}

The hadron-like structure of the photon in resolved processes gives rise to the possibility of multi-parton interactions (MPIs) at HERA.  In the MPI picture, more than one pair of partons takes part in the hard interaction. A schematic of an MPI event is shown in Fig.~\ref{fig:MPI}.  The secondary scatters generate additional hadronic energy flow in the event, the topology and magnitude of which is poorly understood theoretically.  Potentially, this energy flow may lead to the formation of jets and so MPIs may constitute a source of multi-jets in the final state.  Multi-parton interactions have been studied before at the Tevatron~\cite{MPIsTEV4, MPIsTEV1, MPIsTEV2, MPIsTEV5}, in dijet photoproduction events at HERA~\cite{Art:zeusxg, MPIsHERA1} and are expected to be prevalent at the LHC~\cite{MPIsLHC1, MPIsLHC2}.

Three-jet photoproduction events have been studied before at HERA in the $M_{3j}\ge50$~GeV~\cite{Art:zeusPP1} region of three-jet invariant mass.  Multi-jet final states have also been studied in deep inelastic scattering (DIS) at HERA~\cite{Art:h1mjdis, Art:h13jdis, Art:zeus3jalpha, Art:zeus3jdiff}.  In this paper, differential cross sections are presented for the three-jet photoproduction final state, in a wider $M_{3j}$ region and with over seven times the luminosity of the previous ZEUS publication~\cite{Art:zeusPP1}. The four-jet photoproduction cross sections are presented here for the first time. Also examined is the description of the data by two Monte Carlo (MC) models and whether or not it is improved by the introduction of simulated MPIs.  In addition, an $\mathcal{O}(\alpha\alpha_s^2)$ prediction has been compared to the three-jet data.

%%%%%%%%%%%%%%%%%%%%%%%%%%%%%%%%%%%%%%%%%%%%%%%%%%%%%%%%%%%%%%%%%%%%%%%%%%%%%%

\section{Experimental conditions}

The data were collected using the ZEUS detector during the 1996 to 2000 running periods. In 1996 and 1997, HERA collided positrons with protons, with energies of $E_e=27.5$~GeV and $E_p=820$~GeV, respectively, corresponding to a beam centre-of-mass energy, $\sqrt{s}$, of 300~GeV.  From 1998 onwards the proton beam energy was raised to $E_p=920$~GeV ($\sqrt{s}=318$~GeV). Furthermore, a subsample of these later data consists of electron-proton collisions.  The total sample corresponds to an integrated luminosity of $(121\pm2)~{\rm~pb^{\rm -1}}$, of which $82~{\rm~pb^{\rm -1}}$ were collected at $\sqrt{s}=318$~GeV. A detailed description of the ZEUS detector can be found elsewhere~\cite{pl:b297:404,zeus:1993:bluebook}. A brief outline of the components that are of most relevance to this analysis is given below.

The high-resolution uranium--scintillator calorimeter (CAL)~\cite{nim:a309:77,*nim:a309:101,*nim:a336:23,*nim:a321:356} consists of three parts: the forward, the barrel and the rear calorimeters. Each part is subdivided transversely into towers and longitudinally into one electromagnetic and either one (in the rear) or two (in the barrel and forward) hadronic sections. The smallest subdivision of the calorimeter is called a cell. The CAL relative energy resolutions, as measured under test-beam conditions, are $0.18/\sqrt{E}$ for electrons and $0.35/\sqrt{E}$ for hadrons, with $E$ in GeV.

Charged particles are measured in the central tracking detector (CTD)~\cite{nim:a279:290,*npps:b32:181,*nim:a338:254}, which operates in a magnetic field of 1.43~T provided by a thin superconducting coil. The CTD covers the polar-angle\footnote{The ZEUS coordinate system is a right-handed Cartesian system, with the $Z$ axis pointing in the proton beam direction, referred to as the ``forward direction'', and the $X$ axis pointing left towards the centre of HERA. The coordinate origin is at the nominal interaction point.} region \mbox{$15^\circ<\theta<164^\circ$}. The relative transverse-momentum resolution for full-length tracks is $\sigma_{p_T}/p_T=0.0058p_T \oplus 0.0065 \oplus 0.0014/p_T$, with $p_T$ in GeV. Both tracking and calorimetry were used to reconstruct the transverse energy and direction of jets as described in Section~\ref{eventrecon}.

The luminosity was measured from the rate of the bremsstrahlung process $ep~\rightarrow~e\gamma p$, where the photon is detected in a lead--scintillator calorimeter~\cite{desy-92-066-tim,*zfp:c63:391,*desy-01-141} placed in the HERA tunnel at $Z=-107$~m.

%%%%%%%%%%%%%%%%%%%%%%%%%%%%%%%%%%%%%%%%%%%%%%%%%%%%%%%%%%%%%%%%%%%%%%%%%%%%%%

\section{Event reconstruction and selection}
\label{eventrecon}

A three-level trigger system was used to select events online~\cite{zeus:1993:bluebook,epj:c1:109}. At the first two levels, general characteristics of photoproduction collisions were required and background from beam-gas events was rejected.  At the third level, jets were reconstructed by applying a cone jet algorithm to the CAL cells.  Events with at least two jets satisfying $E_T^{\rm jet}\ge4.5$~GeV and $\eta^{\rm jet}\le2.5$ were accepted, where $E_T^{\rm jet}$ and $\eta^{\rm jet}$ are the transverse energy and pseudorapidity of the jet evaluated in the laboratory frame.

In the offline analysis, the hadronic final state was reconstructed using energy-flow objects~\cite{Art:efo1, thesis:hefa} (EFOs), which are formed from a combination of track and calorimeter information.  This approach optimises the energy resolution and the one-to-one correspondence between the detector-level objects and the hadrons.  The EFOs were corrected~\cite{Art:efocorr, Art:ZEUSEFO} to account for energy losses in the dead material and were forced to be massless by setting the energy component equal to the magnitude of the three-momentum.

The jets were reconstructed using the $k_T$ cluster algorithm~\cite{Art:ktclus} in the longitudinally invariant inclusive mode~\cite{Art:ktclusinc} using the $p_T^2$ recombination scheme.  At the detector level, jets were formed from the EFOs.  The energy and angular reconstruction of the detector-level jets exhibited no systematic bias with respect to those defined at the hadron level.

The data sample was selected by requiring the following:

\begin{itemize}
\item the longitudinal position of the reconstructed vertex was in the range $|Z_{\rm vtx}|\le 40$~cm;
\item $y_{\rm JB}\ge 0.2$, to reduce the contamination from beam-gas events, where $y_{\rm JB}$ is the Jacquet--Blondel estimator~\cite{proc:epfacility:1979:391} of the inelasticity, $y$, the fraction of the incoming electron momentum carried by the photon;
\item no scattered electron was observed in the CAL with $y_e\le0.85$, to remove background due to DIS, where $y_e$ is the electron-method estimator of $y$;
\item $y_{\rm JB}\le 0.85$, which further removed contamination from DIS;
\item $P_T^{\rm miss}/\sqrt{E_T}\le 2$~GeV$^{1/2}$, where $P_T^{\rm miss}$ is the missing transverse momentum and $E_T$ is the total transverse energy.  This removed any background from charged current, cosmic and halo-muon events;
\item at least three (or four, depending on the specific sample) jets were found with $|\eta^{\rm jet}|\le2.4$ and $E_T^{\rm jet}\ge 6~\rm{GeV}$.
\end{itemize}

Furthermore, the three- and four-jet analyses were conducted in low- and high-mass regions given by $25\le M_{nj}\le50$~GeV and $M_{nj}\ge50$~GeV, respectively, where $M_{nj}$ is the invariant mass of the $n$-jet system.

After all of the selection criteria had been applied, the low- and high-mass, three-jet data samples had 291646 and 38098 events, respectively, while the equivalent four-jet data samples had 31533 and 12525 events.

Photoproduction events are characterised by the low virtuality, $Q^2$, of the quasi-real exchanged photon, where $Q^2$ is the negative of the photon four-momentum squared.  The kinematics are specified by $y$ and $x_\gamma$, the fraction of the photon momentum taking part in the interaction.  The variable $x_\gamma$ can be approximated using the observable $x_\gamma^{\rm obs}$, defined as
\begin{equation}
x_\gamma^{\rm obs}=\frac{\sum_{i=1}^{\rm n}E_{T,i}^{\rm jet}\exp(-\eta_i^{\rm jet})}{2yE_e},
\label{equ:xgammaobs}
\end{equation}
where the sum runs over the $n$-jets considered in the event and $E_e=27.5$~GeV denotes the energy of the incoming electron.

\begin{figure}
\begin{center}
\includegraphics*[angle=0,scale=0.5]{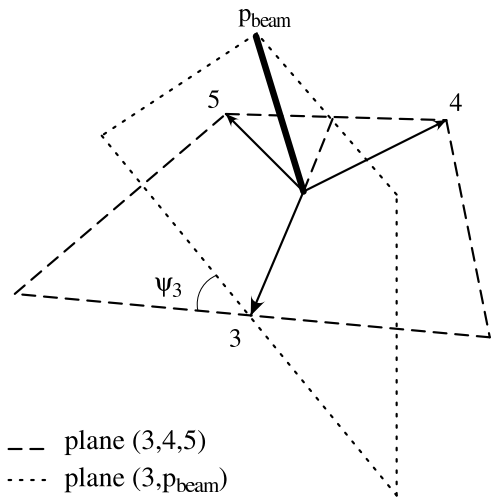}
\caption{
A schematic representation of the three-jet centre-of-mass system and the $\psi_3$ angle.
\label{fig:angles}}
\end{center}
\end{figure}

An angular variable used to analyse three-jet events is $\psi_3$~\cite{Art:multijetkine}, which, as shown in Fig.~\ref{fig:angles}, is the angle in the three-jet centre-of-mass frame between the plane containing the highest energy jet and the beam, and the plane containing the three jets.  The beam is described by the three-vector ${\bf p}_{\rm beam}={\bf p}_p-{\bf p}_e$, where ${\bf p}_p$ and ${\bf p}_e$ are the momenta of the proton and electron beams, respectively. The variable may be written as
\begin{equation}
\cos(\psi_{\rm 3})=\frac{({\bf p}_{\rm beam}\times{\bf p}_{\rm 3})\cdot({\bf p}_{\rm 4}\times {\bf p}_{\rm 5})}{|{\bf p}_{\rm beam}\times{\bf p}_{\rm 3}||{\bf p}_{\rm 4}\times{\bf p}_{\rm 5}|},
\label{equ:cos_psi3}
\end{equation}
where it is conventional to number the three jets, 3, 4 and 5, in order of decreasing energy.  The $\psi_3$ angle reflects the orientation of the lowest-energy jet.  In the case where this jet arises from initial-state gluon radiation, the coherence property of QCD will tend to orient the third jet close to the incoming proton or photon direction. The two planes shown in Fig.~\ref{fig:angles} will therefore tend to coincide leading to a $\psi_3$ distribution that peaks toward 0 and $\pi$.

\section{Monte Carlo models}

Two MC generators were used to simulate photoproduction events, {\sc Herwig} 6.505~\cite{Art:herwig, *Art:herwig2, *Art:herwig3} and {\sc Pythia} 6.206~\cite{Art:pythia, *Art:pythia2}. Both models include the LO ($2\rightarrow2$) matrix elements, approximate higher-order processes using initial-state and final-state parton showers and simulate hadronisation.  Direct and resolved photoproduction samples were generated separately.  The implementation of the parton-showers and hadronisation models in the generators differs~\cite{Art:pythia, *Art:pythia2, Art:hrwPS1, Art:hrwPS2, Art:hrwPS3}.  Hadronisation in {\sc Herwig} is simulated using the cluster model~\cite{Art:hrwHad1} while {\sc Pythia} uses the Lund string model~\cite{Art:pytHad1, Art:pytHad2}.

\subsection{Underlying-event models}

In the {\sc Herwig} samples, MPIs were simulated using a separate program called {\sc Jimmy} 4.0~\cite{Art:butterworthMPI3, Art:butterworthMPI1, Art:butterworthMPI2, unp:JIMMY4}.  The latter is based on a simple eikonal model that approximates the interacting hadrons as disks that are extended in the transverse plane.  An impact parameter quantifies the degree to which the two disks overlap during the collision.

The {\sc Pythia} MPI model that was used in this paper is known as the ``simple model''~\cite{Art:pythia, *Art:pythia2}.  It estimates the average number of MPIs per event, $\bar n$, as $\bar n = \sigma_H(\hat{p}_T^{\rm min})/\sigma_{\rm ND}(s)$, where $\sigma_H$ and $\sigma_{\rm ND}$ are the inclusive-hard and non-diffractive, inelastic cross sections, respectively, and $\hat{p}_T^{\rm min}$ is a minimal constraint applied to the $p_T$ of the partonic collision.  The number of MPIs is assumed to obey a Poisson distribution.  The probability that the secondary collisions occur with transverse momentum $p_T$ is given by $P(p_T)=(1/\sigma_{\rm ND}(s))({\rm d}\sigma_H/{\rm d}p_T)$. The simple model proceeds by ordering the secondary interactions in terms of $p_T$ and then calculates the probability of each successive scatter.

\subsection{Monte Carlo samples}

The MC samples were used for two purposes: to correct the data for detector effects and to compare to the measured hadron-level cross sections. The MC samples used to correct the data included a full {\sc Geant} 3.13~\cite{unp:geant} simulation of the ZEUS detector and three-level trigger. The resolved and direct MC samples were combined in the ratio that gave the best fit to the $x_\gamma^{\rm obs}$ distribution in the data, significantly improving the overall description of the data by the MC. The resolved and direct samples used to compare to the measured cross sections, however, were combined in the ratio predicted by the MC models.

The {\sc Herwig} samples that were compared to the data were generated with $\hat{p}_T^{\rm min}$ set to 4~GeV and with the CTEQ5L~\cite{Art:cteq5l} and GRV-G LO~\cite{Art:grvgloTN} parameterisations for the proton and photon parton density functions (PDFs), respectively. All other {\sc Herwig} parameters were set to default.  To simulate the {\sc Herwig} MPIs, the {\sc Jimmy} model was run with the minimum $p_T$ of the secondary scatters, $\hat{p}_T^{\rm mpi}$, set to 2.2~GeV, the probability that the photon would resolve via a large hadron-like fluctuation set to $1/340$ and the effective transverse radius of the photon squared was increased by a factor of 4 from its default value. These adjusted MPI settings were the result of simultaneous tuning to the three- and four-jet $x_\gamma^{\rm obs}$ cross sections measured in this analysis, in the low- and high-mass regions.

The {\sc Pythia} samples that were compared to the data were generated with $\hat{p}_T^{\rm min}=4.5$~GeV, $\hat{p}_T^{\rm mpi}=1.9$~GeV, which is the default value, and with the CTEQ5L and GRV-G LO parameterisations for the proton and photon PDFs, respectively.  All other {\sc Pythia} and simple-model parameters were set to default.

%%%%%%%%%%%%%%%%%%%%%%%%%%%%%%%%%%%%%%%%%%%%%%%%%%%%%%%%%%%%%%%%%%%%%%%%

\subsection{Monte Carlo scaling factors}
\label{sec:MCmodels}

Both LO MC simulations underestimated the magnitude of the measured cross sections.  To compare with the shape of the differential cross sections, the MC predictions were either area normalised to the equivalent data set or scaled by the factor required to area normalise the high-mass MC cross sections to the high-mass data, where MPIs are expected to be less influential.  The choice of normalisation procedure will be clearly stated in Section~\ref{sec:results} and in the figure captions.

The scaling factors that were applied to the {\sc Herwig} and {\sc Pythia} cross sections are shown in Table~\ref{tab:mc_scl}.  The scaling factors for the {\sc Pythia} model were significantly larger than for {\sc Herwig}.  Within each model, the factors applied to the three-jet predictions were similar whether MPIs were included or not, indicating that both models expect relatively little effect from MPIs in the high-mass three-jet sample.  In the four-jet case, however, the scaling factors for the high-mass MPI samples were significantly smaller than those for the nominal samples, showing that both models predict a sizable contribution from MPIs in the high-mass four-jet sample.

\section{The $\mathcal{O}(\alpha\alpha_s^2)$ pQCD calculation}

The three-jet differential cross sections were calculated at $\mathcal{O}(\alpha\alpha_s^2)$ using the program by Klasen, Kleinwort and Kramer~\cite{Art:klasen1,Art:klasen2}. The calculation is LO for this process. The renormalisation, $\mu_r$, and factorisation, $\mu_f$, scales were set to $\mu_r=\mu_f=\mu=E_T^{\rm max}$, the $E_T$ of the hardest parton. The theoretical scale uncertainty was evaluated by varying $\mu$, setting it to $2^{\rm \pm1}E_T^{\rm max}$.  The value of $\alpha_s$  was calculated with one-loop precision and assuming five active flavours.  A value of $\Lambda^{(5)}_{\rm \overline{MS}}=181$~MeV was taken.  The CTEQ6L~\cite{Art:cteq6l} proton and GRV-G LO~\cite{Art:grvgnlo, *Art:grvglo} photon PDFs were used for the calculation.

The theoretical calculations were corrected for both hadronisation and MPI effects. The correction factors were obtained using the {\sc Herwig} and {\sc Pythia} models. The hadronisation corrections, $\mathcal{C}_{\rm had}$,  were calculated by taking the bin-by-bin ratio of the MC cross sections at the hadron ($\sigma_{\rm HL}$) and parton levels ($\sigma_{\rm PL}$), $\mathcal{C}_{\rm had}=\sigma_{\rm HL}/\sigma_{\rm PL}$. To obtain $\sigma_{\rm PL}$, the $k_T$ cluster algorithm was run over all partons produced by the parton shower, prior to hadronisation.  The MPI corrections, $\mathcal{C}_{\rm MPI}$, were calculated by taking the bin-by-bin ratio of the hadron-level cross sections, with ($\sigma_{\rm HL}^{\rm MPI}$) and without MPIs ($\sigma_{\rm HL}^{\rm noMPI}$), $\mathcal{C}_{\rm MPI}=\sigma_{\rm HL}^{\rm MPI}/\sigma_{\rm HL}^{\rm noMPI}$. The hadronisation and MPI corrections that were applied to the calculation represented the average corrections taken from the two MC models.  A symmetric uncertainty was associated with the correction factors equal to half the difference between the predictions of the two MC models.  The hadronisation corrections were found to typically reduce the cross section by $\mathcal{O}(20\%)$, whereas the MPI corrections increased the cross section by $\mathcal{O}(50\%)$ in the low-mass region and by $\mathcal{O}(10\%)$ in the high-mass sample.  The uncertainty associated with the MPI corrections was larger than that estimated for the hadronisation factors.

%%%%%%%%%%%%%%%%%%%%%%%%%%%%%%%%%%%%%%%%%%%%%%%%%%%%%%%%%%%%%%%%%%%%%%%%%%%%%%

\section{Acceptance corrections}
\label{sec:unfold}

Detector effects and trigger inefficiencies were accounted for by applying bin-by-bin correction factors, $\mathcal{C}_{\rm det}=\sigma_{\rm HL}/\sigma_{\rm DL}$, where $\sigma_{\rm DL}$ is the predicted detector-level cross section.  The correction factors were extracted separately from the {\sc Herwig} and {\sc Pythia} MC samples with MPIs.  Before calculating the corrections, the resolved components of the MC samples were reweighted in $y_{\rm JB}$ to improve the overall description of the data~\cite{thesis:tim}. The typical bin-by-bin corrections applied to the low- and high-mass three-(four-)jet data were approximately $0.75$ ($0.45$) and $1.3$ ($0.95$), respectively.  The largest correction factors were found in the lowest $E_T^{\rm jet}$ and the extreme $\eta^{\rm jet}$ bins.

Once corrected to the hadron level, the measured cross sections, $\sigma_{\rm \sqrt{s}}$, at $\sqrt{s}=300$~GeV and $\sqrt{s}=318$~GeV were combined using the following formula:

\begin{equation}
\sigma = \frac{\sigma_{\rm 300}\cdot(\sigma_{\rm 318}^{\rm mc}/\sigma_{\rm 300}^{\rm mc})\cdot\mathcal{L}_{\rm 300} + \sigma_{\rm 318}\cdot\mathcal{L}_{\rm 318}}{\mathcal{L}_{\rm 300}+\mathcal{L}_{\rm 318}},
\label{equ:epcom}
\end{equation}

where $\mathcal{L}_{\rm \sqrt{s}}$ is the integrated luminosity and $\sigma_{\rm \sqrt{s}}^{\rm mc}$, the predicted cross section at $\sqrt{s}$. As such, the measured cross sections presented here correspond to $\sqrt{s}=318$~GeV.  The $\sigma_{\rm 318}^{\rm mc}/\sigma_{\rm 300}^{\rm mc}$ ratios were $\sim 1.1$ for all but the low-mass four-jet sample, where it was around 10\% larger.

The results presented here represent the average hadron-level cross sections obtained when the data is treated with {\sc Herwig} or {\sc Pythia}, with half the spread interpreted as symmetric systematic uncertainty.  This systematic uncertainty was added in quadrature to those described below.

%%%%%%%%%%%%%%%%%%%%%%%%%%%%%%%%%%%%%%%%%%%%%%%%%%%%%%%%%%%%%%%%%%%%%%%%%%%%%%

\section{Systematic uncertainties}
\label{sec:syst}
A detailed study of the sources of systematic uncertainty associated with the measurement was performed using the {\sc Herwig} MC sample~\cite{thesis:tim}.  The sources contributing to the quoted systematic uncertainties are,

\begin{itemize}
\item the CAL energy scale uncertainty is $\pm3\%$. The energy scale was changed in the MC simulation accordingly;
\item the uncertainty due to the dependence of the measured cross sections on the detector-level selection criteria, which were varied up and down by the detector resolution for the data and the MC samples together.  More specifically,
\begin{itemize}
\item the $E_T^{\rm jet}$ cut on each of the jets was changed by $\pm1$~GeV;
\item the $|\eta^{\rm jet}|$ cut on each of the jets was changed by $\pm0.04$;
\item the $M_{nj}$ selection criteria were changed by $\pm10\%$;
\item the lower and upper cut on $y_{\rm JB}$ was changed by $\pm0.03$ and $\pm0.05$, respectively;
\item the cut on $y_e$, used to differentiate between real and fake scattered-electron candidates, was changed by $\pm0.05$.
\end{itemize}
\end{itemize}

In addition, it was verified that the bin-by-bin acceptance corrections were not sensitive to variations in the relative amount of direct and resolved events or the exact shape of the $y_{\rm JB}$ distribution in the combined MC sample.

All the systematic uncertainties were added in quadrature except that associated with the CAL energy scale.  The CAL energy scale uncertainty is highly correlated between bins and is displayed separately in the plots presented here.

The largest systematic uncertainty in the cross sections, except for the high-mass three-jet case, was that associated with the CAL energy scale, which led to an uncertainty of approximately $\pm10\%$ in both the low- and high-mass three-jet cross sections, and $\pm20\%$ and $\pm15\%$ in the low- and high-mass four-jet cross sections, respectively.  The largest uncertainty in the high-mass three-jet cross section came from the choice of MC model used to calculate the acceptance corrections and was approximately $\pm10\%$.  In the low-mass three-jet sample, the uncertainty due to the MC model was $\pm5\%$ and in the four-jet low- and high-mass samples the values were $\pm3\%$ and $\pm4\%$, respectively. Another significant source of systematic uncertainty in the low-mass samples was that associated with the $E_T^{\rm jet}$ selection criteria, which generated a $\pm4\%$ and $\pm8\%$ effect in the three- and four-jet cases, respectively. In the high-mass samples, varying the $M_{nj}$ selection criteria generated a $\pm8\%$ and $\pm6\%$ effect in the three- and four-jet cross sections, respectively.  The high-mass four-jet cross section was also sensitive to varying the $E_T^{\rm jet}$ cut, which changed its value by about $\pm4\%$.

%%%%%%%%%%%%%%%%%%%%%%%%%%%%%%%%%%%%%%%%%%%%%%%%%%%%%%%%%%%%%%%%%%%%%%%%%%%%%%

\section{Results and discussion}
\label{sec:results}

The three- and four-jet photoproduction cross sections are presented here at the hadron level for jets with $E_T^{\rm jet}\ge6$~GeV and $|\eta^{\rm jet}|\le 2.4$, in the kinematic region given by $Q^2<1~{\rm GeV^2}$ and $0.2\le y\le0.85$, and in low- ($25\le M_{nj}<50$~GeV) and high-mass ($M_{nj}\ge50$~GeV) regions.

\subsection{The ${\rm d}\sigma/{\rm d}M_{nj}$ cross section}

\begin{figure}
\begin{center}
\vspace{-2.0cm}
\hspace{0.7cm}
\includegraphics*[angle=0,scale=0.5]{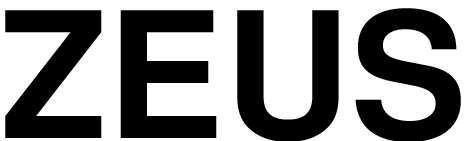}
\vspace{0.0cm}
\hspace{-0.7cm}

\includegraphics*[angle=0,scale=0.5]{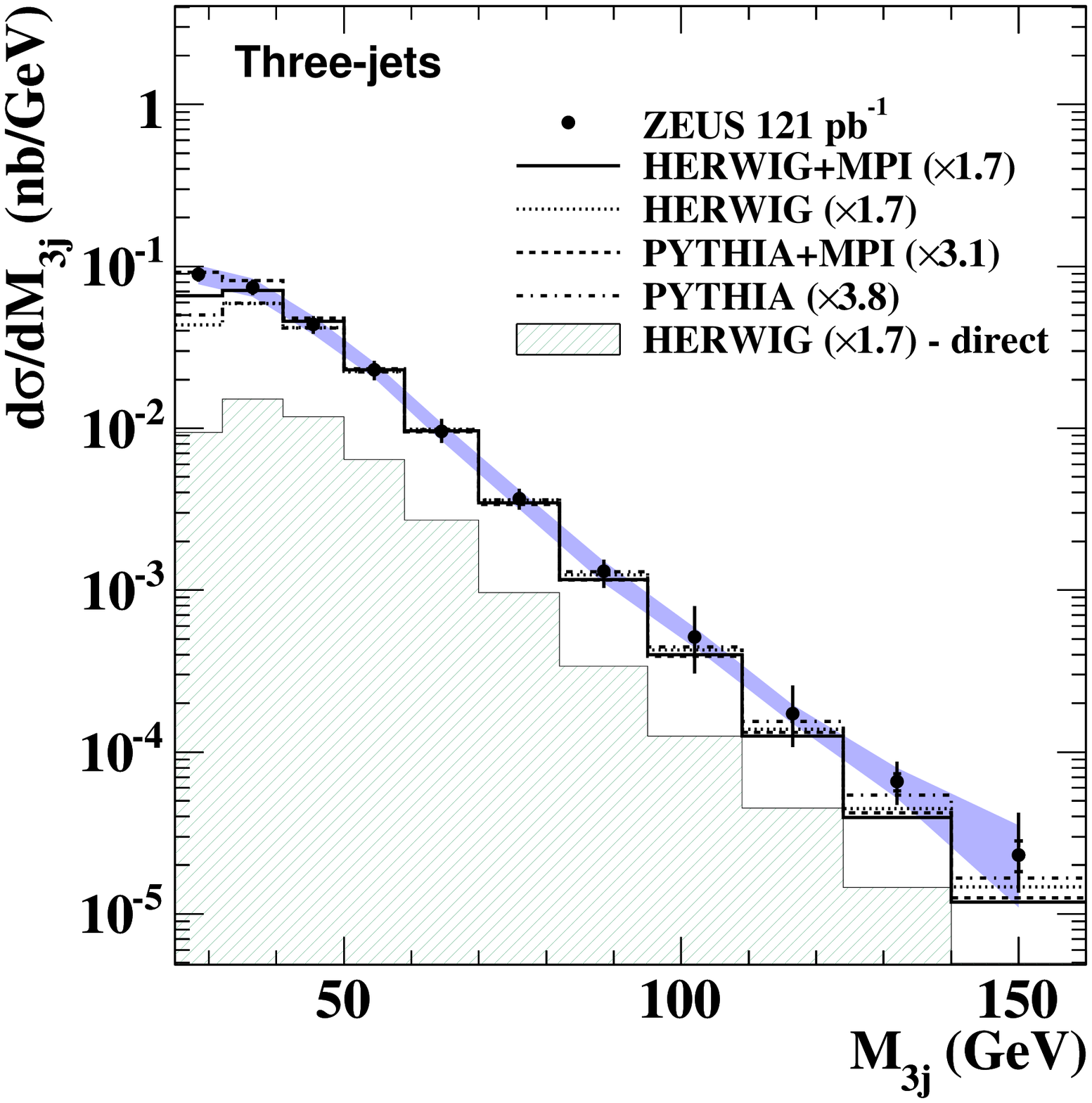}

\vspace{-9.2cm}\hspace{7.6cm}(a)\vspace{8.4cm}

\includegraphics*[angle=0,scale=0.5]{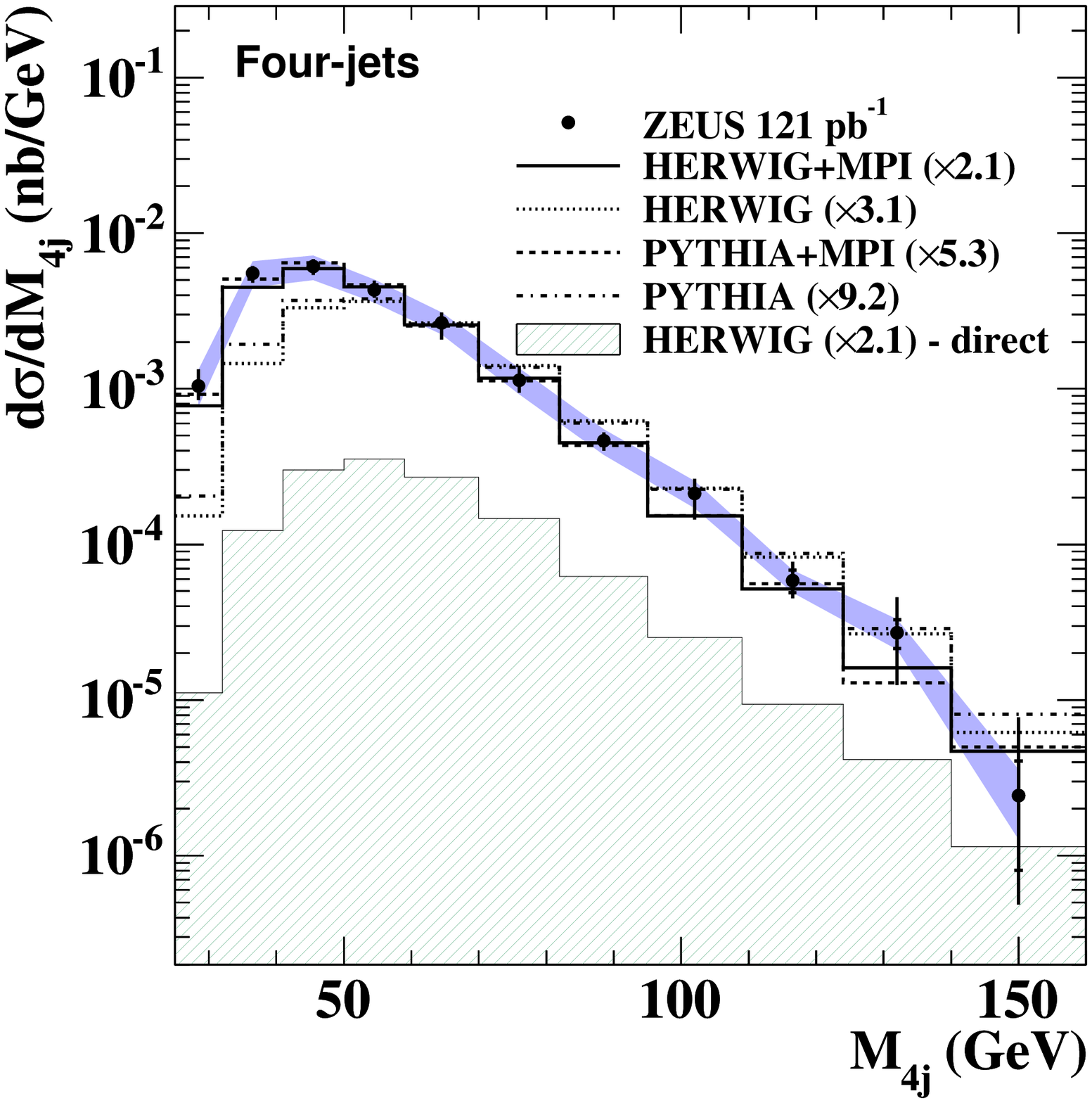}

\vspace{-9.2cm}\hspace{7.6cm}(b)\vspace{8.4cm}
\caption{
Measured cross section as a function of (a) $M_{3j}$ and (b) $M_{4j}$ for the three- and four-jet samples (solid circles). The inner error bars represent the statistical uncertainty.  The outer error bars represent the statistical plus systematic uncertainties added in quadrature. The shaded band represents the calorimeter energy scale uncertainty. The predictions from the {\sc Herwig} and {\sc Pythia} models are also shown both with and without MPIs, as is the direct component of the {\sc Herwig} cross sections.  Each Monte Carlo cross section has been area normalised to the high-mass ($M_{nj}\ge50$~GeV) measured cross section by scaling the predictions by the factors indicated in the legend.
\label{fig:xsec:Mnj}}
\end{center}
\end{figure}

The three- and four-jet cross sections are given as a function of $M_{nj}$ in Fig.~\ref{fig:xsec:Mnj} and Tables~\ref{tab:Mnj_3j_} and \ref{tab:Mnj_4j_}.  In general, both cross sections decrease exponentially with increasing $M_{nj}$.  Also shown in Fig.~\ref{fig:xsec:Mnj} are the {\sc Herwig} and {\sc Pythia} predictions without MPIs, normalised to the high-mass region ($M_{nj}\ge50$~GeV).  Both models incorrectly describe the $M_{nj}$ dependence of the cross section and significantly underestimate the low mass data. The discrepancy is larger in the four-jet case.  With the inclusion of MPIs, both scaled MC predictions give a reasonably good description of the data over the full $M_{nj}$ ranges.

\subsection{The ${\rm d}\sigma/{\rm d}x_\gamma^{\rm obs}$ cross section}

\begin{figure}
\begin{center}
\vspace{-0.4cm}
\hspace{0.6cm}
\includegraphics*[angle=0,scale=0.5]{DESY-07-102_0.eps}
\vspace{0.4cm}
\hspace{-0.6cm}

\includegraphics*[angle=0,scale=0.40]{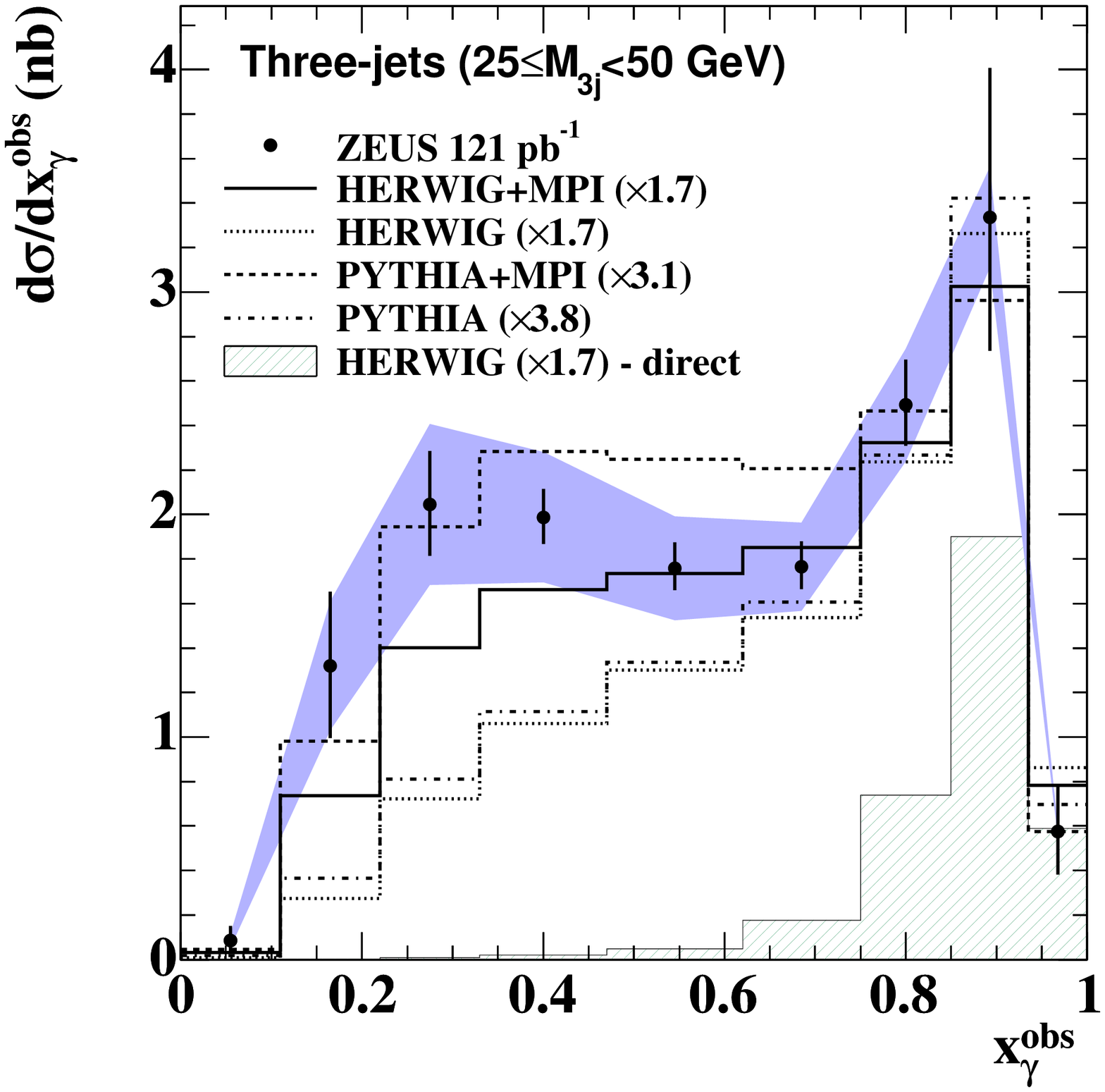}
\hspace{-0.5cm}
\includegraphics*[angle=0,scale=0.40]{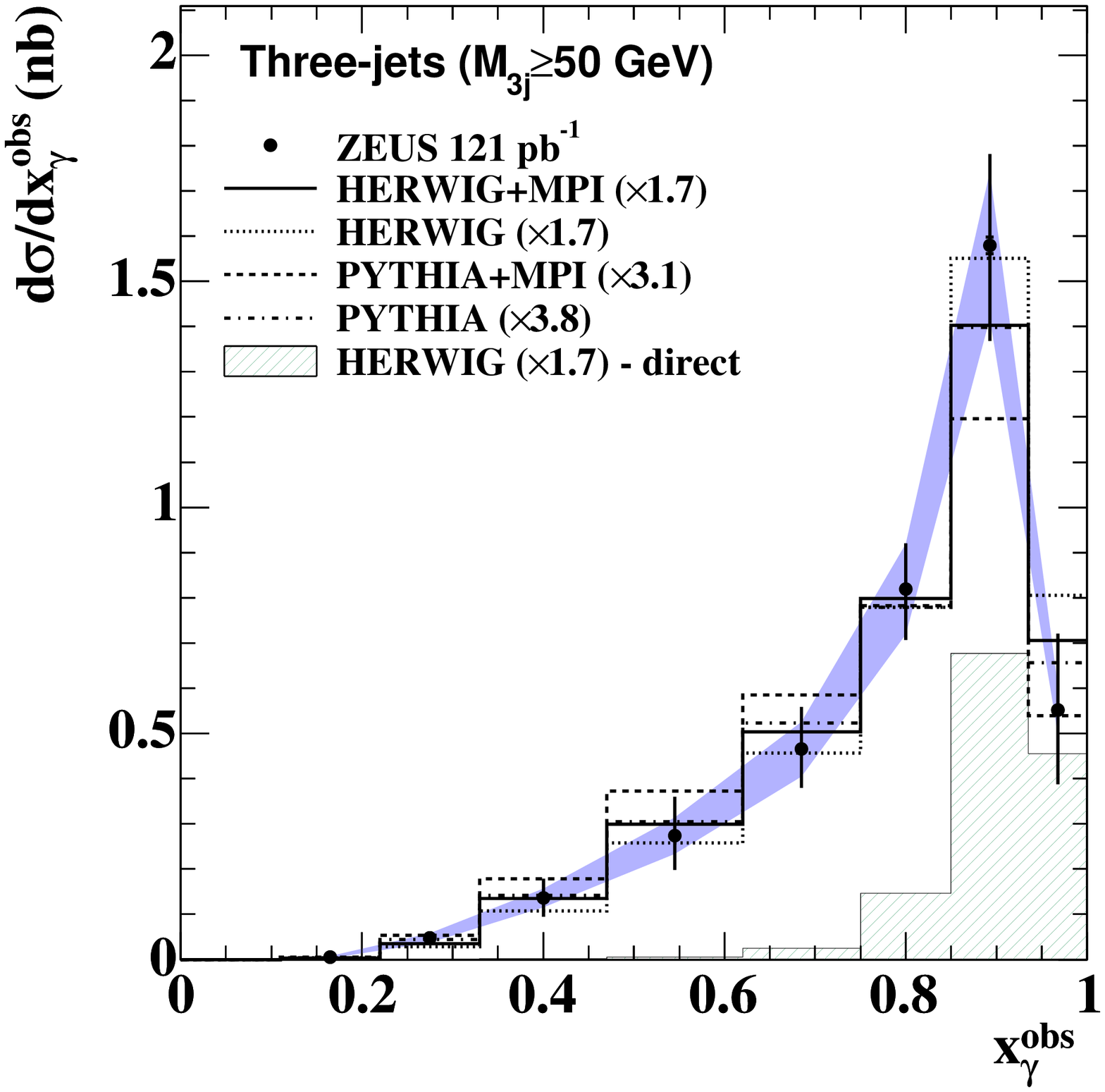}

\vspace{-7.4cm}\hspace{6.2cm}(a)\hspace{7.3cm}(b)\vspace{6.6cm}

\includegraphics*[angle=0,scale=0.40]{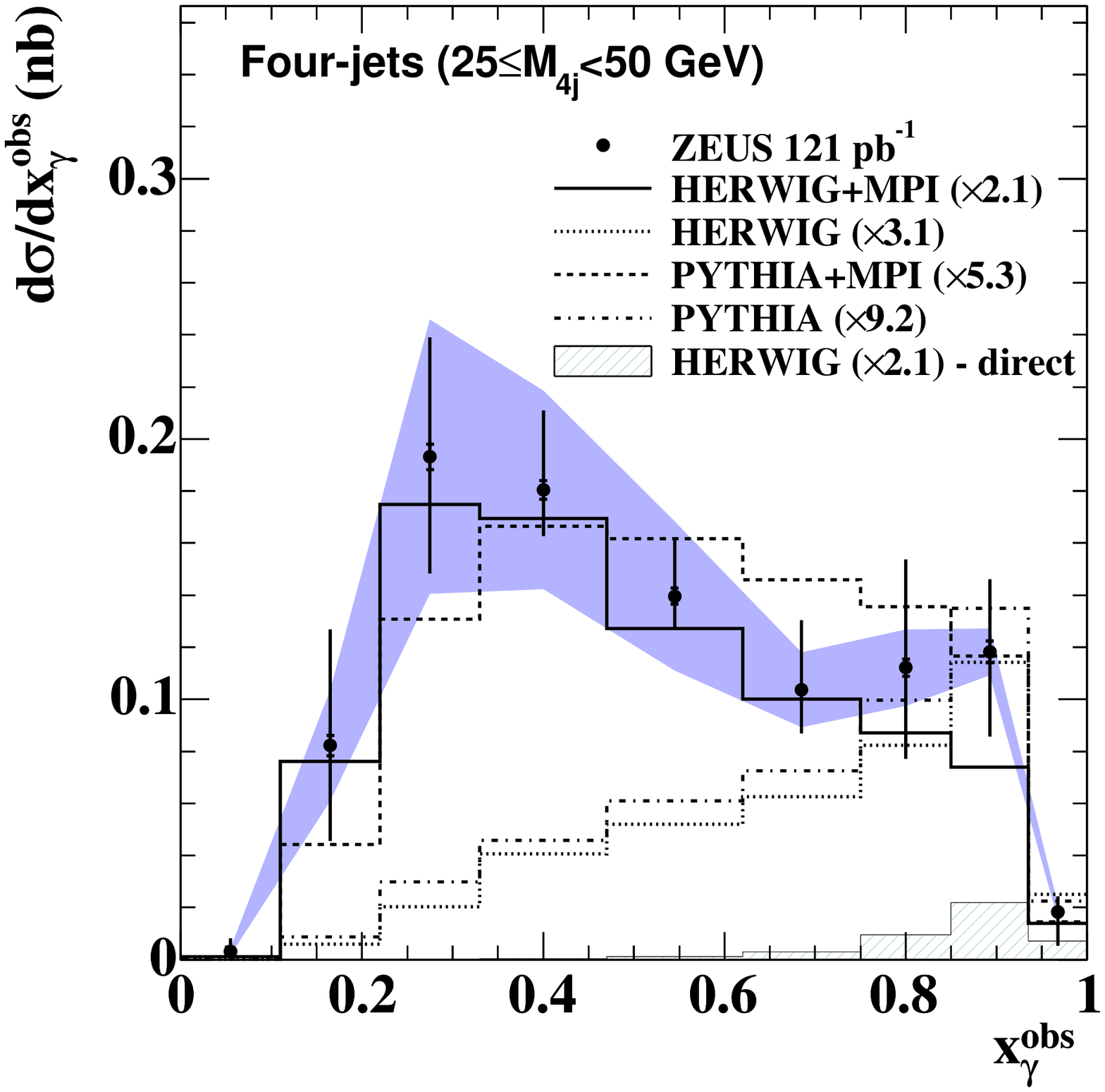}
\hspace{-0.5cm}
\includegraphics*[angle=0,scale=0.40]{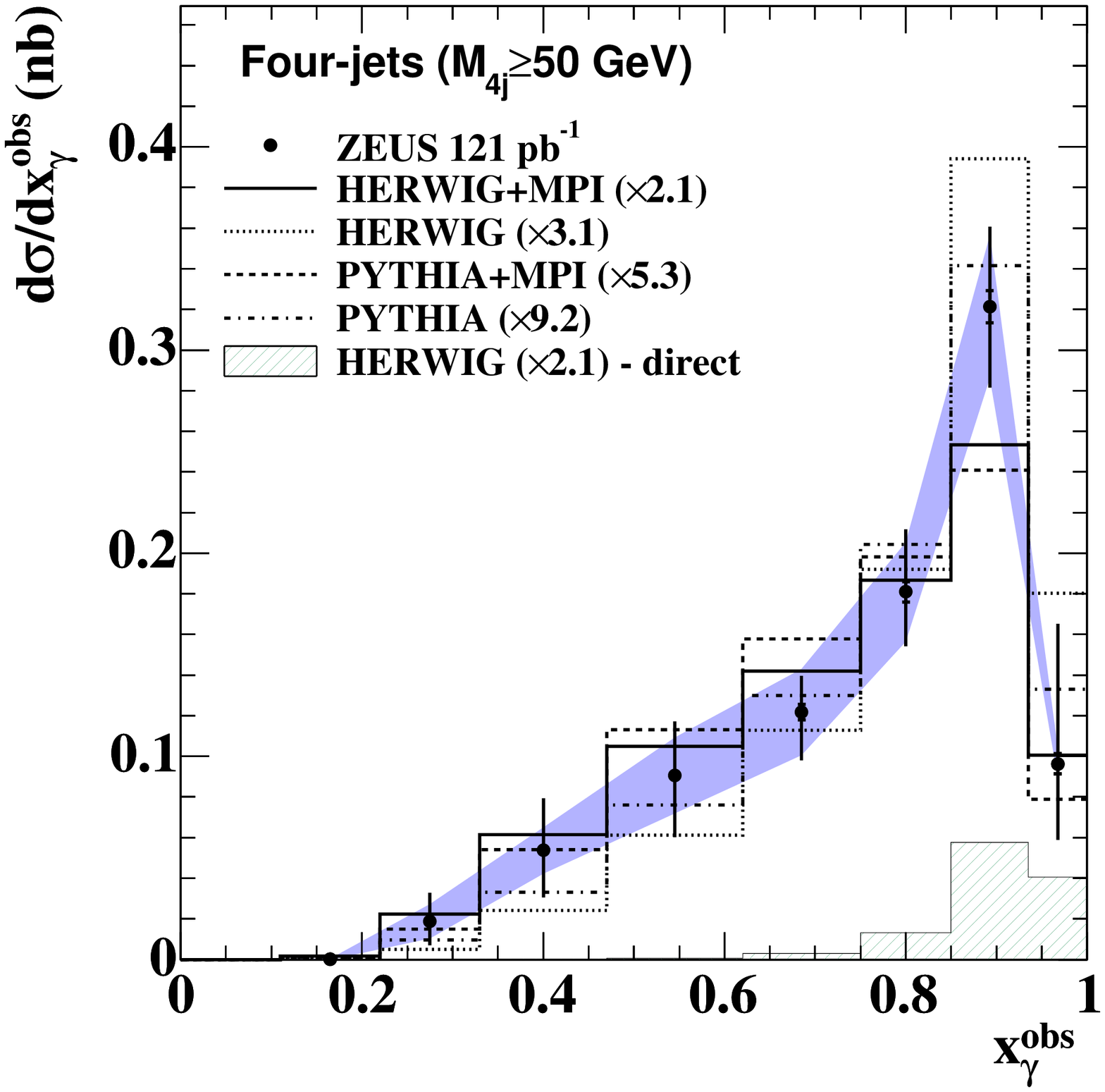}

\vspace{-7.4cm}\hspace{6.2cm}(c)\hspace{7.3cm}(d)\vspace{6.6cm}
\caption{
Measured cross section as a function of $x_\gamma^{obs}$ in the three-jet (a) low- and (b) high-mass samples and the four-jet, (c) low- and (d) high-mass samples. Other details as in the caption to Fig.~\ref{fig:xsec:Mnj}.
\label{fig:xsec:Xgam}}
\end{center}
\end{figure}

The three- and four-jet cross sections are given as a function of $x_\gamma^{\rm obs}$ for the low- and high-mass samples in Fig.~\ref{fig:xsec:Xgam} and Tables~\ref{tab:X_gamma_3j_} and \ref{tab:X_gamma_4j_}.  The distributions exhibit a peak at $x_\gamma^{\rm obs}\approx0.9$ in all but the low-mass four-jet sample.  The low-mass cross sections show enhancement at low $x_\gamma^{\rm obs}$ compared to those at high-mass, which is a consequence of the tighter kinematic constraints at high mass.  The degree to which the low $x_\gamma^{\rm obs}$ region is enhanced is larger in the four-jet case.  Shown also in Fig.~\ref{fig:xsec:Xgam} are the scaled MC predictions compared to the data.

The MC predictions show that, in the absence of MPIs, the three- and four-jet cross sections are expected to decrease with decreasing $x_\gamma^{\rm obs}$ in both mass regions.  Moreover, the low-$x_\gamma^{\rm obs}$ suppression is expected to be marginally stronger in the four-jet case.  The low-mass data, however, are in contradiction to both predictions.  These data therefore suggest that some mechanism in addition to the processes modelled in the MC without MPIs, is contributing to the low-mass cross sections.  One possible mechanism is MPIs.

A comparison of the data with the MC predictions, normalised to the high-mass data, shows that both those with and without MPIs are in reasonable agreement with the high-mass cross sections (Figs.~\ref{fig:xsec:Xgam}b and \ref{fig:xsec:Xgam}d).  In the low-mass low-$x_\gamma^{\rm obs}$ regions that are poorly described by the nominal MC models, the introduction of MPIs to the simulations aids the description of the data.  More specifically, the {\sc Herwig} prediction with tuned MPIs describes the data reasonably well in all of the samples. The default {\sc Pythia} model tends to overestimate the data in the mid-$x_\gamma^{\rm obs}$ region ($0.4\lesssim x_\gamma^{\rm obs}\lesssim 0.8$) in the low-mass samples but describes the cross section reasonably well elsewhere.

Shown also in Fig.~\ref{fig:xsec:Xgam} is the contribution to the {\sc Herwig} predictions attributed to direct processes, as defined in that LO model with parton showers.  In all four samples, direct events are predicted to be found solely at high $x_\gamma^{\rm obs}$, although resolved processes are expected to make a significant, and in the four-jet case dominant, contribution even at high $x_\gamma^{\rm obs}$.  This is in contrast to the behaviour of dijet photoproduction, where the high-$x_\gamma^{\rm obs}$ region is dominated by direct events~\cite{Art:zeusxg}.

\subsection{The ${\rm d}\sigma/{\rm d}y$ cross section}

\begin{figure}
\begin{center}
\vspace{-0.4cm}
\hspace{0.6cm}
\includegraphics*[angle=0,scale=0.5]{DESY-07-102_0.eps}
\vspace{0.4cm}
\hspace{-0.6cm}

\includegraphics*[angle=0,scale=0.40]{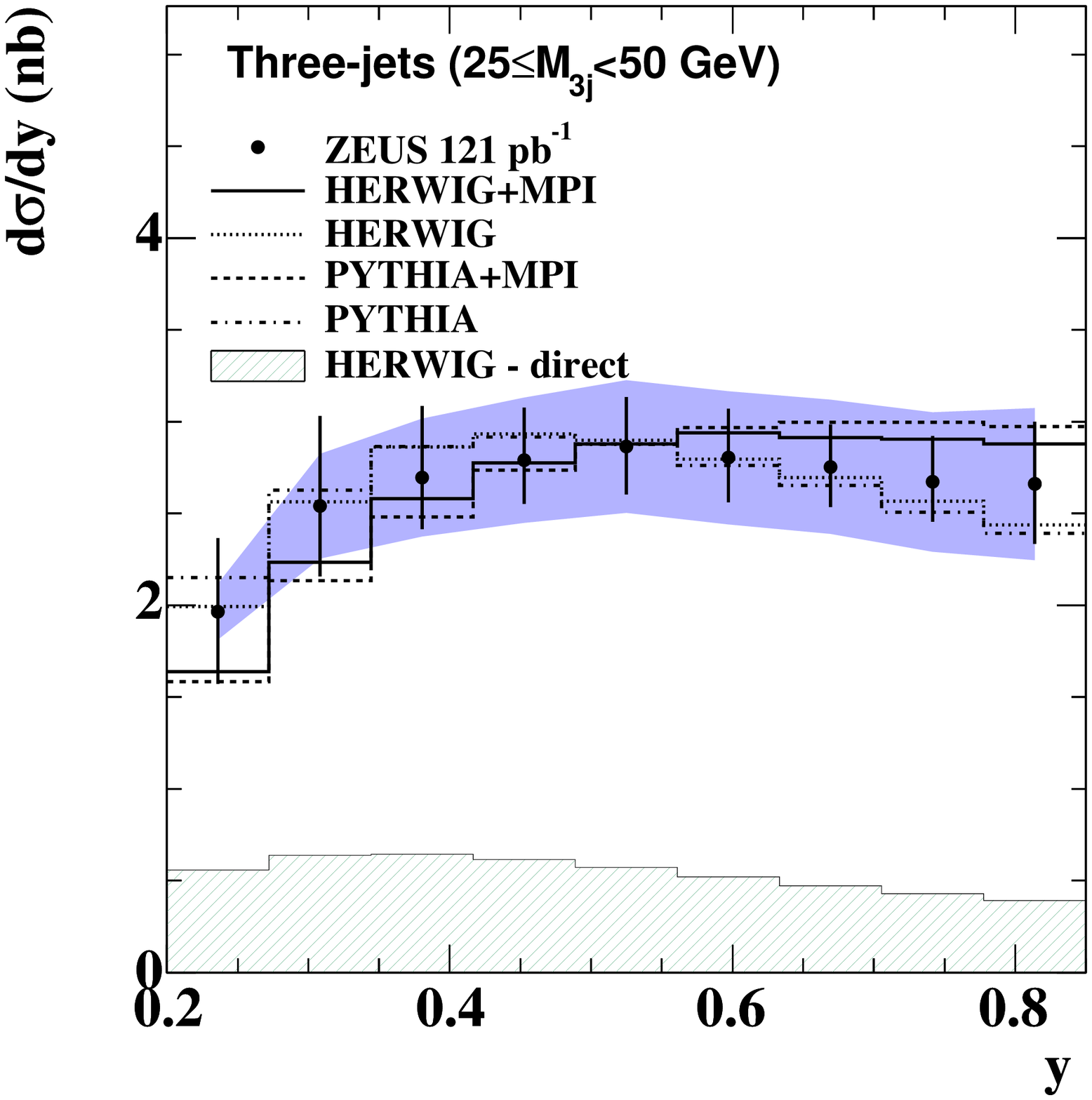}
\hspace{-0.5cm}
\includegraphics*[angle=0,scale=0.40]{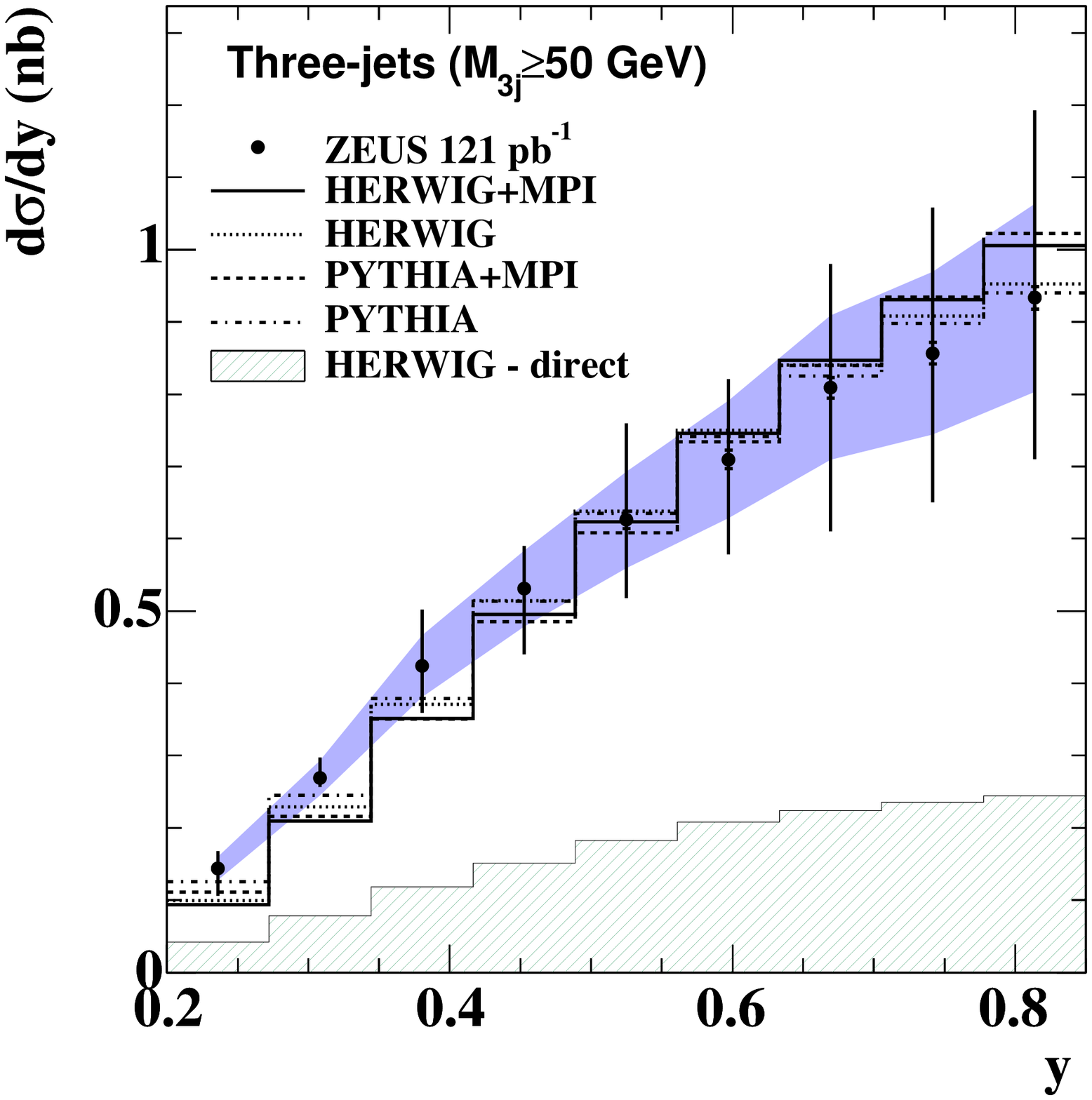}

\vspace{-7.4cm}\hspace{6.2cm}(a)\hspace{7.3cm}(b)\vspace{6.6cm}

\includegraphics*[angle=0,scale=0.40]{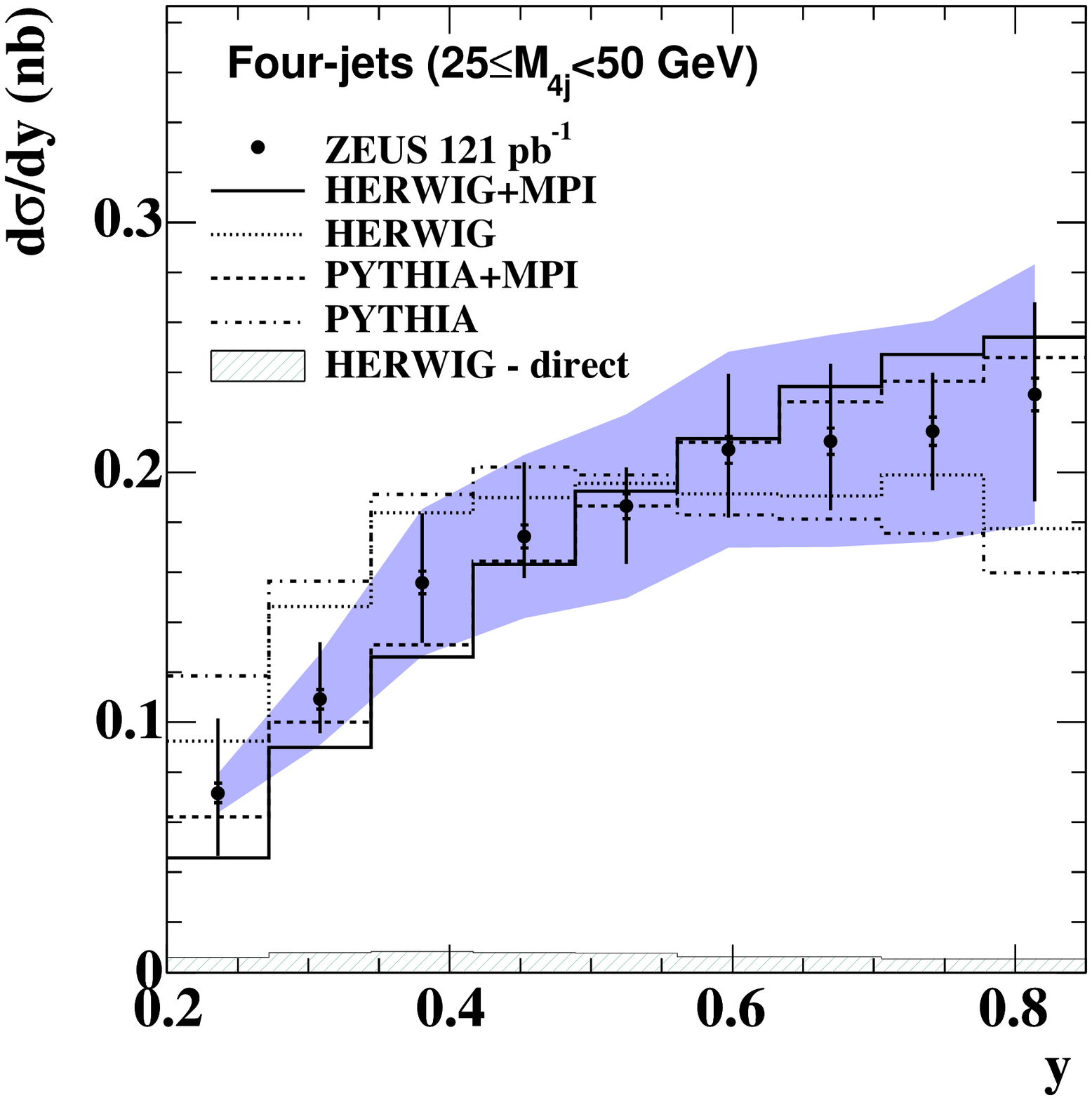}
\hspace{-0.5cm}
\includegraphics*[angle=0,scale=0.40]{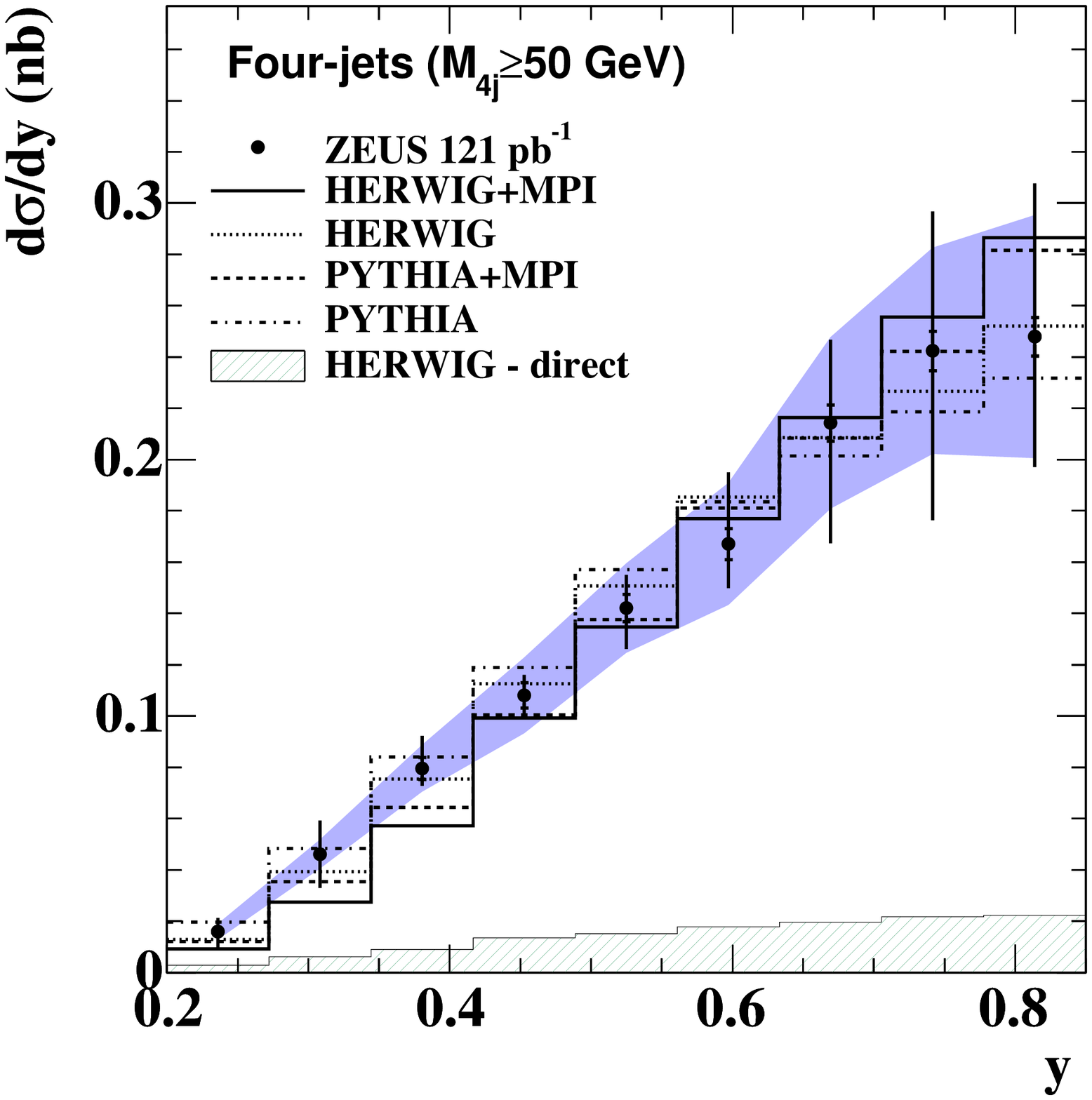}

\vspace{-7.4cm}\hspace{6.2cm}(c)\hspace{7.3cm}(d)\vspace{6.6cm}
\caption{
Measured cross section as a function of $y$ in the three-jet (a) low- and (b) high-mass samples and the four-jet, (c) low- and (d) high-mass samples. The predictions from the {\sc Herwig} and {\sc Pythia} models are also shown both with and without MPIs, as is the direct component of the {\sc Herwig} cross sections.  Each Monte Carlo cross section has been area normalised to the data.  Other details as in the caption to Fig.~\ref{fig:xsec:Mnj}.
\label{fig:xsec:yJB}}
\end{center}
\end{figure}

The three- and four-jet cross sections are given as a function of $y$ in the low- and high-mass samples in Fig.~\ref{fig:xsec:yJB} and Tables~\ref{tab:Yjb_3j_} and \ref{tab:Yjb_4j_}. The cross sections are observed to increase steeply with $y$ in all but the low-mass three-jet sample where it is approximately constant.  The shapes of the distributions are governed by the available kinematic phase space and the energy distribution of the photon flux: while the phase space increases with $y$, the photon flux decreases with increasing $y$.

Also shown in Fig.~\ref{fig:xsec:yJB} are the predictions from {\sc Herwig} and {\sc Pythia}, with and without MPIs, which have been area normalised to the data.  At high mass, the predictions with and without MPIs are similar and describe the data well.  In the low-mass samples, MPIs are predicted to cause a more steeply increasing cross section.  However, the data are not precise enough to differentiate between the various models and all of the MC predictions roughly describe the data.

\subsection{The ${\rm d}\sigma/{\rm d}E_T^{\rm jet}$ and ${\rm d}\sigma/{\rm d}\eta^{\rm jet}$ cross sections}

\begin{figure}[thb]
\begin{center}
\vspace{-2.0cm}
\hspace{0.7cm}
\includegraphics*[angle=0,scale=0.5]{DESY-07-102_0.eps}
\vspace{0.0cm}
\hspace{-0.7cm}

\includegraphics*[angle=0,scale=0.5]{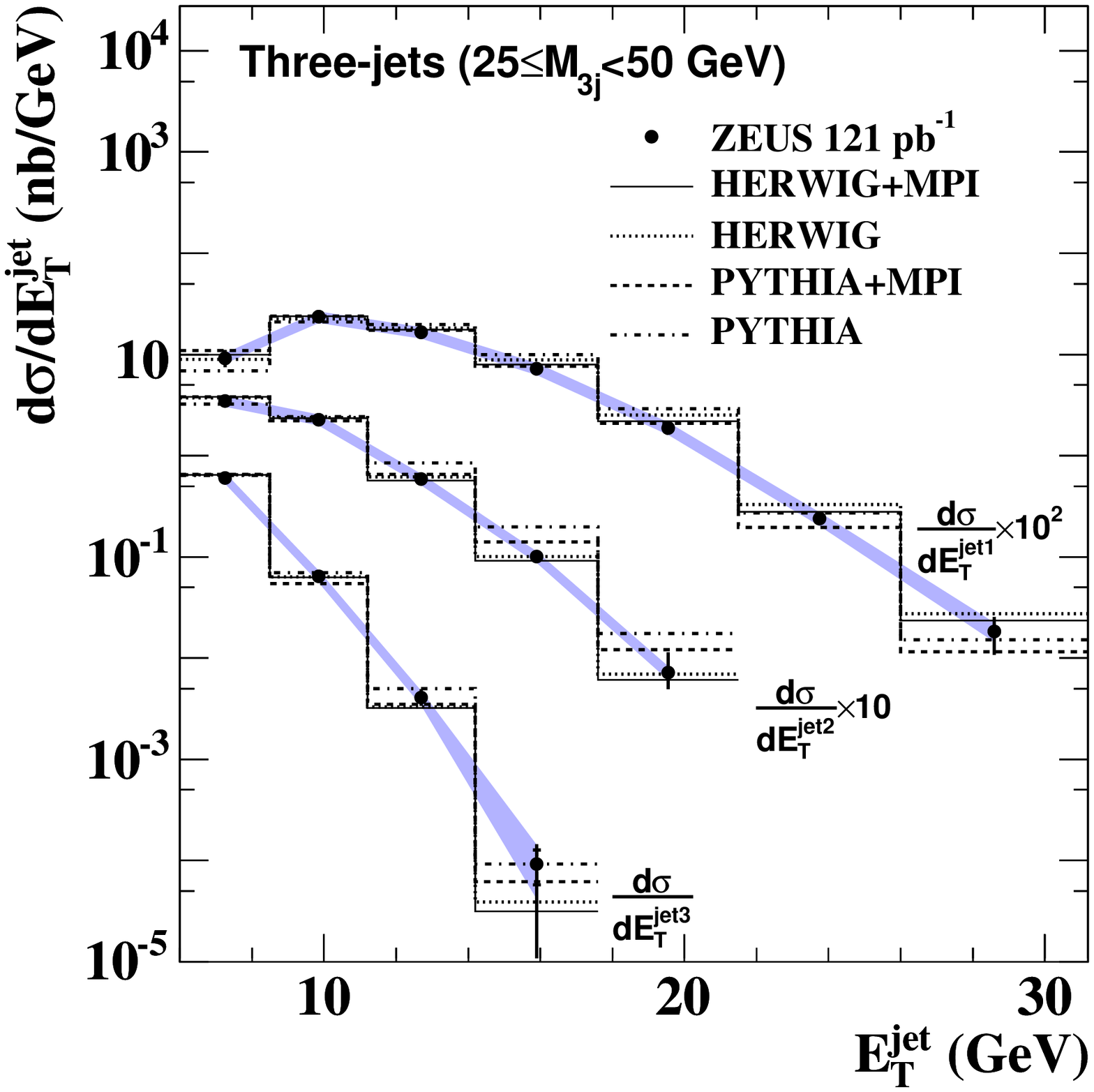}

\vspace{-9.2cm}\hspace{7.6cm}(a)\vspace{8.4cm}

\includegraphics*[angle=0,scale=0.5]{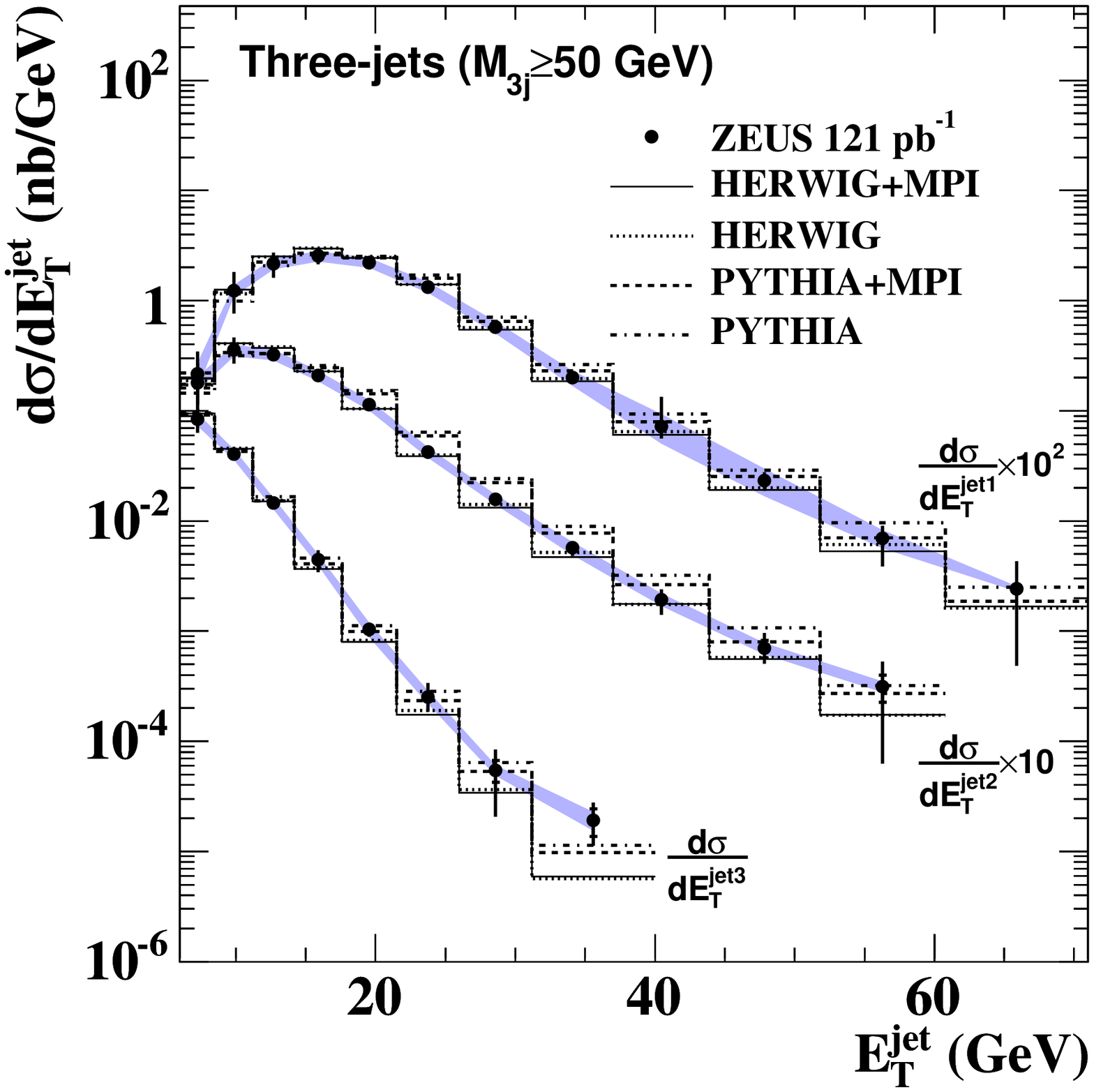}

\vspace{-9.2cm}\hspace{7.6cm}(b)\vspace{8.4cm}
\caption{
Measured cross section as a function of $E_T^{jet}$ in the three-jet (a) low- and (b) high-mass samples. The jets are ordered such that $E_T^{jet1}>E_T^{jet2}>E_T^{jet3}$.  Each cross section has been multiplied by $10^n$, where $n$ is given on the plot, to aid visibility.  Other details as in the caption to Fig.~\ref{fig:xsec:yJB}.
\label{fig:xsec:Et_3j}}
\end{center}
\end{figure}

\begin{figure}[thb]
\begin{center}
\vspace{-2.0cm}
\hspace{0.7cm}
\includegraphics*[angle=0,scale=0.5]{DESY-07-102_0.eps}
\vspace{0.0cm}
\hspace{-0.7cm}

\includegraphics*[angle=0,scale=0.5]{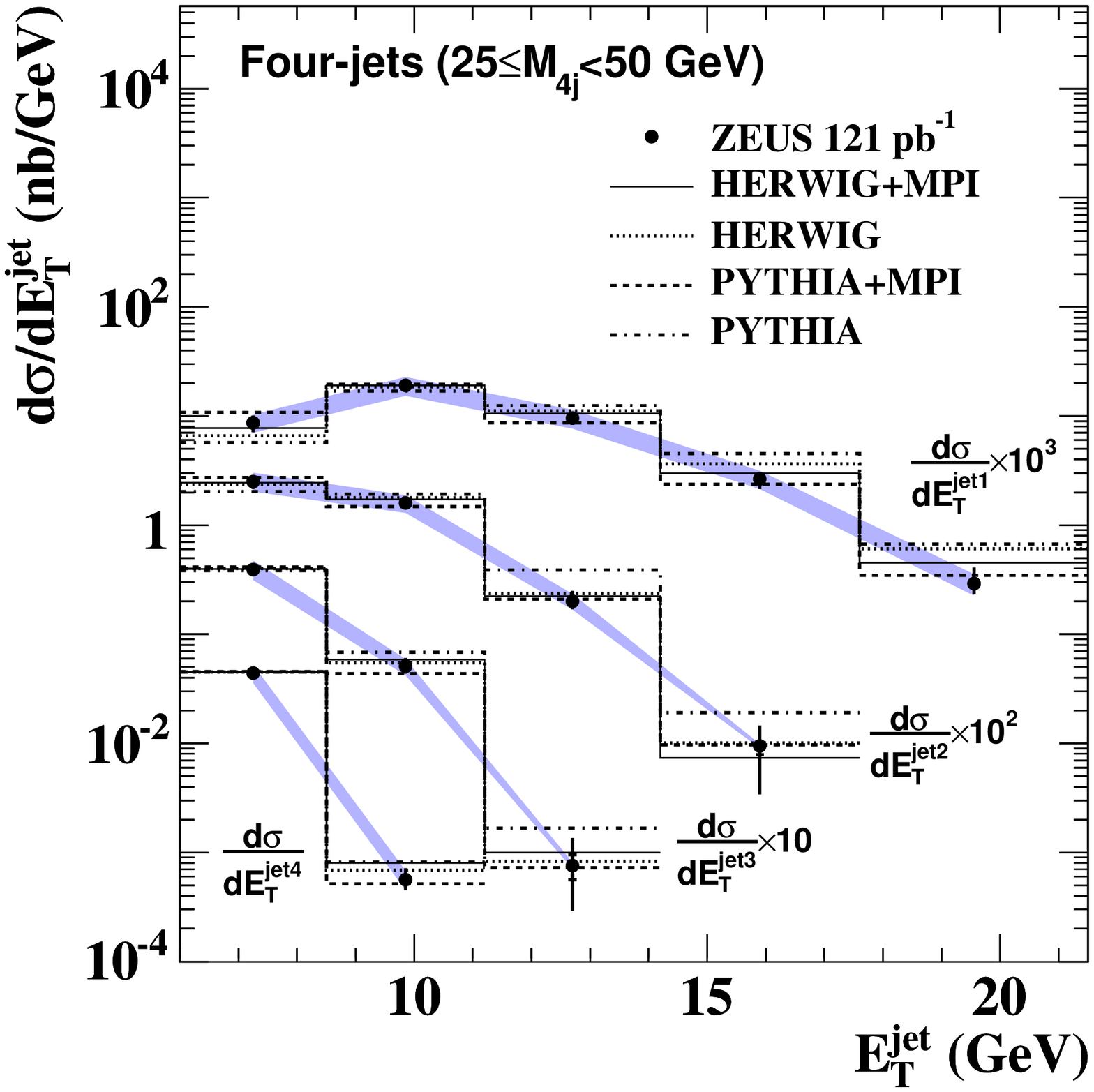}

\vspace{-9.2cm}\hspace{7.6cm}(a)\vspace{8.4cm}

\includegraphics*[angle=0,scale=0.5]{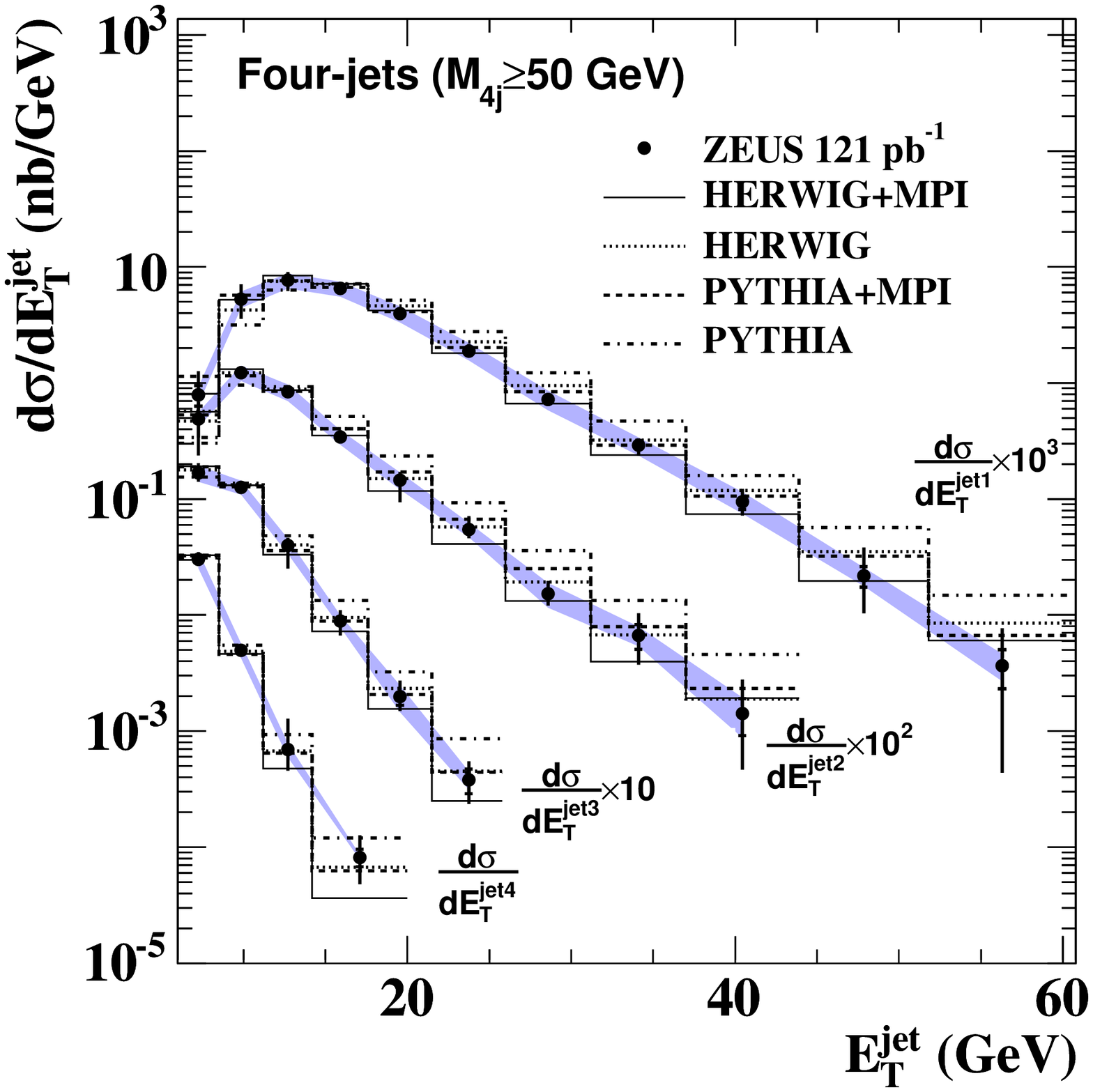}

\vspace{-9.2cm}\hspace{7.6cm}(b)\vspace{8.4cm}
\caption{
Measured cross section as a function of $E_T^{jet}$ in the four-jet (a) low- and (b) high-mass samples. The jets are ordered such that $E_T^{jet1}>E_T^{jet2}>E_T^{jet3}>E_T^{jet4}$. Each cross section has been multiplied by $10^n$, where $n$ is given on the plot, to aid visibility.  Other details as in the caption to Fig.~\ref{fig:xsec:Et_3j}.
\label{fig:xsec:Et_4j}}
\end{center}
\end{figure}

\begin{figure}[thb]
\begin{center}
\vspace{-2.0cm}
\hspace{0.7cm}
\includegraphics*[angle=0,scale=0.5]{DESY-07-102_0.eps}
\vspace{0.0cm}
\hspace{-0.7cm}

\includegraphics*[angle=0,scale=0.5]{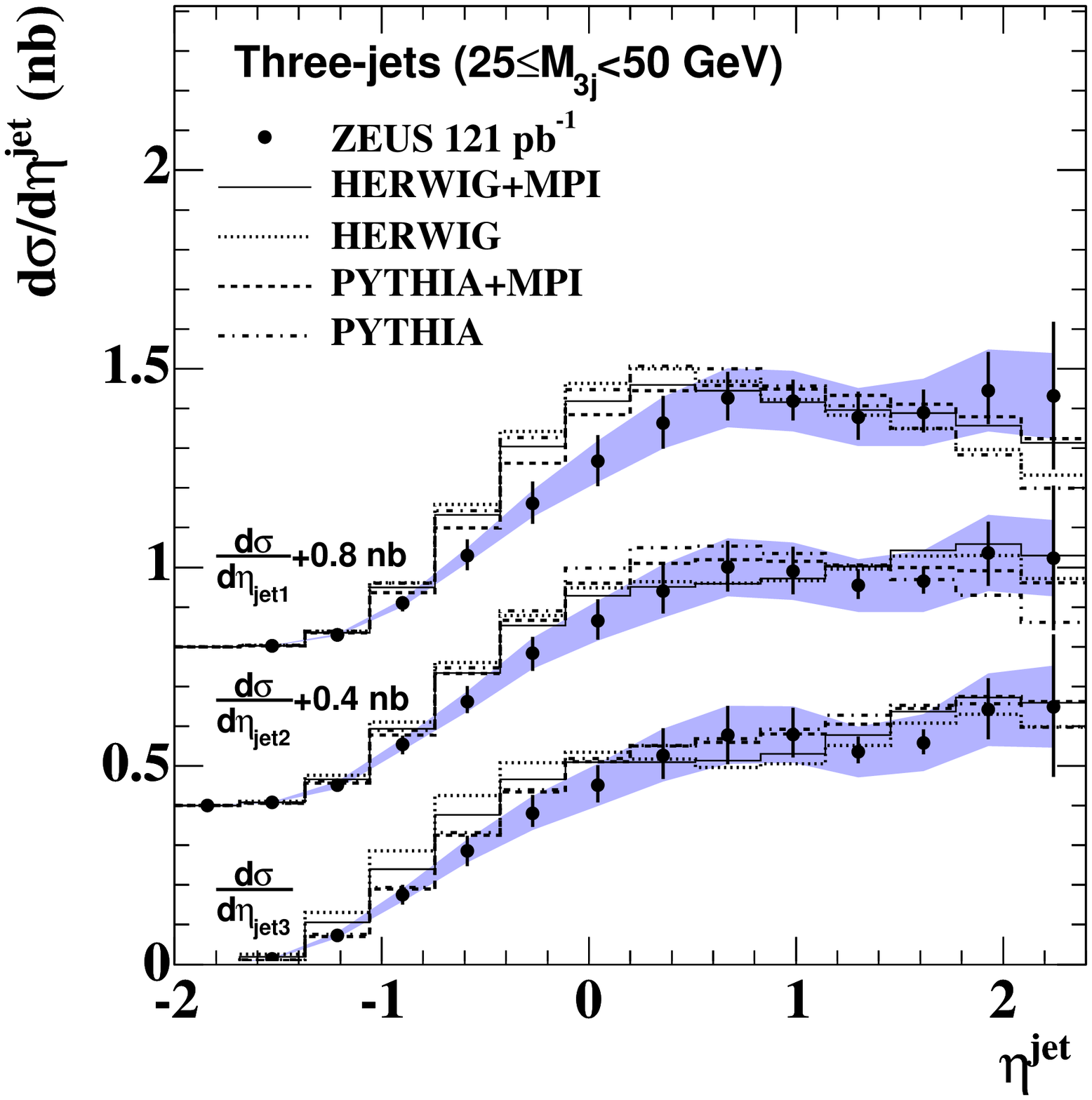}

\vspace{-9.2cm}\hspace{7.6cm}(a)\vspace{8.4cm}

\includegraphics*[angle=0,scale=0.5]{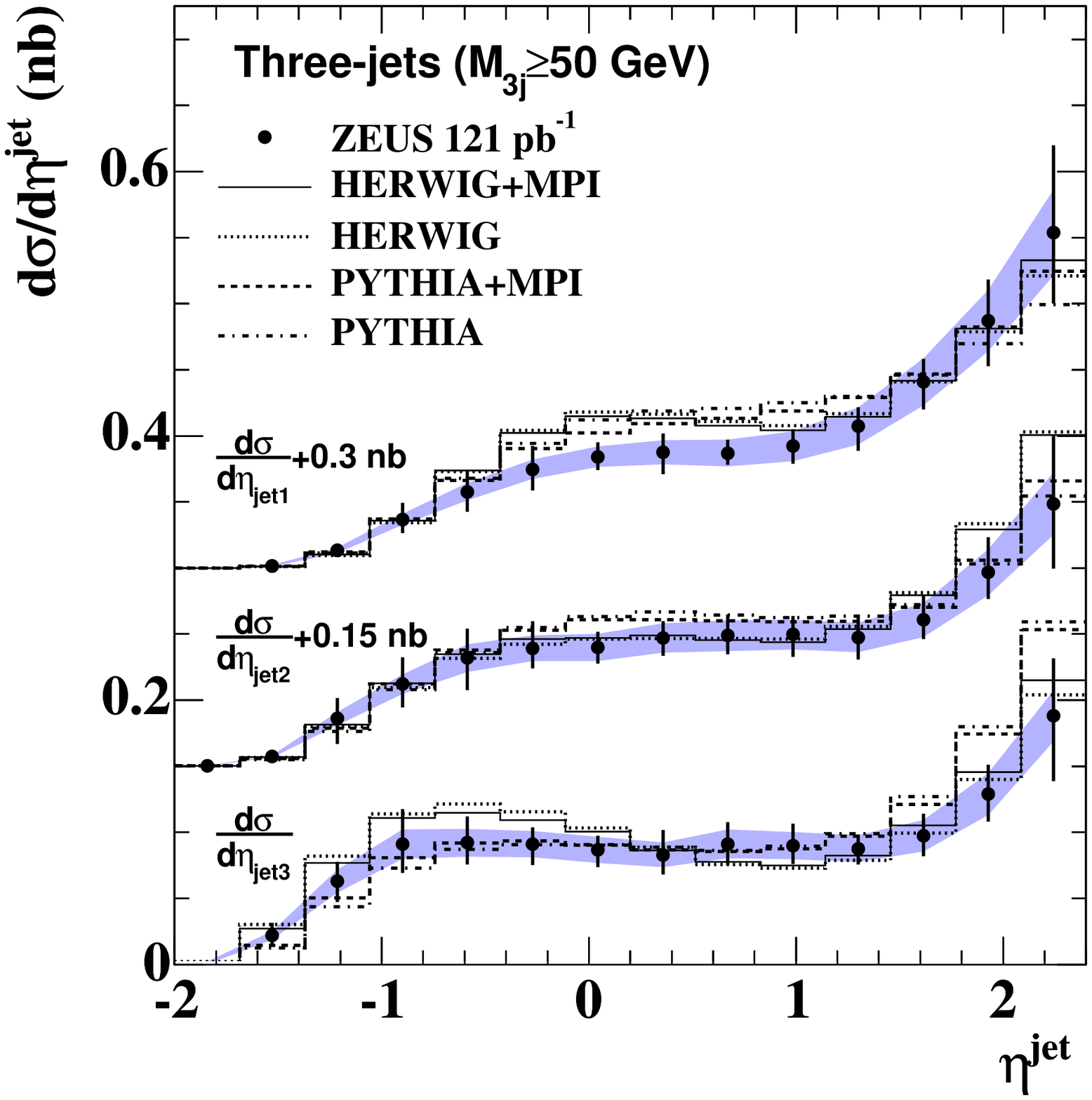}

\vspace{-9.2cm}\hspace{7.6cm}(b)\vspace{8.4cm}
\caption{
Measured cross section as a function of $\eta^{jet}$ in the three-jet (a) low- and (b) high-mass samples. Each cross section has been shifted upwards by the amount given on the plot, to aid visibility.  Other details as in the caption to Fig.~\ref{fig:xsec:Et_3j}.
\label{fig:xsec:eta_3j}}
\end{center}
\end{figure}

\begin{figure}[thb]
\begin{center}
\vspace{-2.0cm}
\hspace{0.7cm}
\includegraphics*[angle=0,scale=0.5]{DESY-07-102_0.eps}
\vspace{0.0cm}
\hspace{-0.7cm}

\includegraphics*[angle=0,scale=0.5]{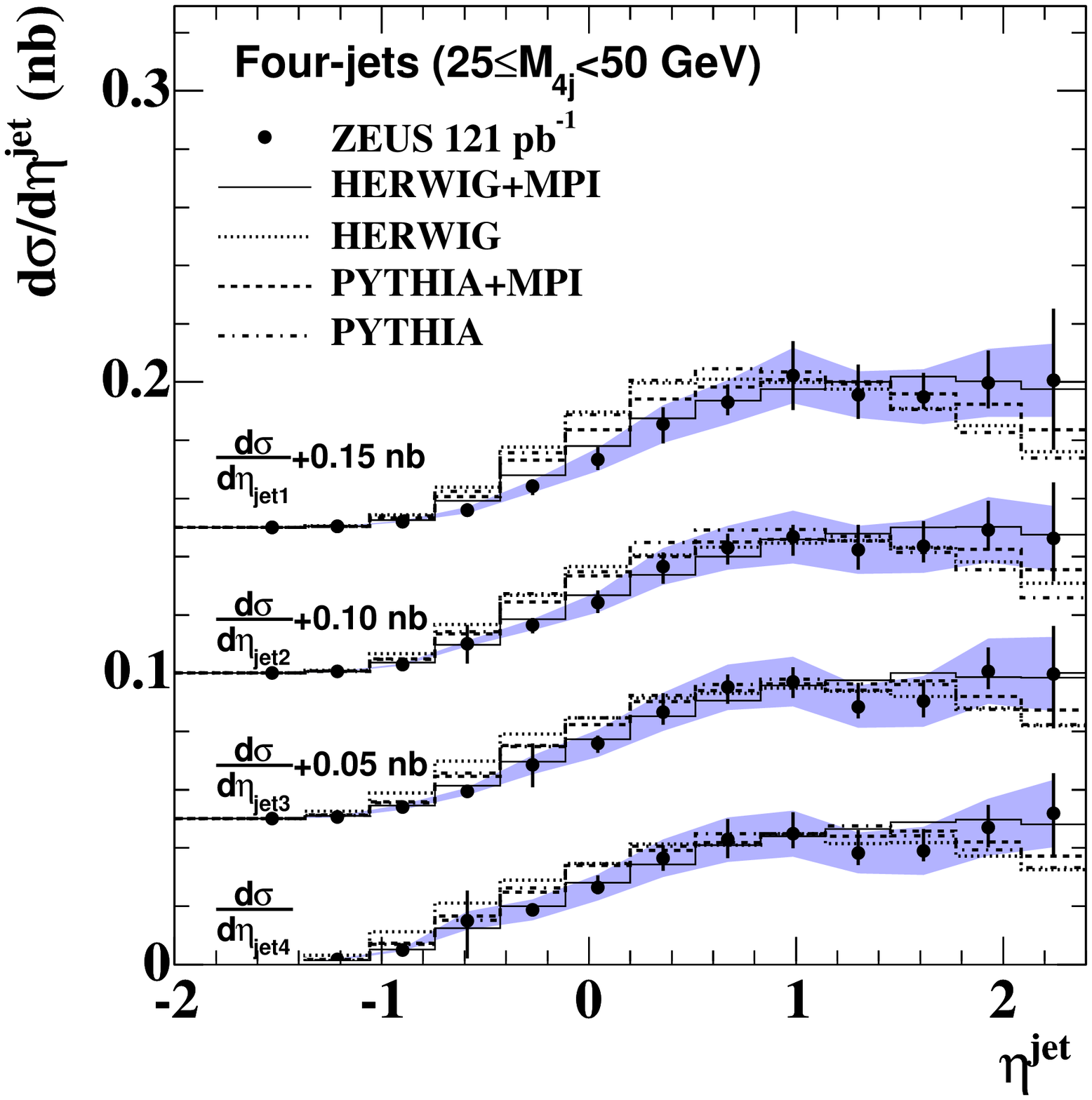}

\vspace{-9.2cm}\hspace{7.6cm}(a)\vspace{8.4cm}

\includegraphics*[angle=0,scale=0.5]{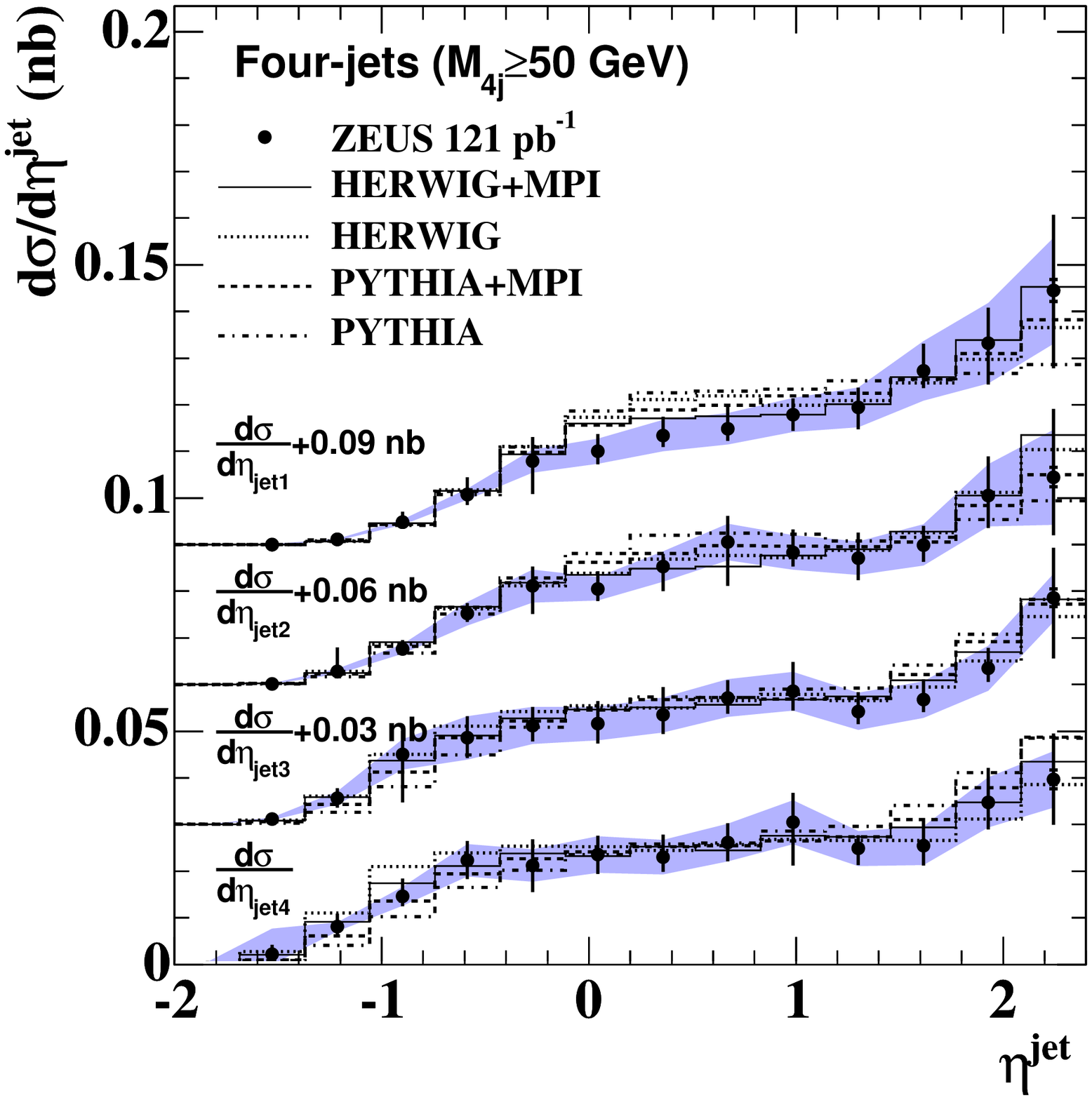}

\vspace{-9.2cm}\hspace{7.6cm}(b)\vspace{8.4cm}
\caption{
Measured cross section as a function of $\eta^{jet}$ in the four-jet (a) low- and (b) high-mass samples. Other details as in the caption to Fig.~\ref{fig:xsec:eta_3j}.
\label{fig:xsec:eta_4j}}
\end{center}
\end{figure}

The cross sections are given as a function of the $E_T^{\rm jet}$ of each of the jets in both three-jet samples in Fig.~\ref{fig:xsec:Et_3j} and Tables~\ref{tab:Et3_3j_} to \ref{tab:Et5_3j_} and in both four-jet samples in Fig.~\ref{fig:xsec:Et_4j} and Tables~\ref{tab:Et3_4j_} to \ref{tab:Et6_4j_}.  Similarly, the ${\rm d}\sigma/{\rm d}\eta^{\rm jet}$ cross sections are given in Figs.~\ref{fig:xsec:eta_3j} and \ref{fig:xsec:eta_4j} and Tables~\ref{tab:Eta3_3j_} to \ref{tab:Eta6_4j_}. In each case the jets have been ordered in descending $E_T^{\rm jet}$.

The $E_T^{\rm jet}$ distributions of all of the jets in all of the samples are observed to fall off approximately exponentially with increasing $E_T^{\rm jet}$. All of the samples are similarly distributed when binned in terms of $\eta^{\rm jet}$, regardless of the position of the jet in the $E_T^{\rm jet}$ ordering. The generic trend is an increasing cross section from $\eta^{\rm jet}\approx-1.4$ followed by a plateau and, in the high-mass samples, a continued rise above $\eta^{\rm jet}\approx1.4$.  In the low-mass samples, the plateaux begin at $\eta^{\rm jet}\approx0.6$ and extend to the upper edge of the measured $\eta$ region.  In the high-mass samples, the lower bounds of the $\eta^{\rm jet}$ ranges covered by the plateaux decrease with decreasing $E_T^{\rm jet}$, while the upper bounds all lie at $\eta^{\rm jet}\approx1.4$.

Shown also in Figs.~\ref{fig:xsec:Et_3j} and \ref{fig:xsec:Et_4j} are ${\rm d}\sigma/{\rm d}E_T^{\rm jet}$ predictions from the {\sc Pythia} and {\sc Herwig} models, with and without MPIs, normalised to the data. The description of the data by each of the MC models is generally good for all of the jets in all of the samples, although the MC predictions do vary somewhat. In all four samples, the {\sc Pythia} cross sections without MPIs give the poorest description, in particular of ${\rm d}\sigma/{\rm d}E_T^{\rm jet2}$.  Overall, the description of the data is improved by the inclusion of MPIs.

Figures~\ref{fig:xsec:eta_3j} and \ref{fig:xsec:eta_4j} show the comparison between ${\rm d}\sigma/{\rm d}\eta^{\rm jet}$ in the data and as predicted by the MC models.  Each MC model largely describes both sets of high-mass data.  The best description of the low-mass data is given by the {\sc Herwig} model with tuned MPIs.  The {\sc Pythia} prediction with default MPIs also generally describes the data.  The poorest description is of the three-jet $\eta^{\rm jet1}$ distribution.  The introduction of MPIs into the simulations certainly aids the description.  In general, the MC models without MPIs predict a cross section that falls off at high $\eta^{\rm jet}$ but MPIs reduce this effect in line with what is observed.

\subsection{The ${\rm d}\sigma/{\rm d}\cos(\psi_3)$ cross section}

\begin{figure}[thb]
\begin{center}
\vspace{-2.0cm}
\hspace{0.7cm}
\includegraphics*[angle=0,scale=0.5]{DESY-07-102_0.eps}
\vspace{0.0cm}
\hspace{-0.7cm}

\includegraphics*[angle=0,scale=0.5]{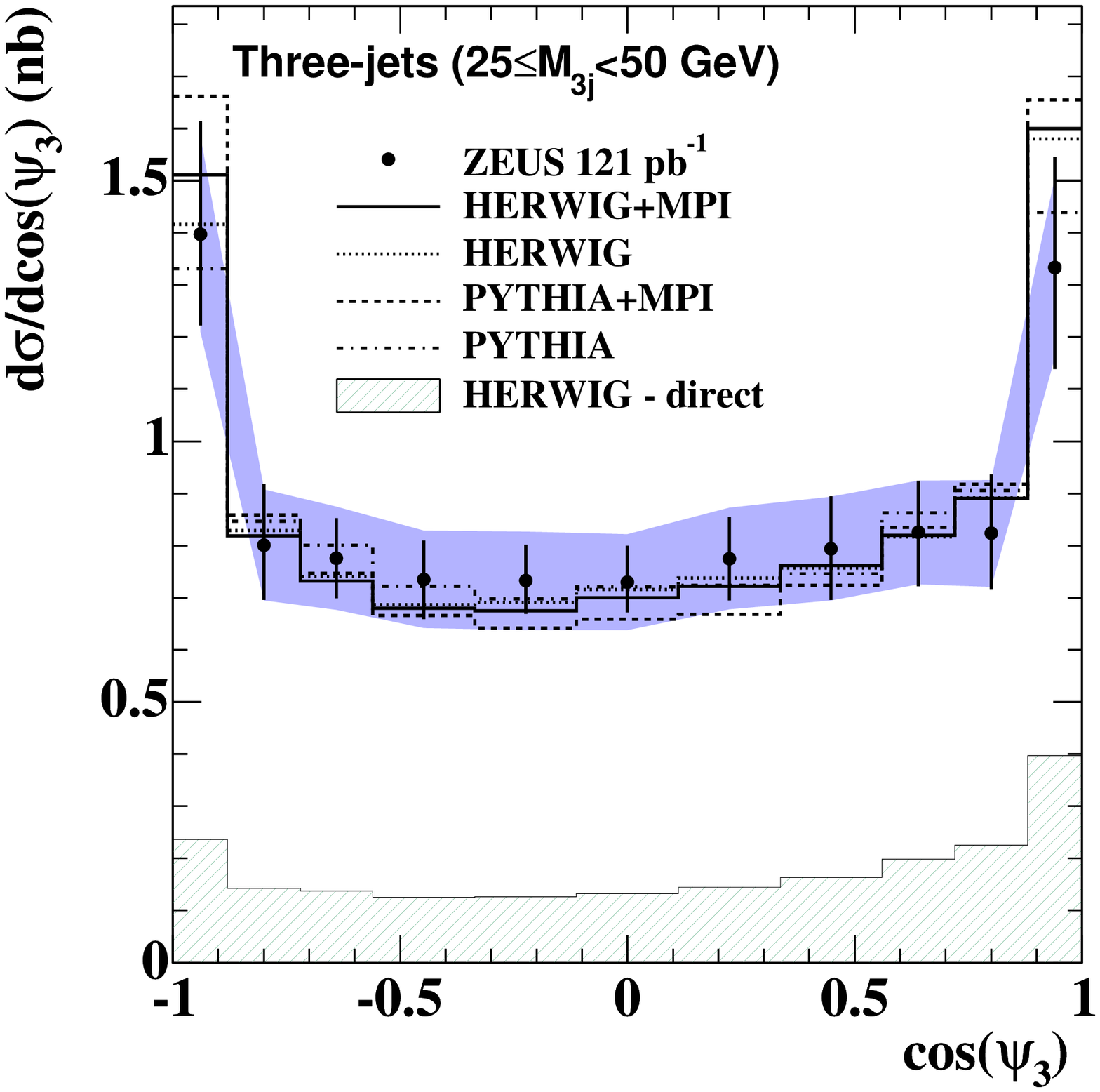}

\vspace{-9.2cm}\hspace{7.2cm}(a)\vspace{8.4cm}

\includegraphics*[angle=0,scale=0.5]{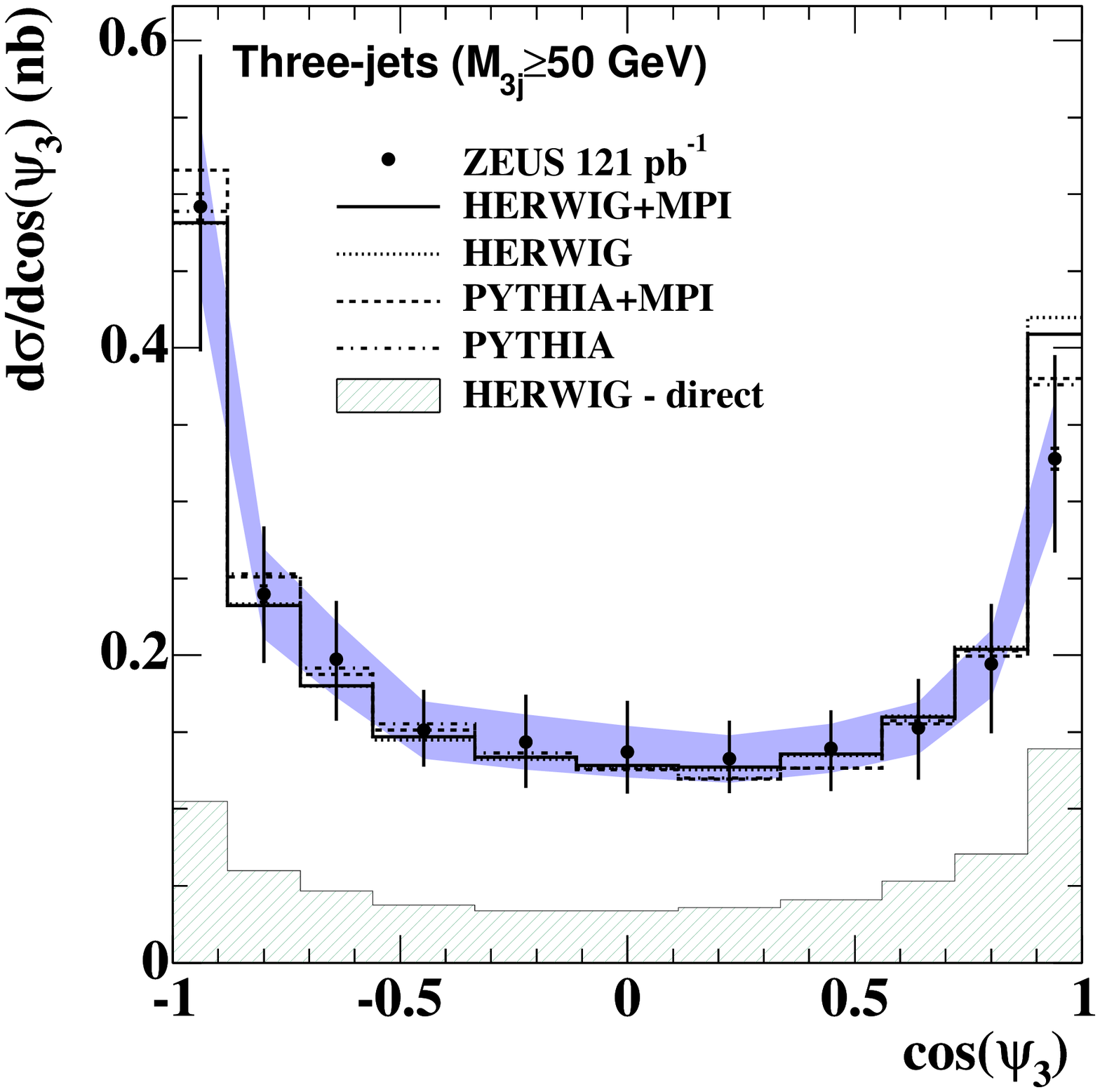}

\vspace{-9.2cm}\hspace{7.6cm}(b)\vspace{8.4cm}
\caption{
Measured cross section as a function of $\cos(\psi_{\rm 3})$ in the three-jet (a) low- and (b) high-mass samples. Other details as in the caption to Fig.~\ref{fig:xsec:yJB}.
\label{fig:xsec:cp3}}
\end{center}
\end{figure}

The three-jet cross section is given as a function of $\cos(\psi_3)$ for the low- and high-mass samples in Fig.~\ref{fig:xsec:cp3} and Table~\ref{tab:cos_psi3_3j_}.  The $\cos(\psi_3)$ cross section has a similar shape in both the low- and high-mass samples; relatively flat in the central $\cos(\psi_3)$ region and increasing rapidly as $|\cos(\psi_3)|\rightarrow 1$. Also shown in Fig.~\ref{fig:xsec:cp3} is a comparison of the MC models with the data.  All of the area normalised MC predictions describe the data well.

\subsection{The ${\rm d}\sigma/{\rm d}M_{3j}$ cross section compared with $\mathcal{O}(\alpha\alpha_s^2)$ pQCD}

\begin{figure}
\begin{center}
\vspace{-1.6cm}
\hspace{0.4cm}
\includegraphics*[angle=0,scale=0.5]{DESY-07-102_0.eps}
\vspace{0.0cm}
\hspace{-0.4cm}

\includegraphics*[angle=0,scale=0.75]{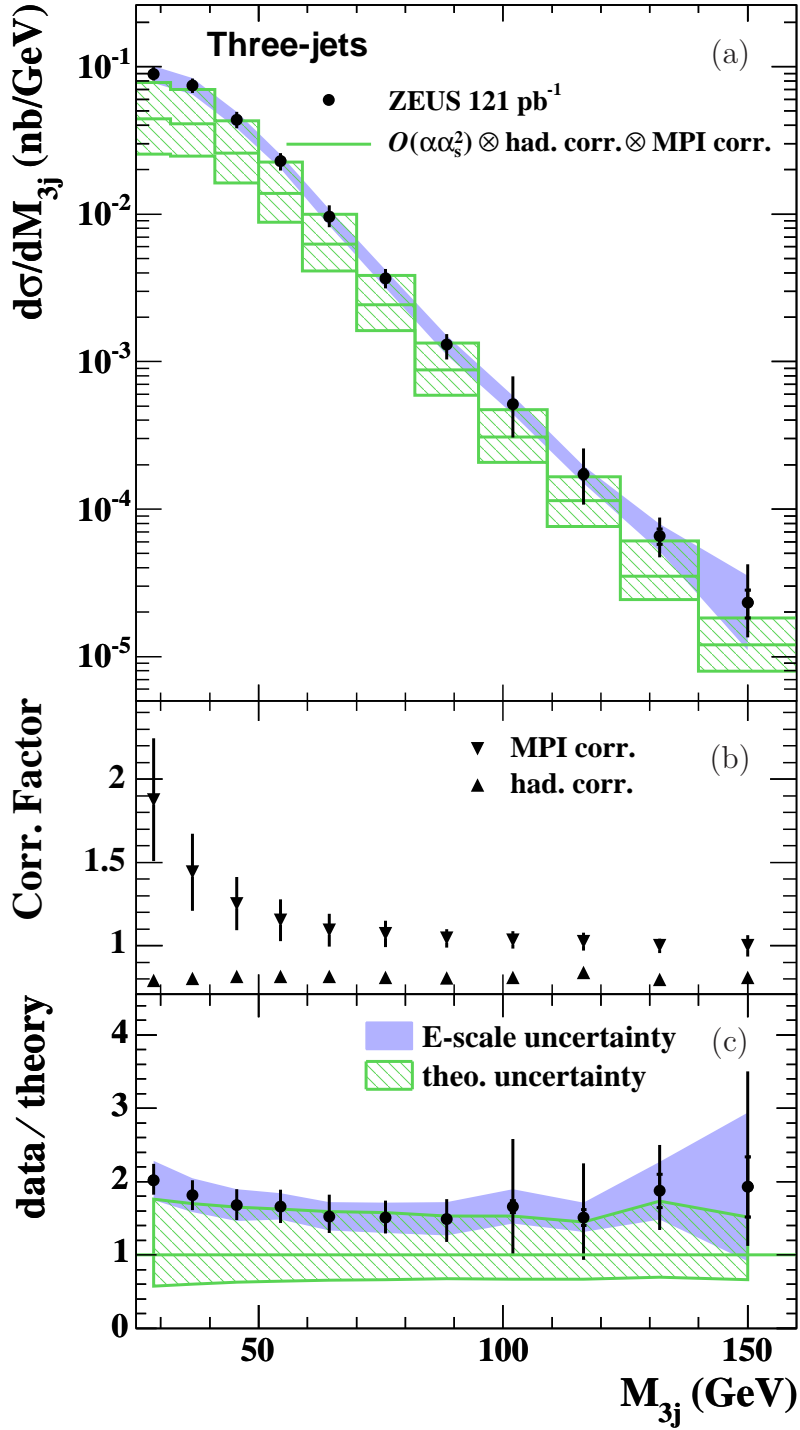}

\vspace{-18.8cm}\hspace{7.6cm}(a)\vspace{18.0cm}

\vspace{-9.2cm}\hspace{7.6cm}(b)\vspace{8.4cm}

\vspace{-5.2cm}\hspace{7.6cm}(c)\vspace{4.8cm}
\caption{
(a) Measured cross section as a function of $M_{3j}$ in the three-jet sample compared with an $\mathcal{O}(\alpha\alpha_s^2)$ prediction, corrected for hadronisation and MPI effects. (b) The hadronisation and MPI corrections as a function of $M_{3j}$. (c) The ratio of the $M_{3j}$ cross section divided by the corrected $\mathcal{O}(\alpha\alpha_s^2)$ prediction. The theoretical uncertainty is represented by the dashed bands.  Other details as in the caption to Fig.~\ref{fig:xsec:Mnj}.
\label{fig:tree:Mnj}}
\end{center}
\end{figure}

Figure~\ref{fig:tree:Mnj} shows an $\mathcal{O}(\alpha\alpha_s^2)$ prediction, corrected for hadronisation effects and MPIs, compared to the measured ${\rm d}\sigma/{\rm d}M_{3j}$ cross section. The hadronisation and MPI corrections, including their estimated uncertainties, are given in Fig.~\ref{fig:tree:Mnj}b and in Table~\ref{tab:Mnj_3j_}.  The hadronisation corrections are constant in $M_{3j}$, while the MPI corrections increase significantly towards low $M_{3j}$. The theoretical uncertainties on both the MPI corrections and the pQCD predictions are large. The magnitude and shape of the calculation is consistent with the data within the large theoretical uncertainties.  This is best seen in the data over theory ratio shown in Fig.~\ref{fig:tree:Mnj}c. The level of consistency between data and theory would be far worse at low $M_{3j}$ if it were not for the large MPI corrections.

\subsection{The ${\rm d}\sigma/{\rm d}E_T^{\rm jet}$ cross section compared with $\mathcal{O}(\alpha\alpha_s^2)$ pQCD}

\begin{figure}[thb]
\begin{center}
\vspace{-2.0cm}
\hspace{0.7cm}
\includegraphics*[angle=0,scale=0.5]{DESY-07-102_0.eps}
\vspace{0.0cm}
\hspace{-0.7cm}

\includegraphics*[angle=0,scale=0.5]{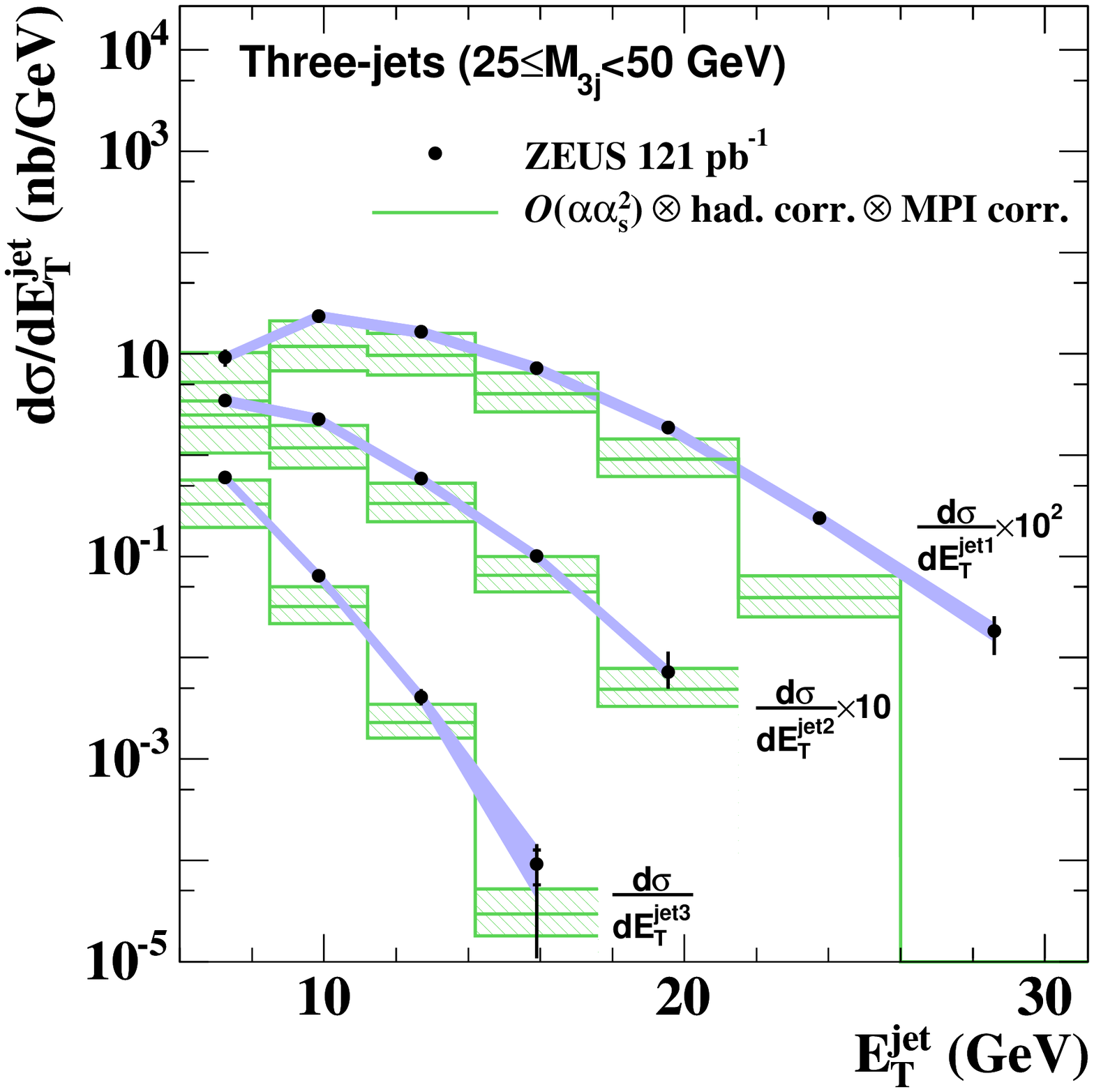}

\vspace{-9.2cm}\hspace{7.6cm}(a)\vspace{8.4cm}

\includegraphics*[angle=0,scale=0.5]{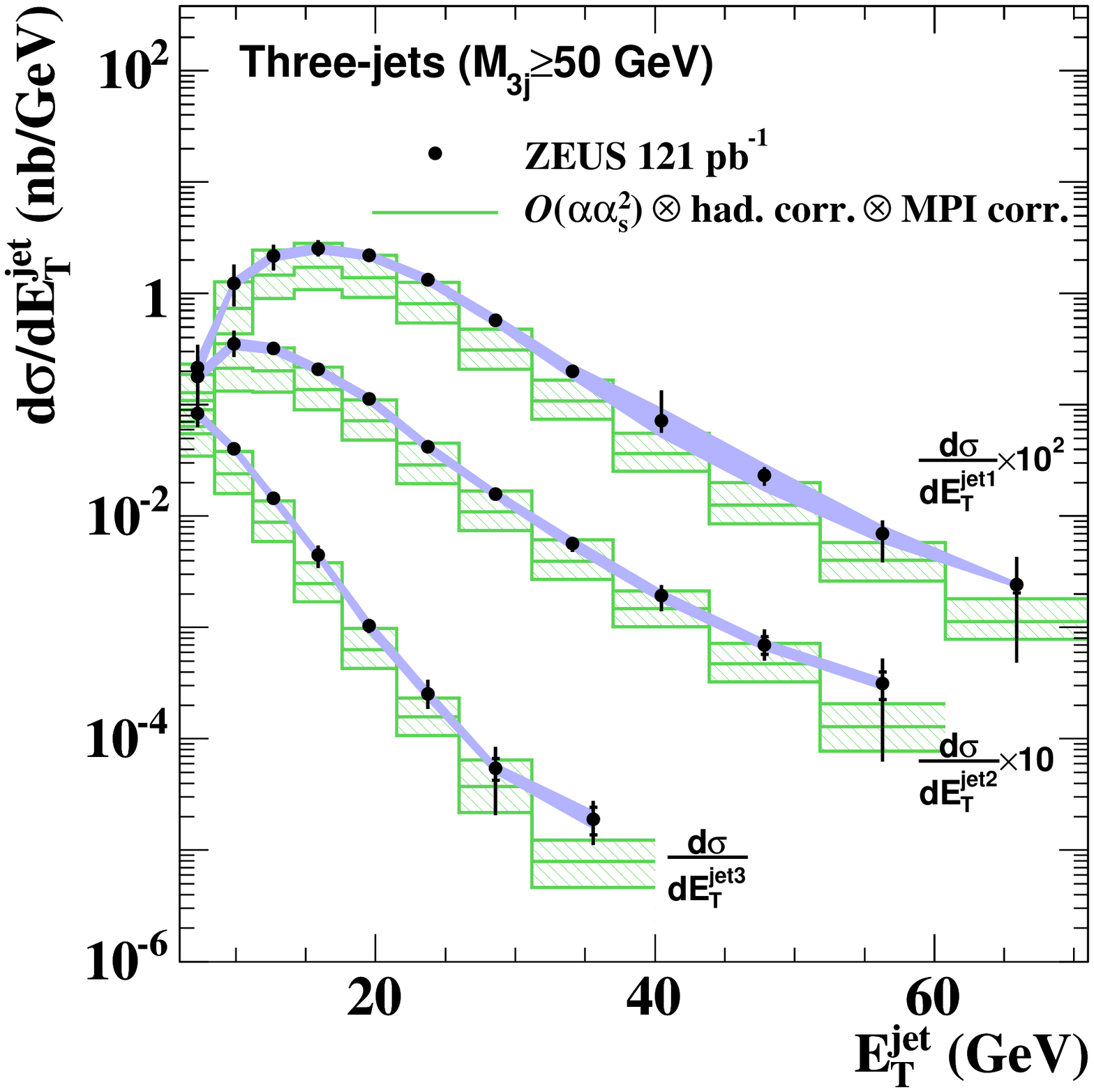}

\vspace{-9.2cm}\hspace{7.6cm}(b)\vspace{8.4cm}
\caption{
Measured cross section as a function of $E_T^{jet}$ for each jet in the three-jet (a) low- and (b) high-mass samples compared with $\mathcal{O}(\alpha\alpha_s^2)$ predictions, corrected for both hadronisation and MPI effects. Each cross section has been multiplied by $10^n$, where $n$ is given on the plot, to aid visibility. Other details as in the caption to Fig.~\ref{fig:tree:Mnj}.
\label{fig:tree:Et_3j}}
\end{center}
\end{figure}

Figure~\ref{fig:tree:Et_3j} shows the comparison between the measured and predicted ${\rm d}\sigma/{\rm d}E_T^{\rm jet}$ cross sections at $\mathcal{O}(\alpha\alpha_s^2)$ for each jet in the low- and high-mass three-jet samples. The pQCD predictions have been corrected for hadronisation effects and MPIs, which are detailed in Tables~\ref{tab:Et3_3j_} to \ref{tab:Et5_3j_} (not shown in Fig.~\ref{fig:tree:Et_3j}).  Hadronisation is predicted to decrease the parton-level cross sections at low $E_T^{\rm jet}$, whereas the MPIs are expected to do the reverse in the low-mass region: at high mass they are small.

The calculation is largely consistent, within the large uncertainties, with both the low- and high-mass data.  The exception to this is the rapid predicted falloff of the $E_T^{\rm jet1}$ cross section at high $E_T^{\rm jet1}$ in the low-mass sample, which is not observed in the data.  This effect is due to the three-parton kinematics within the theory, which precludes the highest momentum parton from carrying more than half of the total centre-of-mass energy.  Thus, for the low-mass sample, the $M_{nj}<50$~GeV criterium translates into an $E_T<25$~GeV constraint for the partons in the theory, as indeed observed.

\subsection{The ${\rm d}\sigma/{\rm d}\eta^{\rm jet}$ cross section compared with $\mathcal{O}(\alpha\alpha_s^2)$ pQCD}

\begin{figure}[thb]
\begin{center}
\vspace{-2.0cm}
\hspace{0.7cm}
\includegraphics*[angle=0,scale=0.5]{DESY-07-102_0.eps}
\vspace{0.0cm}
\hspace{-0.7cm}

\includegraphics*[angle=0,scale=0.5]{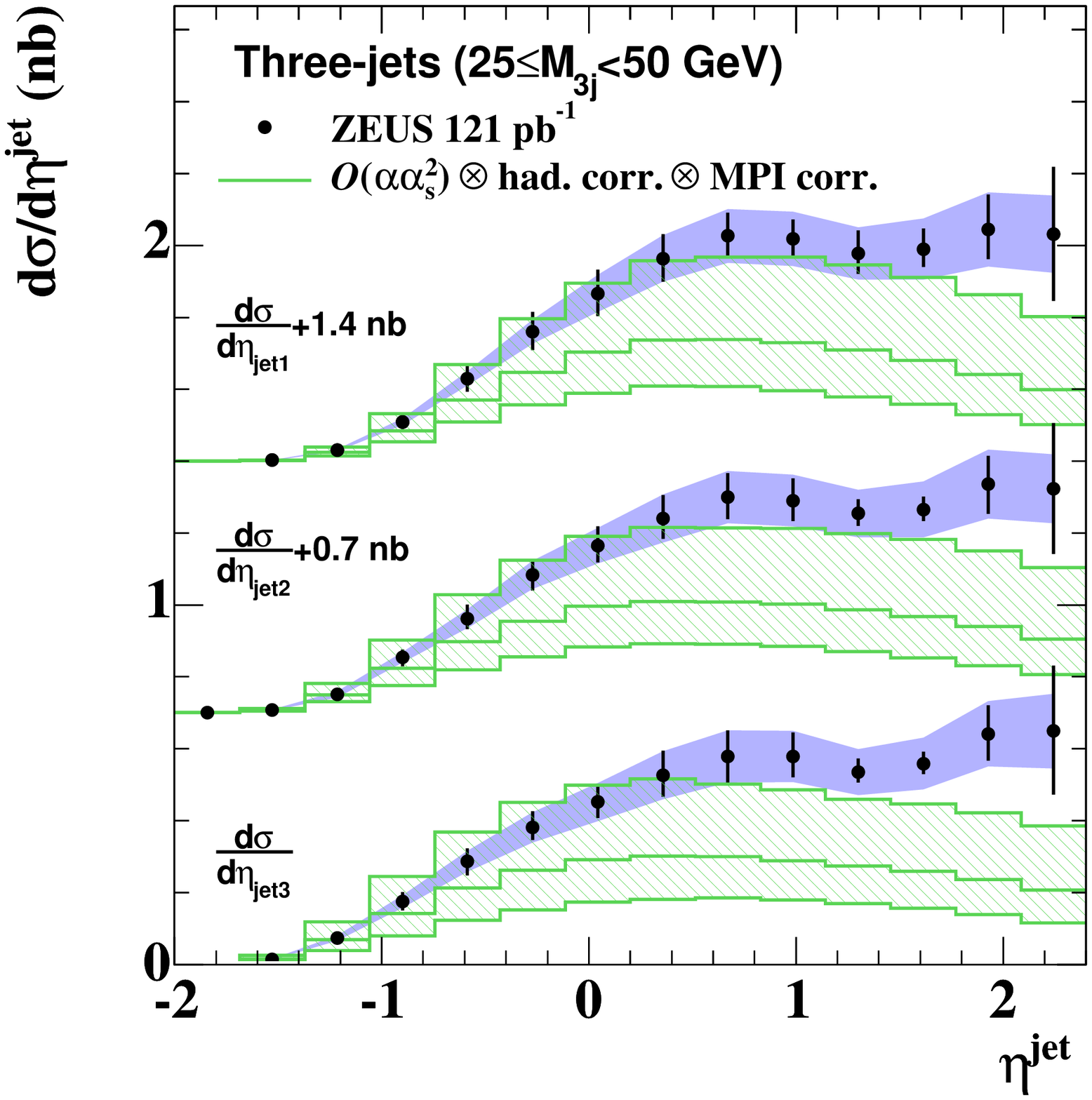}

\vspace{-9.2cm}\hspace{7.6cm}(a)\vspace{8.4cm}

\includegraphics*[angle=0,scale=0.5]{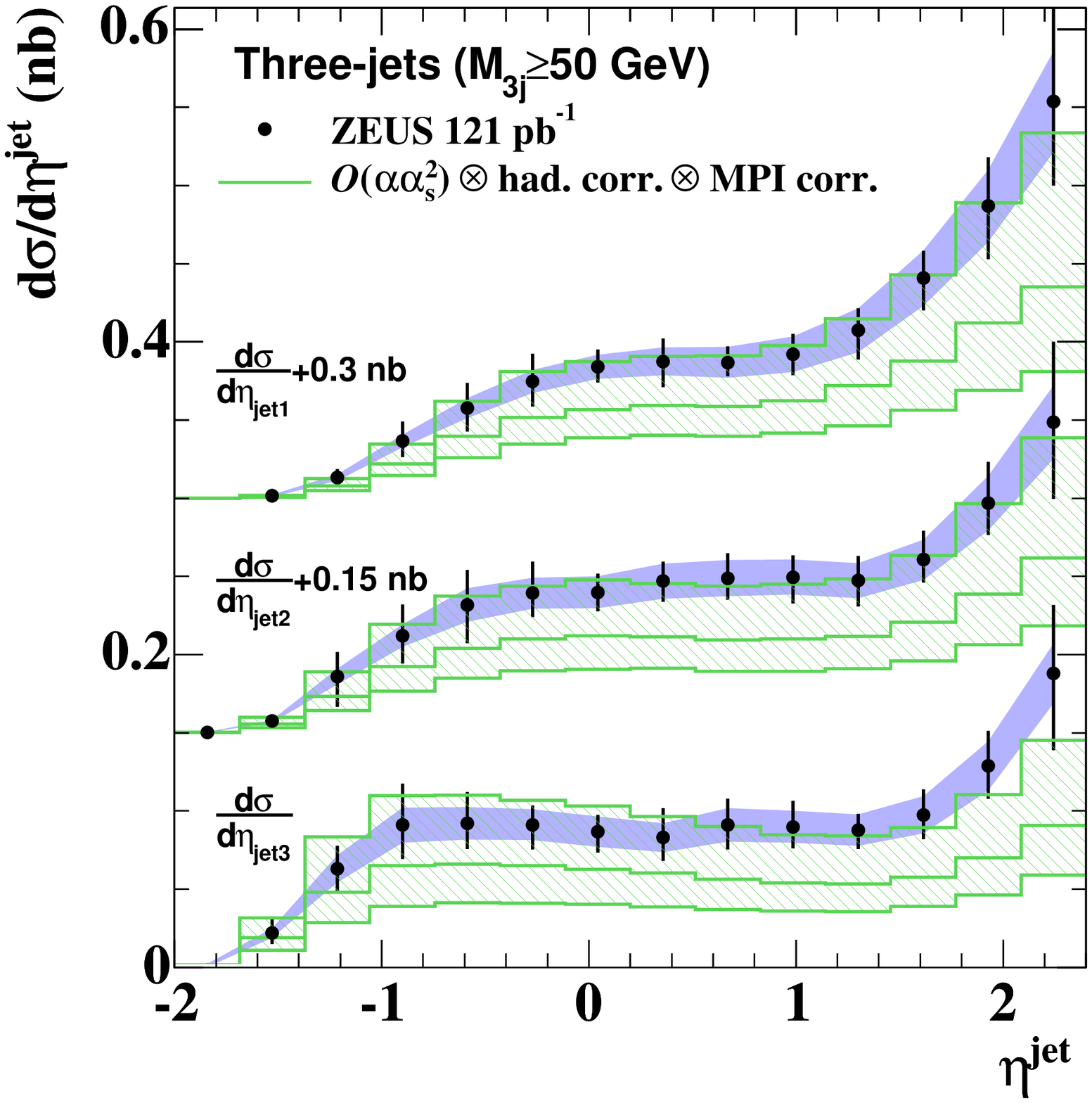}

\vspace{-9.2cm}\hspace{7.2cm}(b)\vspace{8.4cm}
\caption{
Measured cross section as a function of $\eta^{jet}$ for each jet in the three-jet (a) low- and (b) high-mass samples compared with $\mathcal{O}(\alpha\alpha_s^2)$ predictions, corrected for both hadronisation and MPI effects. Each cross section has been shifted upwards by the amount given on the plot, to aid visibility. Other details as in the caption to Fig.~\ref{fig:tree:Mnj}.
\label{fig:tree:eta_3j}}
\end{center}
\end{figure}

Figure~\ref{fig:tree:eta_3j} shows the comparison between the measured and predicted ${\rm d}\sigma/{\rm d}\eta^{\rm jet}$ cross sections at $\mathcal{O}(\alpha\alpha_s^2)$ for each jet in the low- and high-mass three-jet samples. The hadronisation and MPI correction factors are given in Tables~\ref{tab:Eta3_3j_} to \ref{tab:Eta5_3j_} (not shown in Fig.~\ref{fig:tree:eta_3j}). Hadronisation is predicted to cause an overall decrease in the parton-level cross sections, most significantly at low $\eta^{\rm jet}$. The MPIs in the low-mass sample are predicted to cause a small increase in the cross section at low $\eta^{\rm jet}$, becoming larger as $\eta^{\rm jet}$ increases. The calculation is consistent with the $\eta^{\rm jet}$ distributions.  There is some indication of a difference in shape at lower masses, however, the uncertainties are again large.  The description of the low-mass data would be worse if it were not for the MPI corrections.

\subsection{The ${\rm d}\sigma/{\rm d}\cos(\psi_3)$ cross section compared with $\mathcal{O}(\alpha\alpha_s^2)$ pQCD}

\begin{figure}
\begin{center}
\vspace{-1.6cm}
\hspace{0.4cm}
\includegraphics*[angle=0,scale=0.5]{DESY-07-102_0.eps}
\vspace{0.0cm}
\hspace{-0.4cm}

\includegraphics*[angle=0,scale=0.75]{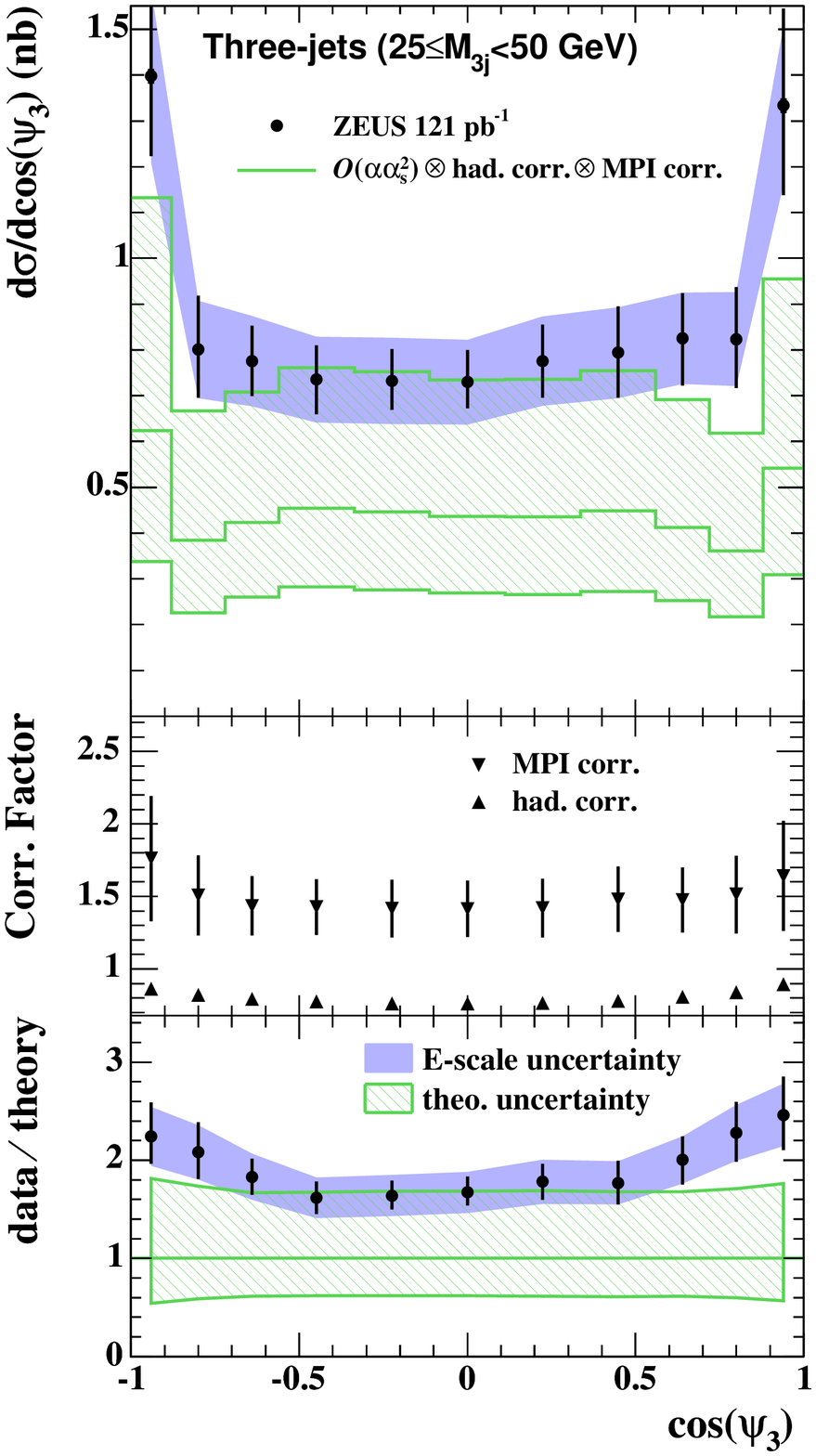}

\vspace{-18.8cm}\hspace{7.6cm}(a)\vspace{18.0cm}

\vspace{-9.2cm}\hspace{7.6cm}(b)\vspace{8.4cm}

\vspace{-5.2cm}\hspace{7.6cm}(c)\vspace{4.8cm}
\caption{
(a) Measured cross section as a function of $\cos(\psi_3)$ in the low-mass three-jet sample.  Other details as in the caption to Fig.~\ref{fig:tree:Mnj}.
\label{fig:tree:cp3_i}}
\end{center}
\end{figure}

\begin{figure}
\begin{center}
\vspace{-1.6cm}
\hspace{0.4cm}
\includegraphics*[angle=0,scale=0.5]{DESY-07-102_0.eps}
\vspace{0.0cm}
\hspace{-0.4cm}

\includegraphics*[angle=0,scale=0.75]{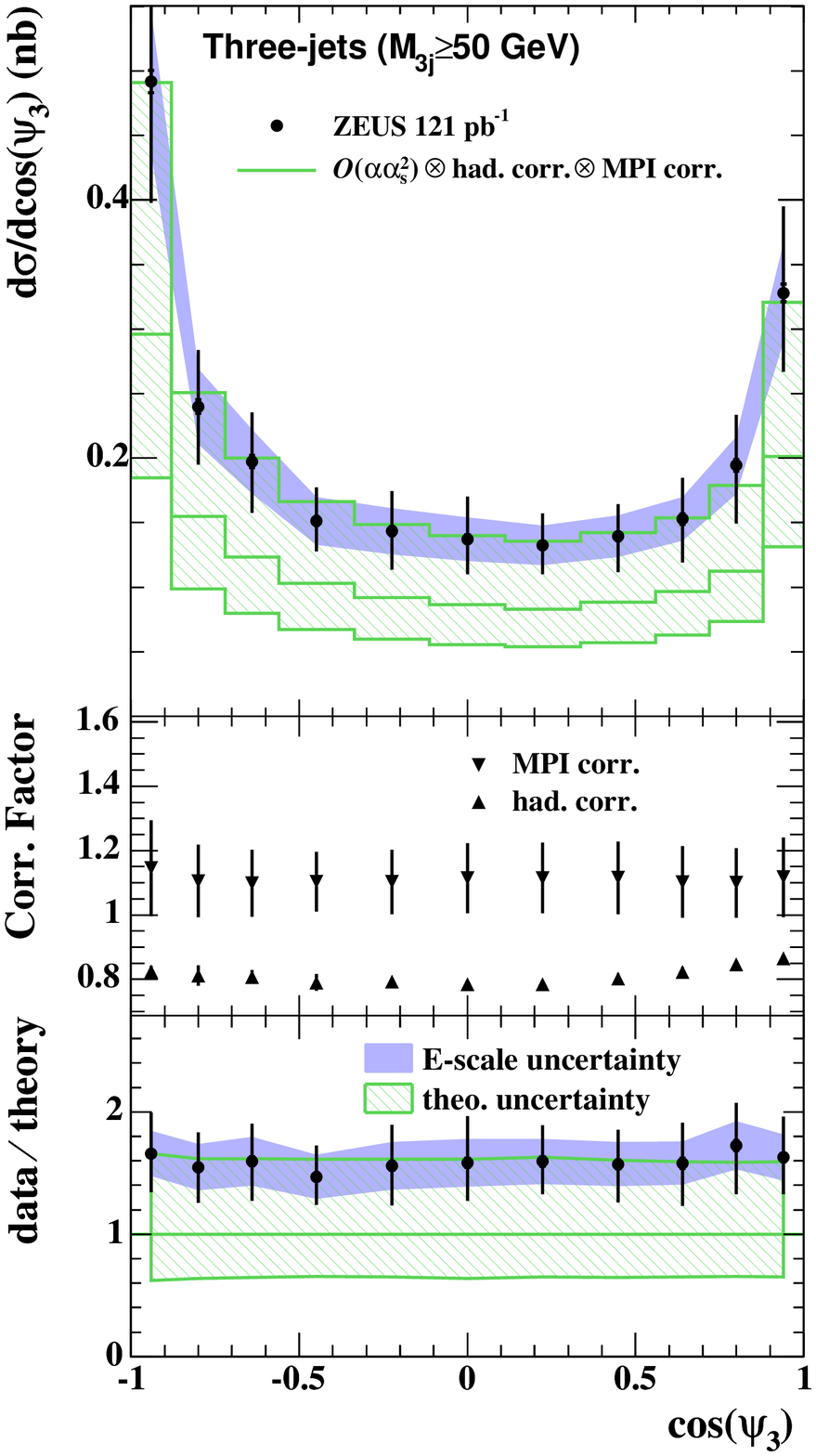}

\vspace{-18.8cm}\hspace{7.6cm}(a)\vspace{18.0cm}

\vspace{-9.2cm}\hspace{7.6cm}(b)\vspace{8.4cm}

\vspace{-5.2cm}\hspace{7.6cm}(c)\vspace{4.8cm}
\caption{
(a) Measured cross section as a function of $\cos(\psi_3)$ in the high-mass three-jet sample.  Other details as in the caption to Fig.~\ref{fig:tree:Mnj}.
\label{fig:tree:cp3_h}}
\end{center}
\end{figure}

The comparison between the measured and the predicted ${\rm d}\sigma/{\rm d}\cos(\psi_3)$ cross sections at $\mathcal{O}(\alpha\alpha_s^2)$ is shown for the low-mass sample in Fig.~\ref{fig:tree:cp3_i} and for the high-mass region in Fig.~\ref{fig:tree:cp3_h}. The ratio of the low-mass data divided by the calculation is shown along with the hadronisation and MPI corrections, which are also given in Table~\ref{tab:cos_psi3_3j_}.  In both low- and high-mass regions, the QCD calculation is consistent with the data. There is some indication of a difference in shape at low masses; however, the theory has large uncertainties.

%%%%%%%%%%%%%%%%%%%%%%%%%%%%%%%%%%%%%%%%%%%%%%%%%%%%%%%%%%%%%%%%%%%%%%%%%%%%%%

\section{Summary}

Three- and four-jet states have been measured in hard $\gamma p$ collisions at HERA, using an integrated luminosity of $121~{\rm pb^{\rm -1}}$. The three- and four-jet cross sections have been measured for jets with $E_T^{\rm jet}>6$~GeV and $|\eta^{\rm jet}|<2.4$, in the kinematic region given by $Q^2<1~\rm{GeV^2}$ and $0.2\le y\le0.85$ and in two mass regions defined as $25\le M_{nj}<50$~GeV and $M_{nj}\ge50$~GeV. The three-jet events have been measured with over seven times the luminosity of the previous ZEUS publication and in a wider $M_{3j}$ region.  The four-jet process described here has been measured for the first time at HERA.

In the high-mass regions, the shape of the three- and four-jet cross sections are reasonably well described by both the {\sc Pythia} and {\sc Herwig} models without MPIs. In the low-mass region, the MC models without MPIs underestimate the data when normalised to the measured high-mass cross section.  When MPIs are added to the MC simulations, the agreement between the models and data is generally improved.  Although the data have large uncertainties, the measured cross sections are potentially useful in the testing and tuning of MPI and underlying-event models.

The $\mathcal{O}(\alpha\alpha_s^2)$ pQCD calculation was compared to the three-jet data and is consistent within the large uncertainties.  The MPI corrections typically improved the description of the data by the pQCD calculation.  This data will provide a testing ground for higher-order calculations in photoproduction.

\section{Acknowledgments}

The support and encouragement of the DESY Directorate has been invaluable and we are indebted to the HERA machine group for their diligent efforts.  The design, construction and installation of the ZEUS detector involved many people both in and outside of DESY that are not listed as authors here but their contributions are acknowledged with great appreciation nonetheless.

\newpage 

\providecommand{\etal}{et al.\xspace}
\providecommand{\coll}{Coll.\xspace}
\catcode`\@=11
\def\@bibitem#1{%
\ifmc@bstsupport
  \mc@iftail{#1}%
    {;\newline\ignorespaces}%
    {\ifmc@first\else.\fi\orig@bibitem{#1}}
  \mc@firstfalse
\else
  \mc@iftail{#1}%
    {\ignorespaces}%
    {\orig@bibitem{#1}}%
\fi}%
\catcode`\@=12

\newpage 

\begin{table}
\begin{center}
\begin{tabular}{|c|c|c|c|}
\hline
\multicolumn{4}{|c|}{{\sc Herwig} scaling factors} \\
\hline
3-jet no MPIs & 3-jet with MPIs & 4-jet no MPIs & 4-jet with MPIs\\
\hline
1.7 & 1.7 & 3.1 & 2.1 \\
\hline
\hline
\multicolumn{4}{|c|}{{\sc Pythia} scaling factors} \\
\hline
3-jet no MPIs & 3-jet with MPIs & 4-jet no MPIs & 4-jet with MPIs\\
\hline
3.8 & 3.1 & 9.2 & 5.3 \\
\hline
\end{tabular}
\caption{The scaling factors applied to the {\sc Herwig} and {\sc
    Pythia} $x_\gamma^{\rm obs}$ and $M_{nj}$ cross sections.
\label{tab:mc_scl}}
\end{center}
\end{table}
\begin{table}
\begin{center}
\begin{tabular}{|l c l||l l l l||l l l l|}
\hline
\multicolumn{3}{|c||}{$M_{3j}$ range} & $d\sigma/dM_{3j}$ & $\pm$stat. & $\pm$syst. & $\pm$E-scale & \multicolumn{2}{c}{had. corr.}&\multicolumn{2}{c|}{MPI corr.}\\
\multicolumn{3}{|c||}{(GeV)} & \multicolumn{4}{c||}{(pb/GeV)} & & & &\\
\hline
25 & - & 32 & 94 & $\pm$ 0 & $_{-9}^{+10}$ & $_{-12}^{+12}$ &0.793& $\pm$0.004&1.88& $\pm$0.37\\
32 & - & 41 & 80.4 & $\pm$ 0.3 & $_{-8.2}^{+8.5}$ & $_{-9.5}^{+9.5}$ &0.805& $\pm$0.010&1.44& $\pm$0.23\\
41 & - & 50 & 48.4 & $\pm$ 0.3 & $_{-5.4}^{+5.5}$ & $_{-5.6}^{+5.6}$ &0.815& $\pm$0.015&1.25& $\pm$0.16\\
50 & - & 59 & 25.2 & $\pm$ 0.2 & $_{-3.1}^{+3.1}$ & $_{-2.5}^{+2.5}$ &0.815& $\pm$0.017&1.15& $\pm$0.13\\
59 & - & 70 & 10.7 & $\pm$ 0.1 & $_{-1.4}^{+1.9}$ & $_{-1.2}^{+1.2}$ &0.817& $\pm$0.008&1.09& $\pm$0.10\\
70 & - & 82 & 4.00 & $\pm$ 0.08 & $_{-0.52}^{+0.56}$ & $_{-0.50}^{+0.50}$ &0.810& $\pm$0.009&1.07& $\pm$0.08\\
82 & - & 95 & 1.44 & $\pm$ 0.04 & $_{-0.27}^{+0.24}$ & $_{-0.20}^{+0.20}$ &0.808& $\pm$0.007&1.04& $\pm$0.05\\
95 & - & 109 & 0.61 & $\pm$ 0.02 & $_{-0.21}^{+0.28}$ & $_{-0.07}^{+0.07}$ &0.811& $\pm$0.014&1.03& $\pm$0.05\\
109 & - & 124 & 0.179 & $\pm$ 0.012 & $_{-0.066}^{+0.084}$ & $_{-0.023}^{+0.023}$ &0.840& $\pm$0.019&1.03& $\pm$0.05\\
124 & - & 140 & 0.066 & $\pm$ 0.008 & $_{-0.019}^{+0.022}$ & $_{-0.014}^{+0.014}$ &0.798& $\pm$0.008&1.00& $\pm$0.04\\
140 & - & 160 & 0.023 & $\pm$ 0.005 & $_{-0.010}^{+0.019}$ & $_{-0.012}^{+0.012}$ &0.809& $\pm$0.012&1.00& $\pm$0.06\\
\hline
\end{tabular}
\caption{The measured three-jet differential cross-section $d\sigma/dM_{3j}$. The statistical, systematic and calorimeter energy scale (E-scale) uncertainties are shown separately. Also shown are the hadronisation and MPI corrections that were applied to the $\mathcal{O}(\alpha\alpha_s^2)$ prediction.
\label{tab:Mnj_3j_}}
\end{center}
\end{table}

\begin{table}
\begin{center}
\begin{tabular}{|l c l||l l l l|}
\hline
\multicolumn{3}{|c||}{$M_{4j}$ range} & $d\sigma/dM_{4j}$ & $\pm$stat. & $\pm$syst. & $\pm$E-scale\\
\multicolumn{3}{|c||}{(GeV)} & \multicolumn{4}{c|}{(pb/GeV)}\\
\hline
25 & - & 32 & 1.16 & $\pm$ 0.04 & $_{-0.20}^{+0.29}$ & $_{-0.28}^{+0.28}$\\
32 & - & 41 & 5.8 & $\pm$ 0.1 & $_{-0.7}^{+0.6}$ & $_{-1.1}^{+1.1}$\\
41 & - & 50 & 6.3 & $\pm$ 0.1 & $_{-0.7}^{+0.7}$ & $_{-1.1}^{+1.1}$\\
50 & - & 59 & 4.49 & $\pm$ 0.09 & $_{-0.48}^{+0.65}$ & $_{-0.73}^{+0.73}$\\
59 & - & 70 & 2.85 & $\pm$ 0.07 & $_{-0.59}^{+0.43}$ & $_{-0.43}^{+0.43}$\\
70 & - & 82 & 1.25 & $\pm$ 0.05 & $_{-0.20}^{+0.27}$ & $_{-0.22}^{+0.22}$\\
82 & - & 95 & 0.488 & $\pm$ 0.027 & $_{-0.063}^{+0.064}$ & $_{-0.087}^{+0.087}$\\
95 & - & 109 & 0.237 & $\pm$ 0.018 & $_{-0.068}^{+0.052}$ & $_{-0.042}^{+0.042}$\\
109 & - & 124 & 0.060 & $\pm$ 0.009 & $_{-0.013}^{+0.019}$ & $_{-0.010}^{+0.010}$\\
124 & - & 140 & 0.027 & $\pm$ 0.006 & $_{-0.015}^{+0.019}$ & $_{-0.006}^{+0.006}$\\
140 & - & 160 & 0.0024 & $\pm$ 0.0016 & $_{-0.0019}^{+0.0053}$ & $_{-0.0012}^{+0.0012}$\\
\hline
\end{tabular}
\caption{The measured four-jet differential cross-section $d\sigma/dM_{4j}$. Other details as described in the caption to Table~\ref{tab:Mnj_3j_}.
\label{tab:Mnj_4j_}}
\end{center}
\end{table}
\clearpage

\begin{table}
\begin{center}
\begin{tabular}{|l c l||l l l l|}
\hline
\multicolumn{3}{|c||}{$x_{\gamma}^{{\rm obs}}$ range} & $d\sigma/dx_{\gamma}^{{\rm obs}}$ & $\pm$stat. & $\pm$syst. & $\pm$E-scale\\
\multicolumn{3}{|c||}{} & \multicolumn{4}{c|}{(pb)}\\
\hline
\multicolumn{7}{|c|}{$25\le M_{3j}<50$~GeV}\\
\hline
0.00 & - & 0.11 & 109 & $\pm$ 7 & $_{-53}^{+66}$ & $_{-31}^{+31}$\\
0.11 & - & 0.22 & 1449 & $\pm$ 15 & $_{-324}^{+334}$ & $_{-289}^{+289}$\\
0.22 & - & 0.33 & 2102 & $\pm$ 15 & $_{-230}^{+240}$ & $_{-361}^{+361}$\\
0.33 & - & 0.47 & 1989 & $\pm$ 13 & $_{-119}^{+128}$ & $_{-292}^{+292}$\\
0.47 & - & 0.62 & 1774 & $\pm$ 12 & $_{-97}^{+117}$ & $_{-234}^{+234}$\\
0.62 & - & 0.75 & 1752 & $\pm$ 13 & $_{-99}^{+115}$ & $_{-199}^{+199}$\\
0.75 & - & 0.85 & 2579 & $\pm$ 19 & $_{-184}^{+204}$ & $_{-253}^{+253}$\\
0.850 & - & 0.935 & 3803 & $\pm$ 26 & $_{-600}^{+672}$ & $_{-227}^{+227}$\\
0.935 & - & 1.000 & 730 & $\pm$ 12 & $_{-194}^{+203}$ & $_{-22}^{+22}$\\
\hline
\multicolumn{7}{|c|}{$M_{3j}\ge50$~GeV}\\
\hline
0.11 & - & 0.22 & 6.9 & $\pm$ 1.0 & $_{-5.0}^{+8.0}$ & $_{-2.0}^{+2.0}$\\
0.22 & - & 0.33 & 51 & $\pm$ 3 & $_{-13}^{+10}$ & $_{-10}^{+10}$\\
0.33 & - & 0.47 & 153 & $\pm$ 4 & $_{-42}^{+42}$ & $_{-21}^{+21}$\\
0.47 & - & 0.62 & 310 & $\pm$ 6 & $_{-75}^{+87}$ & $_{-40}^{+40}$\\
0.62 & - & 0.75 & 504 & $\pm$ 8 & $_{-86}^{+93}$ & $_{-60}^{+60}$\\
0.75 & - & 0.85 & 871 & $\pm$ 12 & $_{-113}^{+101}$ & $_{-100}^{+100}$\\
0.850 & - & 0.935 & 1695 & $\pm$ 18 & $_{-211}^{+202}$ & $_{-161}^{+161}$\\
0.935 & - & 1.000 & 666 & $\pm$ 14 & $_{-164}^{+169}$ & $_{-52}^{+52}$\\
\hline
\end{tabular}
\caption{The measured three-jet differential cross-section $d\sigma/dx_{\gamma}^{{\rm obs}}$ in both the low- and high-mass regions. Other details as described in the caption to Table~\ref{tab:Mnj_3j_}.
\label{tab:X_gamma_3j_}}
\end{center}
\end{table}
\clearpage

\begin{table}
\begin{center}
\begin{tabular}{|l c l||l l l l|}
\hline
\multicolumn{3}{|c||}{$x_{\gamma}^{{\rm obs}}$ range} & $d\sigma/dx_{\gamma}^{{\rm obs}}$ & $\pm$stat. & $\pm$syst. & $\pm$E-scale\\
\multicolumn{3}{|c||}{} & \multicolumn{4}{c|}{(pb)}\\
\hline
\multicolumn{7}{|c|}{$25\le M_{4j}<50$~GeV}\\
\hline
0.00 & - & 0.11 & 4.9 & $\pm$ 1.7 & $_{-3.0}^{+5.1}$ & $_{-1.4}^{+1.4}$\\
0.11 & - & 0.22 & 100 & $\pm$ 4 & $_{-37}^{+44}$ & $_{-21}^{+21}$\\
0.22 & - & 0.33 & 210 & $\pm$ 5 & $_{-45}^{+46}$ & $_{-53}^{+53}$\\
0.33 & - & 0.47 & 177 & $\pm$ 4 & $_{-18}^{+31}$ & $_{-38}^{+38}$\\
0.47 & - & 0.62 & 136 & $\pm$ 3 & $_{-13}^{+21}$ & $_{-29}^{+29}$\\
0.62 & - & 0.75 & 94 & $\pm$ 3 & $_{-17}^{+27}$ & $_{-14}^{+14}$\\
0.75 & - & 0.85 & 113 & $\pm$ 3 & $_{-35}^{+42}$ & $_{-15}^{+15}$\\
0.850 & - & 0.935 & 125 & $\pm$ 4 & $_{-33}^{+28}$ & $_{-9}^{+9}$\\
0.935 & - & 1.000 & 22 & $\pm$ 2 & $_{-13}^{+6}$ & $_{-3}^{+3}$\\
\hline
\multicolumn{7}{|c|}{$M_{4j}\ge50$~GeV}\\
\hline
0.11 & - & 0.22 & 0.4 & $\pm$ 0.2 & $_{-0.2}^{+2.3}$ & $_{-0.2}^{+0.2}$\\
0.22 & - & 0.33 & 25 & $\pm$ 2 & $_{-12}^{+14}$ & $_{-9}^{+9}$\\
0.33 & - & 0.47 & 64 & $\pm$ 3 & $_{-23}^{+26}$ & $_{-11}^{+11}$\\
0.47 & - & 0.62 & 103 & $\pm$ 3 & $_{-30}^{+26}$ & $_{-19}^{+19}$\\
0.62 & - & 0.75 & 126 & $\pm$ 4 & $_{-24}^{+18}$ & $_{-21}^{+21}$\\
0.75 & - & 0.85 & 170 & $\pm$ 5 & $_{-27}^{+31}$ & $_{-24}^{+24}$\\
0.850 & - & 0.935 & 307 & $\pm$ 8 & $_{-40}^{+39}$ & $_{-35}^{+35}$\\
0.935 & - & 1.000 & 122 & $\pm$ 5 & $_{-37}^{+69}$ & $_{-7}^{+7}$\\
\hline
\end{tabular}
\caption{The measured four-jet differential cross-section $d\sigma/dx_{\gamma}^{{\rm obs}}$ in both the low- and high-mass regions. Other details as described in the caption to Table~\ref{tab:Mnj_3j_}.
\label{tab:X_gamma_4j_}}
\end{center}
\end{table}
\clearpage

\begin{table}
\begin{center}
\begin{tabular}{|l c l||l l l l|}
\hline
\multicolumn{3}{|c||}{$y$ range} & $d\sigma/dy$ & $\pm$stat. & $\pm$syst. & $\pm$E-scale\\
\multicolumn{3}{|c||}{} & \multicolumn{4}{c|}{(pb)}\\
\hline
\multicolumn{7}{|c|}{$25\le M_{3j}<50$~GeV}\\
\hline
0.200 & - & 0.272 & 2264 & $\pm$ 22 & $_{-392}^{+400}$ & $_{-152}^{+152}$\\
0.272 & - & 0.344 & 2794 & $\pm$ 22 & $_{-382}^{+491}$ & $_{-286}^{+286}$\\
0.344 & - & 0.417 & 2854 & $\pm$ 22 & $_{-282}^{+390}$ & $_{-321}^{+321}$\\
0.417 & - & 0.489 & 2900 & $\pm$ 22 & $_{-238}^{+288}$ & $_{-340}^{+340}$\\
0.489 & - & 0.561 & 2973 & $\pm$ 22 & $_{-261}^{+270}$ & $_{-361}^{+361}$\\
0.561 & - & 0.633 & 2918 & $\pm$ 22 & $_{-242}^{+268}$ & $_{-363}^{+363}$\\
0.633 & - & 0.706 & 2833 & $\pm$ 22 & $_{-218}^{+232}$ & $_{-366}^{+366}$\\
0.706 & - & 0.778 & 2739 & $\pm$ 22 & $_{-216}^{+250}$ & $_{-380}^{+380}$\\
0.778 & - & 0.850 & 2797 & $\pm$ 23 & $_{-326}^{+340}$ & $_{-415}^{+415}$\\
\hline
\multicolumn{7}{|c|}{$M_{3j}\ge50$~GeV}\\
\hline
0.200 & - & 0.272 & 156 & $\pm$ 7 & $_{-38}^{+24}$ & $_{-16}^{+16}$\\
0.272 & - & 0.344 & 271 & $\pm$ 7 & $_{-13}^{+28}$ & $_{-24}^{+24}$\\
0.344 & - & 0.417 & 455 & $\pm$ 10 & $_{-65}^{+78}$ & $_{-43}^{+43}$\\
0.417 & - & 0.489 & 565 & $\pm$ 11 & $_{-91}^{+59}$ & $_{-52}^{+52}$\\
0.489 & - & 0.561 & 684 & $\pm$ 12 & $_{-108}^{+134}$ & $_{-67}^{+67}$\\
0.561 & - & 0.633 & 768 & $\pm$ 13 & $_{-131}^{+111}$ & $_{-81}^{+81}$\\
0.633 & - & 0.706 & 905 & $\pm$ 14 & $_{-199}^{+171}$ & $_{-100}^{+100}$\\
0.706 & - & 0.778 & 961 & $\pm$ 15 & $_{-206}^{+201}$ & $_{-113}^{+113}$\\
0.778 & - & 0.850 & 1058 & $\pm$ 16 & $_{-224}^{+259}$ & $_{-130}^{+130}$\\
\hline
\end{tabular}
\caption{The measured three-jet differential cross-section $d\sigma/dy$ in both the low- and high-mass regions. Other details as described in the caption to Table~\ref{tab:Mnj_3j_}.
\label{tab:Yjb_3j_}}
\end{center}
\end{table}
\clearpage

\begin{table}
\begin{center}
\begin{tabular}{|l c l||l l l l|}
\hline
\multicolumn{3}{|c||}{$y$ range} & $d\sigma/dy$ & $\pm$stat. & $\pm$syst. & $\pm$E-scale\\
\multicolumn{3}{|c||}{} & \multicolumn{4}{c|}{(pb)}\\
\hline
\multicolumn{7}{|c|}{$25\le M_{4j}<50$~GeV}\\
\hline
0.200 & - & 0.272 & 90 & $\pm$ 4 & $_{-25}^{+30}$ & $_{-8}^{+8}$\\
0.272 & - & 0.344 & 113 & $\pm$ 4 & $_{-14}^{+23}$ & $_{-18}^{+18}$\\
0.344 & - & 0.417 & 154 & $\pm$ 5 & $_{-24}^{+28}$ & $_{-29}^{+29}$\\
0.417 & - & 0.489 & 171 & $\pm$ 5 & $_{-17}^{+30}$ & $_{-33}^{+33}$\\
0.489 & - & 0.561 & 186 & $\pm$ 5 & $_{-23}^{+15}$ & $_{-37}^{+37}$\\
0.561 & - & 0.633 & 210 & $\pm$ 5 & $_{-27}^{+30}$ & $_{-39}^{+39}$\\
0.633 & - & 0.706 & 215 & $\pm$ 5 & $_{-28}^{+31}$ & $_{-42}^{+42}$\\
0.706 & - & 0.778 & 218 & $\pm$ 6 & $_{-24}^{+23}$ & $_{-44}^{+44}$\\
0.778 & - & 0.850 & 246 & $\pm$ 7 & $_{-43}^{+37}$ & $_{-52}^{+52}$\\
\hline
\multicolumn{7}{|c|}{$M_{4j}\ge50$~GeV}\\
\hline
0.200 & - & 0.272 & 16.8 & $\pm$ 2.4 & $_{-6.8}^{+5.3}$ & $_{-3.1}^{+3.1}$\\
0.272 & - & 0.344 & 40 & $\pm$ 3 & $_{-13}^{+13}$ & $_{-6}^{+6}$\\
0.344 & - & 0.417 & 81 & $\pm$ 4 & $_{-7}^{+13}$ & $_{-9}^{+9}$\\
0.417 & - & 0.489 & 108 & $\pm$ 5 & $_{-9}^{+8}$ & $_{-15}^{+15}$\\
0.489 & - & 0.561 & 136 & $\pm$ 5 & $_{-16}^{+13}$ & $_{-17}^{+17}$\\
0.561 & - & 0.633 & 171 & $\pm$ 6 & $_{-17}^{+28}$ & $_{-24}^{+24}$\\
0.633 & - & 0.706 & 228 & $\pm$ 7 & $_{-47}^{+33}$ & $_{-33}^{+33}$\\
0.706 & - & 0.778 & 266 & $\pm$ 8 & $_{-66}^{+54}$ & $_{-40}^{+40}$\\
0.778 & - & 0.850 & 266 & $\pm$ 8 & $_{-51}^{+60}$ & $_{-47}^{+47}$\\
\hline
\end{tabular}
\caption{The measured four-jet differential cross-section $d\sigma/dy$ in both the low- and high-mass regions. Other details as described in the caption to Table~\ref{tab:Mnj_3j_}.
\label{tab:Yjb_4j_}}
\end{center}
\end{table}
\clearpage

\begin{table}
\begin{center}
\begin{tabular}{|l c l||l l l l||l l l l|}
\hline
\multicolumn{3}{|c||}{$E_{T}^{{\rm jet1}}$ range} & $d\sigma/dE_{T}^{{\rm jet1}}$ & $\pm$stat. & $\pm$syst. & $\pm$E-scale & \multicolumn{2}{c}{had. corr.}&\multicolumn{2}{c|}{MPI corr.}\\
\multicolumn{3}{|c||}{(GeV)} & \multicolumn{4}{c||}{(pb/GeV)} & & & &\\
\hline
\multicolumn{11}{|c|}{$25\le M_{3j}<50$~GeV}\\
\hline
6.0 & - & 8.5 & 107 & $\pm$ 1 & $_{-17}^{+19}$ & $_{-11}^{+11}$ &0.744& $\pm$0.026&2.09& $\pm$0.70\\
8.5 & - & 11.2 & 248 & $\pm$ 1 & $_{-20}^{+20}$ & $_{-29}^{+29}$ &0.791& $\pm$0.012&1.67& $\pm$0.34\\
11.2 & - & 14.2 & 169 & $\pm$ 1 & $_{-10}^{+11}$ & $_{-21}^{+21}$ &0.819& $\pm$0.009&1.38& $\pm$0.17\\
14.2 & - & 17.6 & 75.2 & $\pm$ 0.6 & $_{-7.2}^{+6.9}$ & $_{-9.7}^{+9.7}$ &0.842& $\pm$0.003&1.24& $\pm$0.12\\
17.6 & - & 21.5 & 19.9 & $\pm$ 0.3 & $_{-2.3}^{+2.6}$ & $_{-2.6}^{+2.6}$ &0.834& $\pm$0.017&1.17& $\pm$0.09\\
21.5 & - & 26.0 & 2.54 & $\pm$ 0.09 & $_{-0.32}^{+0.38}$ & $_{-0.37}^{+0.37}$ &0.774& $\pm$0.064&1.16& $\pm$0.11\\
26.0 & - & 31.2 & 0.200 & $\pm$ 0.022 & $_{-0.077}^{+0.071}$ & $_{-0.040}^{+0.040}$ &0.725& $\pm$0.080&1.21& $\pm$0.20\\
\hline
\multicolumn{11}{|c|}{$M_{3j}\ge50$~GeV}\\
\hline
6.0 & - & 8.5 & 2.7 & $\pm$ 0.2 & $_{-1.0}^{+1.3}$ & $_{-0.6}^{+0.6}$ &0.515& $\pm$0.052&1.42& $\pm$0.29\\
8.5 & - & 11.2 & 14.4 & $\pm$ 0.4 & $_{-4.7}^{+5.8}$ & $_{-1.5}^{+1.5}$ &0.646& $\pm$0.044&1.28& $\pm$0.20\\
11.2 & - & 14.2 & 24.5 & $\pm$ 0.4 & $_{-5.6}^{+5.5}$ & $_{-2.5}^{+2.5}$ &0.759& $\pm$0.024&1.17& $\pm$0.16\\
14.2 & - & 17.6 & 27.8 & $\pm$ 0.4 & $_{-4.0}^{+4.7}$ & $_{-2.7}^{+2.7}$ &0.839& $\pm$0.018&1.13& $\pm$0.13\\
17.6 & - & 21.5 & 23.1 & $\pm$ 0.3 & $_{-2.1}^{+2.2}$ & $_{-2.2}^{+2.2}$ &0.878& $\pm$0.007&1.09& $\pm$0.09\\
21.5 & - & 26.0 & 14.1 & $\pm$ 0.2 & $_{-1.3}^{+1.5}$ & $_{-1.5}^{+1.5}$ &0.864& $\pm$0.016&1.07& $\pm$0.08\\
26.0 & - & 31.2 & 6.07 & $\pm$ 0.12 & $_{-0.58}^{+0.59}$ & $_{-0.72}^{+0.72}$ &0.859& $\pm$0.025&1.06& $\pm$0.07\\
31.2 & - & 37.0 & 2.06 & $\pm$ 0.07 & $_{-0.18}^{+0.19}$ & $_{-0.32}^{+0.32}$ &0.848& $\pm$0.034&1.04& $\pm$0.05\\
37.0 & - & 43.9 & 0.79 & $\pm$ 0.04 & $_{-0.16}^{+0.63}$ & $_{-0.22}^{+0.22}$ &0.838& $\pm$0.032&1.02& $\pm$0.04\\
43.9 & - & 51.8 & 0.246 & $\pm$ 0.020 & $_{-0.046}^{+0.044}$ & $_{-0.066}^{+0.066}$ &0.847& $\pm$0.045&1.04& $\pm$0.07\\
51.8 & - & 60.8 & 0.067 & $\pm$ 0.007 & $_{-0.031}^{+0.021}$ & $_{-0.013}^{+0.013}$ &0.881& $\pm$0.039&1.00& $\pm$0.04\\
60.8 & - & 71.0 & 0.035 & $\pm$ 0.004 & $_{-0.019}^{+0.019}$ & $_{-0.001}^{+0.001}$ &0.870& $\pm$0.001&1.02& $\pm$0.07\\
\hline
\end{tabular}
\caption{The measured three-jet differential cross-section $d\sigma/dE_{T}^{{\rm jet1}}$ in both the low- and high-mass regions. Other details as described in the caption to Table~\ref{tab:Mnj_3j_}.
\label{tab:Et3_3j_}}
\end{center}
\end{table}
\clearpage

\begin{table}
\begin{center}
\begin{tabular}{|l c l||l l l l||l l l l|}
\hline
\multicolumn{3}{|c||}{$E_{T}^{{\rm jet2}}$ range} & $d\sigma/dE_{T}^{{\rm jet2}}$ & $\pm$stat. & $\pm$syst. & $\pm$E-scale & \multicolumn{2}{c}{had. corr.}&\multicolumn{2}{c|}{MPI corr.}\\
\multicolumn{3}{|c||}{(GeV)} & \multicolumn{4}{c||}{(pb/GeV)} & & & &\\
\hline
\multicolumn{11}{|c|}{$25\le M_{3j}<50$~GeV}\\
\hline
6.0 & - & 8.5 & 383 & $\pm$ 1 & $_{-47}^{+48}$ & $_{-43}^{+43}$ &0.772& $\pm$0.023&1.68& $\pm$0.39\\
8.5 & - & 11.2 & 229 & $\pm$ 1 & $_{-9}^{+10}$ & $_{-28}^{+28}$ &0.829& $\pm$0.010&1.41& $\pm$0.18\\
11.2 & - & 14.2 & 59.0 & $\pm$ 0.5 & $_{-2.7}^{+3.7}$ & $_{-7.4}^{+7.4}$ &0.868& $\pm$0.018&1.25& $\pm$0.10\\
14.2 & - & 17.6 & 10.4 & $\pm$ 0.2 & $_{-1.0}^{+1.3}$ & $_{-1.3}^{+1.3}$ &0.869& $\pm$0.005&1.19& $\pm$0.06\\
17.6 & - & 21.5 & 0.84 & $\pm$ 0.05 & $_{-0.23}^{+0.42}$ & $_{-0.11}^{+0.11}$ &0.861& $\pm$0.042&1.16& $\pm$0.09\\
\hline
\multicolumn{11}{|c|}{$M_{3j}\ge50$~GeV}\\
\hline
6.0 & - & 8.5 & 22.2 & $\pm$ 0.5 & $_{-8.0}^{+9.0}$ & $_{-2.6}^{+2.6}$ &0.626& $\pm$0.027&1.26& $\pm$0.22\\
8.5 & - & 11.2 & 40 & $\pm$ 1 & $_{-9}^{+11}$ & $_{-4}^{+4}$ &0.759& $\pm$0.019&1.16& $\pm$0.15\\
11.2 & - & 14.2 & 34.3 & $\pm$ 0.4 & $_{-4.0}^{+4.1}$ & $_{-3.6}^{+3.6}$ &0.861& $\pm$0.005&1.11& $\pm$0.12\\
14.2 & - & 17.6 & 21.6 & $\pm$ 0.3 & $_{-1.4}^{+1.5}$ & $_{-2.3}^{+2.3}$ &0.891& $\pm$0.008&1.08& $\pm$0.09\\
17.6 & - & 21.5 & 11.5 & $\pm$ 0.2 & $_{-0.6}^{+0.7}$ & $_{-1.3}^{+1.3}$ &0.898& $\pm$0.011&1.07& $\pm$0.07\\
21.5 & - & 26.0 & 4.48 & $\pm$ 0.12 & $_{-0.54}^{+0.49}$ & $_{-0.47}^{+0.47}$ &0.888& $\pm$0.015&1.06& $\pm$0.07\\
26.0 & - & 31.2 & 1.57 & $\pm$ 0.07 & $_{-0.12}^{+0.18}$ & $_{-0.16}^{+0.16}$ &0.881& $\pm$0.010&1.06& $\pm$0.07\\
31.2 & - & 37.0 & 0.558 & $\pm$ 0.035 & $_{-0.091}^{+0.065}$ & $_{-0.077}^{+0.077}$ &0.892& $\pm$0.028&1.03& $\pm$0.05\\
37.0 & - & 43.9 & 0.214 & $\pm$ 0.021 & $_{-0.053}^{+0.046}$ & $_{-0.025}^{+0.025}$ &0.898& $\pm$0.003&1.01& $\pm$0.03\\
43.9 & - & 51.8 & 0.071 & $\pm$ 0.013 & $_{-0.020}^{+0.026}$ & $_{-0.008}^{+0.008}$ &0.898& $\pm$0.028&1.00& $\pm$0.05\\
51.8 & - & 60.8 & 0.042 & $\pm$ 0.009 & $_{-0.025}^{+0.022}$ & $_{-0.004}^{+0.004}$ &0.765& $\pm$0.092&1.02& $\pm$0.10\\
\hline
\end{tabular}
\caption{The measured three-jet differential cross-section $d\sigma/dE_{T}^{{\rm jet2}}$ in both the low- and high-mass regions. Other details as described in the caption to Table~\ref{tab:Mnj_3j_}.
\label{tab:Et4_3j_}}
\end{center}
\end{table}
\clearpage

\begin{table}
\begin{center}
\begin{tabular}{|l c l||l l l l||l l l l|}
\hline
\multicolumn{3}{|c||}{$E_{T}^{{\rm jet3}}$ range} & $d\sigma/dE_{T}^{{\rm jet3}}$ & $\pm$stat. & $\pm$syst. & $\pm$E-scale & \multicolumn{2}{c}{had. corr.}&\multicolumn{2}{c|}{MPI corr.}\\
\multicolumn{3}{|c||}{(GeV)} & \multicolumn{4}{c||}{(pb/GeV)} & & & &\\
\hline
\multicolumn{11}{|c|}{$25\le M_{3j}<50$~GeV}\\
\hline
6.0 & - & 8.5 & 644 & $\pm$ 2 & $_{-57}^{+68}$ & $_{-76}^{+76}$ &0.799& $\pm$0.011&1.53& $\pm$0.28\\
8.5 & - & 11.2 & 63.2 & $\pm$ 0.6 & $_{-2.8}^{+3.5}$ & $_{-8.5}^{+8.5}$ &0.859& $\pm$0.011&1.30& $\pm$0.08\\
11.2 & - & 14.2 & 3.82 & $\pm$ 0.14 & $_{-0.70}^{+0.82}$ & $_{-0.68}^{+0.68}$ &0.887& $\pm$0.017&1.24& $\pm$0.02\\
14.2 & - & 17.6 & 0.096 & $\pm$ 0.034 & $_{-0.090}^{+0.053}$ & $_{-0.050}^{+0.050}$ &0.834& $\pm$0.058&1.09& $\pm$0.21\\
\hline
\multicolumn{11}{|c|}{$M_{3j}\ge50$~GeV}\\
\hline
6.0 & - & 8.5 & 94 & $\pm$ 1 & $_{-20}^{+20}$ & $_{-11}^{+11}$ &0.767& $\pm$0.010&1.15& $\pm$0.14\\
8.5 & - & 11.2 & 43.1 & $\pm$ 0.5 & $_{-4.4}^{+5.1}$ & $_{-4.4}^{+4.4}$ &0.864& $\pm$0.010&1.08& $\pm$0.09\\
11.2 & - & 14.2 & 15.3 & $\pm$ 0.3 & $_{-1.3}^{+1.5}$ & $_{-1.7}^{+1.7}$ &0.914& $\pm$0.015&1.06& $\pm$0.07\\
14.2 & - & 17.6 & 5.0 & $\pm$ 0.1 & $_{-1.0}^{+1.0}$ & $_{-0.5}^{+0.5}$ &0.915& $\pm$0.020&1.04& $\pm$0.05\\
17.6 & - & 21.5 & 1.08 & $\pm$ 0.06 & $_{-0.15}^{+0.15}$ & $_{-0.16}^{+0.16}$ &0.919& $\pm$0.016&1.04& $\pm$0.06\\
21.5 & - & 26.0 & 0.289 & $\pm$ 0.026 & $_{-0.068}^{+0.086}$ & $_{-0.034}^{+0.034}$ &0.898& $\pm$0.032&1.00& $\pm$0.04\\
26.0 & - & 31.2 & 0.065 & $\pm$ 0.012 & $_{-0.034}^{+0.030}$ & $_{-0.006}^{+0.006}$ &0.887& $\pm$0.145&1.00& $\pm$0.08\\
31.2 & - & 40.0 & 0.0185 & $\pm$ 0.0053 & $_{-0.0078}^{+0.0085}$ & $_{-0.0037}^{+0.0037}$ &0.921& $\pm$0.054&1.05& $\pm$0.15\\
\hline
\end{tabular}
\caption{The measured three-jet differential cross-section $d\sigma/dE_{T}^{{\rm jet3}}$ in both the low- and high-mass regions. Other details as described in the caption to Table~\ref{tab:Mnj_3j_}.
\label{tab:Et5_3j_}}
\end{center}
\end{table}
\clearpage

\begin{table}
\begin{center}
\begin{tabular}{|l c l||l l l l|}
\hline
\multicolumn{3}{|c||}{$E_{T}^{{\rm jet1}}$ range} & $d\sigma/dE_{T}^{{\rm jet1}}$ & $\pm$stat. & $\pm$syst. & $\pm$E-scale\\
\multicolumn{3}{|c||}{(GeV)} & \multicolumn{4}{c|}{(pb/GeV)}\\
\hline
\multicolumn{7}{|c|}{$25\le M_{4j}<50$~GeV}\\
\hline
6.0 & - & 8.5 & 9.6 & $\pm$ 0.2 & $_{-1.6}^{+1.4}$ & $_{-1.7}^{+1.7}$\\
8.5 & - & 11.2 & 19.6 & $\pm$ 0.3 & $_{-2.0}^{+1.9}$ & $_{-3.7}^{+3.7}$\\
11.2 & - & 14.2 & 9.3 & $\pm$ 0.2 & $_{-1.1}^{+1.4}$ & $_{-1.8}^{+1.8}$\\
14.2 & - & 17.6 & 2.66 & $\pm$ 0.10 & $_{-0.52}^{+0.40}$ & $_{-0.50}^{+0.50}$\\
17.6 & - & 21.5 & 0.29 & $\pm$ 0.03 & $_{-0.06}^{+0.12}$ & $_{-0.06}^{+0.06}$\\
\hline
\multicolumn{7}{|c|}{$M_{4j}\ge50$~GeV}\\
\hline
6.0 & - & 8.5 & 1.00 & $\pm$ 0.16 & $_{-0.55}^{+0.48}$ & $_{-0.13}^{+0.13}$\\
8.5 & - & 11.2 & 6.1 & $\pm$ 0.2 & $_{-1.7}^{+1.8}$ & $_{-0.9}^{+0.9}$\\
11.2 & - & 14.2 & 8.3 & $\pm$ 0.2 & $_{-1.4}^{+1.4}$ & $_{-1.1}^{+1.1}$\\
14.2 & - & 17.6 & 6.46 & $\pm$ 0.16 & $_{-0.56}^{+0.71}$ & $_{-0.95}^{+0.95}$\\
17.6 & - & 21.5 & 3.93 & $\pm$ 0.11 & $_{-0.29}^{+0.43}$ & $_{-0.62}^{+0.62}$\\
21.5 & - & 26.0 & 1.93 & $\pm$ 0.07 & $_{-0.21}^{+0.29}$ & $_{-0.33}^{+0.33}$\\
26.0 & - & 31.2 & 0.74 & $\pm$ 0.04 & $_{-0.09}^{+0.09}$ & $_{-0.14}^{+0.14}$\\
31.2 & - & 37.0 & 0.289 & $\pm$ 0.022 & $_{-0.052}^{+0.044}$ & $_{-0.051}^{+0.051}$\\
37.0 & - & 43.9 & 0.100 & $\pm$ 0.014 & $_{-0.023}^{+0.028}$ & $_{-0.020}^{+0.020}$\\
43.9 & - & 51.8 & 0.016 & $\pm$ 0.004 & $_{-0.011}^{+0.017}$ & $_{-0.004}^{+0.004}$\\
51.8 & - & 60.8 & 0.0022 & $\pm$ 0.0014 & $_{-0.0022}^{+0.0040}$ & $_{-0.0010}^{+0.0010}$\\
\hline
\end{tabular}
\caption{The measured four-jet differential cross-section $d\sigma/dE_{T}^{{\rm jet1}}$ in both the low- and high-mass regions. Other details as described in the caption to Table~\ref{tab:Mnj_3j_}.
\label{tab:Et3_4j_}}
\end{center}
\end{table}
\clearpage

\begin{table}
\begin{center}
\begin{tabular}{|l c l||l l l l|}
\hline
\multicolumn{3}{|c||}{$E_{T}^{{\rm jet2}}$ range} & $d\sigma/dE_{T}^{{\rm jet2}}$ & $\pm$stat. & $\pm$syst. & $\pm$E-scale\\
\multicolumn{3}{|c||}{(GeV)} & \multicolumn{4}{c|}{(pb/GeV)}\\
\hline
\multicolumn{7}{|c|}{$25\le M_{4j}<50$~GeV}\\
\hline
6.0 & - & 8.5 & 27.0 & $\pm$ 0.3 & $_{-3.6}^{+3.1}$ & $_{-4.9}^{+4.9}$\\
8.5 & - & 11.2 & 15.3 & $\pm$ 0.3 & $_{-1.8}^{+2.2}$ & $_{-2.9}^{+2.9}$\\
11.2 & - & 14.2 & 2.04 & $\pm$ 0.10 & $_{-0.30}^{+0.50}$ & $_{-0.33}^{+0.33}$\\
14.2 & - & 17.6 & 0.066 & $\pm$ 0.017 & $_{-0.061}^{+0.050}$ & $_{-0.004}^{+0.004}$\\
\hline
\multicolumn{7}{|c|}{$M_{4j}\ge50$~GeV}\\
\hline
6.0 & - & 8.5 & 5.7 & $\pm$ 0.2 & $_{-1.5}^{+1.9}$ & $_{-0.7}^{+0.7}$\\
8.5 & - & 11.2 & 12.7 & $\pm$ 0.3 & $_{-1.4}^{+1.8}$ & $_{-1.9}^{+1.9}$\\
11.2 & - & 14.2 & 8.4 & $\pm$ 0.2 & $_{-0.7}^{+0.5}$ & $_{-1.4}^{+1.4}$\\
14.2 & - & 17.6 & 3.40 & $\pm$ 0.13 & $_{-0.25}^{+0.48}$ & $_{-0.48}^{+0.48}$\\
17.6 & - & 21.5 & 1.55 & $\pm$ 0.08 & $_{-0.51}^{+0.26}$ & $_{-0.24}^{+0.24}$\\
21.5 & - & 26.0 & 0.55 & $\pm$ 0.05 & $_{-0.09}^{+0.17}$ & $_{-0.08}^{+0.08}$\\
26.0 & - & 31.2 & 0.152 & $\pm$ 0.021 & $_{-0.032}^{+0.043}$ & $_{-0.036}^{+0.036}$\\
31.2 & - & 37.0 & 0.058 & $\pm$ 0.016 & $_{-0.029}^{+0.036}$ & $_{-0.013}^{+0.013}$\\
37.0 & - & 43.9 & 0.017 & $\pm$ 0.005 & $_{-0.009}^{+0.014}$ & $_{-0.005}^{+0.005}$\\
\hline
\end{tabular}
\caption{The measured four-jet differential cross-section $d\sigma/dE_{T}^{{\rm jet2}}$ in both the low- and high-mass regions. Other details as described in the caption to Table~\ref{tab:Mnj_3j_}.
\label{tab:Et4_4j_}}
\end{center}
\end{table}
\clearpage

\begin{table}
\begin{center}
\begin{tabular}{|l c l||l l l l|}
\hline
\multicolumn{3}{|c||}{$E_{T}^{{\rm jet3}}$ range} & $d\sigma/dE_{T}^{{\rm jet3}}$ & $\pm$stat. & $\pm$syst. & $\pm$E-scale\\
\multicolumn{3}{|c||}{(GeV)} & \multicolumn{4}{c|}{(pb/GeV)}\\
\hline
\multicolumn{7}{|c|}{$25\le M_{4j}<50$~GeV}\\
\hline
6.0 & - & 8.5 & 40.8 & $\pm$ 0.4 & $_{-4.0}^{+4.7}$ & $_{-7.7}^{+7.7}$\\
8.5 & - & 11.2 & 4.72 & $\pm$ 0.14 & $_{-0.72}^{+0.99}$ & $_{-0.85}^{+0.85}$\\
11.2 & - & 14.2 & 0.054 & $\pm$ 0.019 & $_{-0.047}^{+0.060}$ & $_{-0.005}^{+0.005}$\\
\hline
\multicolumn{7}{|c|}{$M_{4j}\ge50$~GeV}\\
\hline
6.0 & - & 8.5 & 17.5 & $\pm$ 0.4 & $_{-2.6}^{+3.7}$ & $_{-2.6}^{+2.6}$\\
8.5 & - & 11.2 & 13.1 & $\pm$ 0.3 & $_{-1.0}^{+1.4}$ & $_{-1.7}^{+1.7}$\\
11.2 & - & 14.2 & 4.3 & $\pm$ 0.1 & $_{-1.5}^{+0.7}$ & $_{-0.6}^{+0.6}$\\
14.2 & - & 17.6 & 0.99 & $\pm$ 0.07 & $_{-0.22}^{+0.21}$ & $_{-0.10}^{+0.10}$\\
17.6 & - & 21.5 & 0.191 & $\pm$ 0.031 & $_{-0.049}^{+0.075}$ & $_{-0.051}^{+0.051}$\\
21.5 & - & 26.0 & 0.035 & $\pm$ 0.009 & $_{-0.015}^{+0.017}$ & $_{-0.007}^{+0.007}$\\
\hline
\end{tabular}
\caption{The measured four-jet differential cross-section $d\sigma/dE_{T}^{{\rm jet3}}$ in both the low- and high-mass regions. Other details as described in the caption to Table~\ref{tab:Mnj_3j_}.
\label{tab:Et5_4j_}}
\end{center}
\end{table}

\begin{table}
\begin{center}
\begin{tabular}{|l c l||l l l l|}
\hline
\multicolumn{3}{|c||}{$E_{T}^{{\rm jet4}}$ range} & $d\sigma/dE_{T}^{{\rm jet4}}$ & $\pm$stat. & $\pm$syst. & $\pm$E-scale\\
\multicolumn{3}{|c||}{(GeV)} & \multicolumn{4}{c|}{(pb/GeV)}\\
\hline
\multicolumn{7}{|c|}{$25\le M_{4j}<50$~GeV}\\
\hline
6.0 & - & 8.5 & 45.4 & $\pm$ 0.4 & $_{-2.4}^{+5.3}$ & $_{-8.6}^{+8.6}$\\
8.5 & - & 11.2 & 0.53 & $\pm$ 0.05 & $_{-0.11}^{+0.15}$ & $_{-0.11}^{+0.11}$\\
\hline
\multicolumn{7}{|c|}{$M_{4j}\ge50$~GeV}\\
\hline
6.0 & - & 8.5 & 32.0 & $\pm$ 0.5 & $_{-3.8}^{+4.5}$ & $_{-4.9}^{+4.9}$\\
8.5 & - & 11.2 & 5.02 & $\pm$ 0.17 & $_{-0.45}^{+0.37}$ & $_{-0.71}^{+0.71}$\\
11.2 & - & 14.2 & 0.81 & $\pm$ 0.07 & $_{-0.24}^{+0.59}$ & $_{-0.09}^{+0.09}$\\
14.2 & - & 20.0 & 0.086 & $\pm$ 0.014 & $_{-0.034}^{+0.044}$ & $_{-0.004}^{+0.004}$\\
\hline
\end{tabular}
\caption{The measured four-jet differential cross-section $d\sigma/dE_{T}^{{\rm jet4}}$ in both the low- and high-mass regions. Other details as described in the caption to Table~\ref{tab:Mnj_3j_}.
\label{tab:Et6_4j_}}
\end{center}
\end{table}
\clearpage

\begin{table}
\begin{center}
\begin{tabular}{|l c l||l l l l||l l l l|}
\hline
\multicolumn{3}{|c||}{$\eta^{{\rm jet1}}$ range} & $d\sigma/d\eta^{{\rm jet1}}$ & $\pm$stat. & $\pm$syst. & $\pm$E-scale & \multicolumn{2}{c}{had. corr.}&\multicolumn{2}{c|}{MPI corr.}\\
\multicolumn{3}{|c||}{} & \multicolumn{4}{c||}{(pb)} & & & &\\
\hline
\multicolumn{11}{|c|}{$25\le M_{3j}<50$~GeV}\\
\hline
-1.686 & - & -1.371 & 3.3 & $\pm$ 0.4 & $_{-1.7}^{+1.4}$ & $_{-0.5}^{+0.5}$ &0.324& $\pm$0.012&1.15& $\pm$0.21\\
-1.371 & - & -1.057 & 36.0 & $\pm$ 1.4 & $_{-9.9}^{+7.8}$ & $_{-2.9}^{+2.9}$ &0.538& $\pm$0.041&1.34& $\pm$0.23\\
-1.057 & - & -0.743 & 123 & $\pm$ 2 & $_{-20}^{+18}$ & $_{-9}^{+9}$ &0.686& $\pm$0.044&1.33& $\pm$0.18\\
-0.743 & - & -0.429 & 256 & $\pm$ 3 & $_{-37}^{+40}$ & $_{-19}^{+19}$ &0.763& $\pm$0.025&1.35& $\pm$0.19\\
-0.429 & - & -0.114 & 394 & $\pm$ 4 & $_{-51}^{+55}$ & $_{-33}^{+33}$ &0.793& $\pm$0.025&1.35& $\pm$0.19\\
-0.114 & - & 0.200 & 506 & $\pm$ 4 & $_{-63}^{+67}$ & $_{-51}^{+51}$ &0.803& $\pm$0.014&1.37& $\pm$0.21\\
0.200 & - & 0.514 & 600 & $\pm$ 5 & $_{-65}^{+68}$ & $_{-65}^{+65}$ &0.820& $\pm$0.009&1.39& $\pm$0.21\\
0.514 & - & 0.829 & 653 & $\pm$ 5 & $_{-57}^{+65}$ & $_{-74}^{+74}$ &0.816& $\pm$0.008&1.43& $\pm$0.22\\
0.829 & - & 1.143 & 631 & $\pm$ 5 & $_{-49}^{+54}$ & $_{-76}^{+76}$ &0.814& $\pm$0.004&1.49& $\pm$0.25\\
1.143 & - & 1.457 & 556 & $\pm$ 4 & $_{-58}^{+63}$ & $_{-72}^{+72}$ &0.814& $\pm$0.007&1.55& $\pm$0.28\\
1.457 & - & 1.771 & 573 & $\pm$ 4 & $_{-50}^{+58}$ & $_{-85}^{+85}$ &0.821& $\pm$0.003&1.64& $\pm$0.31\\
1.771 & - & 2.086 & 697 & $\pm$ 6 & $_{-84}^{+97}$ & $_{-103}^{+103}$ &0.829& $\pm$0.002&1.75& $\pm$0.36\\
2.086 & - & 2.400 & 790 & $\pm$ 9 & $_{-186}^{+186}$ & $_{-107}^{+107}$ &0.843& $\pm$0.005&1.90& $\pm$0.42\\
\hline
\multicolumn{11}{|c|}{$M_{3j}\ge50$~GeV}\\
\hline
-1.686 & - & -1.371 & 2.4 & $\pm$ 0.5 & $_{-1.6}^{+1.7}$ & $_{-0.6}^{+0.6}$ &0.300& $\pm$0.040&1.19& $\pm$0.25\\
-1.371 & - & -1.057 & 14.7 & $\pm$ 1.1 & $_{-4.0}^{+5.5}$ & $_{-2.0}^{+2.0}$ &0.498& $\pm$0.022&1.20& $\pm$0.15\\
-1.057 & - & -0.743 & 42 & $\pm$ 2 & $_{-11}^{+12}$ & $_{-4}^{+4}$ &0.648& $\pm$0.030&1.13& $\pm$0.10\\
-0.743 & - & -0.429 & 67 & $\pm$ 2 & $_{-15}^{+16}$ & $_{-6}^{+6}$ &0.757& $\pm$0.019&1.10& $\pm$0.10\\
-0.429 & - & -0.114 & 84 & $\pm$ 2 & $_{-16}^{+18}$ & $_{-7}^{+7}$ &0.830& $\pm$0.019&1.08& $\pm$0.09\\
-0.114 & - & 0.200 & 90 & $\pm$ 2 & $_{-10}^{+11}$ & $_{-8}^{+8}$ &0.863& $\pm$0.010&1.06& $\pm$0.06\\
0.200 & - & 0.514 & 99 & $\pm$ 2 & $_{-16}^{+14}$ & $_{-9}^{+9}$ &0.867& $\pm$0.008&1.06& $\pm$0.07\\
0.514 & - & 0.829 & 92 & $\pm$ 2 & $_{-9}^{+10}$ & $_{-10}^{+10}$ &0.840& $\pm$0.006&1.07& $\pm$0.08\\
0.829 & - & 1.143 & 99 & $\pm$ 2 & $_{-13}^{+13}$ & $_{-11}^{+11}$ &0.830& $\pm$0.018&1.08& $\pm$0.09\\
1.143 & - & 1.457 & 115 & $\pm$ 2 & $_{-19}^{+14}$ & $_{-14}^{+14}$ &0.812& $\pm$0.026&1.11& $\pm$0.12\\
1.457 & - & 1.771 & 145 & $\pm$ 2 & $_{-21}^{+17}$ & $_{-18}^{+18}$ &0.808& $\pm$0.019&1.12& $\pm$0.12\\
1.771 & - & 2.086 & 200 & $\pm$ 3 & $_{-34}^{+31}$ & $_{-23}^{+23}$ &0.818& $\pm$0.012&1.16& $\pm$0.16\\
2.086 & - & 2.400 & 283 & $\pm$ 4 & $_{-54}^{+66}$ & $_{-32}^{+32}$ &0.829& $\pm$0.012&1.22& $\pm$0.17\\
\hline
\end{tabular}
\caption{The measured three-jet differential cross-section $d\sigma/d\eta^{{\rm jet1}}$ in both the low- and high-mass regions. Other details as described in the caption to Table~\ref{tab:Mnj_3j_}.
\label{tab:Eta3_3j_}}
\end{center}
\end{table}
\clearpage

\begin{table}
\begin{center}
\begin{tabular}{|l c l||l l l l||l l l l|}
\hline
\multicolumn{3}{|c||}{$\eta^{{\rm jet2}}$ range} & $d\sigma/d\eta^{{\rm jet2}}$ & $\pm$stat. & $\pm$syst. & $\pm$E-scale & \multicolumn{2}{c}{had. corr.}&\multicolumn{2}{c|}{MPI corr.}\\
\multicolumn{3}{|c||}{} & \multicolumn{4}{c||}{(pb)} & & & &\\
\hline
\multicolumn{11}{|c|}{$25\le M_{3j}<50$~GeV}\\
\hline
-1.686 & - & -1.371 & 9.1 & $\pm$ 0.8 & $_{-3.4}^{+7.4}$ & $_{-1.8}^{+1.8}$ &0.382& $\pm$0.005&1.17& $\pm$0.20\\
-1.371 & - & -1.057 & 57 & $\pm$ 2 & $_{-10}^{+11}$ & $_{-9}^{+9}$ &0.592& $\pm$0.019&1.35& $\pm$0.27\\
-1.057 & - & -0.743 & 168 & $\pm$ 3 & $_{-25}^{+21}$ & $_{-15}^{+15}$ &0.720& $\pm$0.024&1.40& $\pm$0.25\\
-0.743 & - & -0.429 & 282 & $\pm$ 4 & $_{-30}^{+39}$ & $_{-26}^{+26}$ &0.766& $\pm$0.015&1.42& $\pm$0.26\\
-0.429 & - & -0.114 & 408 & $\pm$ 4 & $_{-44}^{+41}$ & $_{-39}^{+39}$ &0.780& $\pm$0.012&1.43& $\pm$0.25\\
-0.114 & - & 0.200 & 492 & $\pm$ 4 & $_{-46}^{+54}$ & $_{-50}^{+50}$ &0.808& $\pm$0.007&1.42& $\pm$0.22\\
0.200 & - & 0.514 & 574 & $\pm$ 5 & $_{-57}^{+66}$ & $_{-67}^{+67}$ &0.817& $\pm$0.005&1.44& $\pm$0.22\\
0.514 & - & 0.829 & 628 & $\pm$ 5 & $_{-61}^{+67}$ & $_{-73}^{+73}$ &0.817& $\pm$0.003&1.45& $\pm$0.21\\
0.829 & - & 1.143 & 615 & $\pm$ 5 & $_{-57}^{+62}$ & $_{-72}^{+72}$ &0.816& $\pm$0.003&1.48& $\pm$0.22\\
1.143 & - & 1.457 & 550 & $\pm$ 4 & $_{-34}^{+40}$ & $_{-66}^{+66}$ &0.818& $\pm$0.013&1.50& $\pm$0.25\\
1.457 & - & 1.771 & 564 & $\pm$ 4 & $_{-32}^{+36}$ & $_{-77}^{+77}$ &0.819& $\pm$0.012&1.56& $\pm$0.29\\
1.771 & - & 2.086 & 686 & $\pm$ 6 & $_{-82}^{+79}$ & $_{-96}^{+96}$ &0.833& $\pm$0.015&1.63& $\pm$0.33\\
2.086 & - & 2.400 & 767 & $\pm$ 8 & $_{-181}^{+183}$ & $_{-96}^{+96}$ &0.844& $\pm$0.015&1.75& $\pm$0.38\\
\hline
\multicolumn{11}{|c|}{$M_{3j}\ge50$~GeV}\\
\hline
-2.000 & - & -1.686 & 0.48 & $\pm$ 0.16 & $_{-0.27}^{+0.44}$ & $_{-0.18}^{+0.18}$ &0.085& $\pm$0.007&1.00& $\pm$0.18\\
-1.686 & - & -1.371 & 8.2 & $\pm$ 0.7 & $_{-2.5}^{+2.5}$ & $_{-0.9}^{+0.9}$ &0.347& $\pm$0.021&1.20& $\pm$0.22\\
-1.371 & - & -1.057 & 43 & $\pm$ 2 & $_{-19}^{+16}$ & $_{-5}^{+5}$ &0.571& $\pm$0.025&1.20& $\pm$0.17\\
-1.057 & - & -0.743 & 70 & $\pm$ 2 & $_{-18}^{+20}$ & $_{-7}^{+7}$ &0.710& $\pm$0.038&1.18& $\pm$0.14\\
-0.743 & - & -0.429 & 92 & $\pm$ 2 & $_{-24}^{+22}$ & $_{-11}^{+11}$ &0.783& $\pm$0.014&1.14& $\pm$0.12\\
-0.429 & - & -0.114 & 97 & $\pm$ 2 & $_{-15}^{+20}$ & $_{-10}^{+10}$ &0.822& $\pm$0.013&1.12& $\pm$0.09\\
-0.114 & - & 0.200 & 96 & $\pm$ 2 & $_{-12}^{+12}$ & $_{-10}^{+10}$ &0.845& $\pm$0.006&1.10& $\pm$0.10\\
0.200 & - & 0.514 & 104 & $\pm$ 2 & $_{-13}^{+12}$ & $_{-11}^{+11}$ &0.852& $\pm$0.007&1.09& $\pm$0.08\\
0.514 & - & 0.829 & 106 & $\pm$ 2 & $_{-14}^{+16}$ & $_{-12}^{+12}$ &0.858& $\pm$0.001&1.09& $\pm$0.09\\
0.829 & - & 1.143 & 108 & $\pm$ 2 & $_{-17}^{+14}$ & $_{-11}^{+11}$ &0.851& $\pm$0.009&1.09& $\pm$0.10\\
1.143 & - & 1.457 & 103 & $\pm$ 2 & $_{-16}^{+16}$ & $_{-11}^{+11}$ &0.846& $\pm$0.011&1.09& $\pm$0.10\\
1.457 & - & 1.771 & 113 & $\pm$ 2 & $_{-14}^{+19}$ & $_{-13}^{+13}$ &0.827& $\pm$0.018&1.10& $\pm$0.11\\
1.771 & - & 2.086 & 157 & $\pm$ 3 & $_{-20}^{+26}$ & $_{-17}^{+17}$ &0.829& $\pm$0.013&1.12& $\pm$0.13\\
2.086 & - & 2.400 & 232 & $\pm$ 4 & $_{-49}^{+51}$ & $_{-23}^{+23}$ &0.839& $\pm$0.010&1.14& $\pm$0.15\\
\hline
\end{tabular}
\caption{The measured three-jet differential cross-section $d\sigma/d\eta^{{\rm jet2}}$ in both the low- and high-mass regions. Other details as described in the caption to Table~\ref{tab:Mnj_3j_}.
\label{tab:Eta4_3j_}}
\end{center}
\end{table}
\clearpage

\begin{table}
\begin{center}
\begin{tabular}{|l c l||l l l l||l l l l|}
\hline
\multicolumn{3}{|c||}{$\eta^{{\rm jet3}}$ range} & $d\sigma/d\eta^{{\rm jet3}}$ & $\pm$stat. & $\pm$syst. & $\pm$E-scale & \multicolumn{2}{c}{had. corr.}&\multicolumn{2}{c|}{MPI corr.}\\
\multicolumn{3}{|c||}{} & \multicolumn{4}{c||}{(pb)} & & & &\\
\hline
\multicolumn{11}{|c|}{$25\le M_{3j}<50$~GeV}\\
\hline
-2.000 & - & -1.686 & 0.33 & $\pm$ 0.23 & $_{-0.33}^{+0.49}$ & $_{-0.08}^{+0.08}$ &0.090& $\pm$0.045&1.42& $\pm$0.82\\
-1.686 & - & -1.371 & 15.2 & $\pm$ 1.0 & $_{-3.5}^{+3.4}$ & $_{-1.3}^{+1.3}$ &0.388& $\pm$0.019&1.30& $\pm$0.33\\
-1.371 & - & -1.057 & 80 & $\pm$ 2 & $_{-13}^{+16}$ & $_{-7}^{+7}$ &0.631& $\pm$0.004&1.33& $\pm$0.33\\
-1.057 & - & -0.743 & 187 & $\pm$ 3 & $_{-24}^{+25}$ & $_{-17}^{+17}$ &0.738& $\pm$0.007&1.38& $\pm$0.34\\
-0.743 & - & -0.429 & 308 & $\pm$ 4 & $_{-39}^{+37}$ & $_{-29}^{+29}$ &0.791& $\pm$0.002&1.41& $\pm$0.31\\
-0.429 & - & -0.114 & 396 & $\pm$ 4 & $_{-35}^{+46}$ & $_{-42}^{+42}$ &0.800& $\pm$0.006&1.44& $\pm$0.30\\
-0.114 & - & 0.200 & 474 & $\pm$ 4 & $_{-44}^{+49}$ & $_{-53}^{+53}$ &0.826& $\pm$0.009&1.46& $\pm$0.27\\
0.200 & - & 0.514 & 560 & $\pm$ 5 & $_{-60}^{+67}$ & $_{-67}^{+67}$ &0.829& $\pm$0.015&1.49& $\pm$0.26\\
0.514 & - & 0.829 & 616 & $\pm$ 5 & $_{-74}^{+73}$ & $_{-72}^{+72}$ &0.818& $\pm$0.000&1.51& $\pm$0.22\\
0.829 & - & 1.143 & 605 & $\pm$ 5 & $_{-58}^{+67}$ & $_{-71}^{+71}$ &0.807& $\pm$0.010&1.52& $\pm$0.21\\
1.143 & - & 1.457 & 538 & $\pm$ 4 & $_{-29}^{+38}$ & $_{-64}^{+64}$ &0.806& $\pm$0.017&1.50& $\pm$0.19\\
1.457 & - & 1.771 & 559 & $\pm$ 4 & $_{-30}^{+34}$ & $_{-72}^{+72}$ &0.808& $\pm$0.022&1.52& $\pm$0.21\\
1.771 & - & 2.086 & 685 & $\pm$ 5 & $_{-75}^{+79}$ & $_{-91}^{+91}$ &0.822& $\pm$0.032&1.57& $\pm$0.24\\
2.086 & - & 2.400 & 793 & $\pm$ 8 & $_{-177}^{+183}$ & $_{-103}^{+103}$ &0.835& $\pm$0.030&1.66& $\pm$0.29\\
\hline
\multicolumn{11}{|c|}{$M_{3j}\ge50$~GeV}\\
\hline
-2.000 & - & -1.686 & 1.9 & $\pm$ 0.3 & $_{-1.2}^{+1.2}$ & $_{-1.0}^{+1.0}$ &0.145& $\pm$0.014&1.01& $\pm$0.11\\
-1.686 & - & -1.371 & 26.1 & $\pm$ 1.3 & $_{-6.9}^{+9.5}$ & $_{-3.1}^{+3.1}$ &0.456& $\pm$0.004&1.19& $\pm$0.22\\
-1.371 & - & -1.057 & 70 & $\pm$ 2 & $_{-16}^{+14}$ & $_{-9}^{+9}$ &0.673& $\pm$0.003&1.20& $\pm$0.22\\
-1.057 & - & -0.743 & 104 & $\pm$ 2 & $_{-22}^{+26}$ & $_{-11}^{+11}$ &0.788& $\pm$0.001&1.18& $\pm$0.19\\
-0.743 & - & -0.429 & 100 & $\pm$ 2 & $_{-16}^{+20}$ & $_{-10}^{+10}$ &0.840& $\pm$0.005&1.14& $\pm$0.15\\
-0.429 & - & -0.114 & 97 & $\pm$ 2 & $_{-16}^{+12}$ & $_{-10}^{+10}$ &0.876& $\pm$0.014&1.13& $\pm$0.14\\
-0.114 & - & 0.200 & 92 & $\pm$ 2 & $_{-13}^{+11}$ & $_{-10}^{+10}$ &0.901& $\pm$0.033&1.11& $\pm$0.12\\
0.200 & - & 0.514 & 89 & $\pm$ 2 & $_{-15}^{+19}$ & $_{-9}^{+9}$ &0.900& $\pm$0.036&1.10& $\pm$0.11\\
0.514 & - & 0.829 & 98 & $\pm$ 2 & $_{-16}^{+17}$ & $_{-11}^{+11}$ &0.856& $\pm$0.022&1.12& $\pm$0.10\\
0.829 & - & 1.143 & 95 & $\pm$ 2 & $_{-14}^{+16}$ & $_{-10}^{+10}$ &0.840& $\pm$0.002&1.11& $\pm$0.09\\
1.143 & - & 1.457 & 89 & $\pm$ 2 & $_{-12}^{+11}$ & $_{-10}^{+10}$ &0.819& $\pm$0.012&1.12& $\pm$0.08\\
1.457 & - & 1.771 & 103 & $\pm$ 2 & $_{-15}^{+16}$ & $_{-12}^{+12}$ &0.815& $\pm$0.023&1.11& $\pm$0.06\\
1.771 & - & 2.086 & 138 & $\pm$ 3 & $_{-21}^{+22}$ & $_{-16}^{+16}$ &0.817& $\pm$0.030&1.11& $\pm$0.08\\
2.086 & - & 2.400 & 222 & $\pm$ 4 & $_{-49}^{+44}$ & $_{-19}^{+19}$ &0.820& $\pm$0.042&1.12& $\pm$0.07\\
\hline
\end{tabular}
\caption{The measured three-jet differential cross-section $d\sigma/d\eta^{{\rm jet3}}$ in both the low- and high-mass regions. Other details as described in the caption to Table~\ref{tab:Mnj_3j_}.
\label{tab:Eta5_3j_}}
\end{center}
\end{table}
\clearpage

\begin{table}
\begin{center}
\begin{tabular}{|l c l||l l l l|}
\hline
\multicolumn{3}{|c||}{$\eta^{{\rm jet1}}$ range} & $d\sigma/d\eta^{{\rm jet1}}$ & $\pm$stat. & $\pm$syst. & $\pm$E-scale\\
\multicolumn{3}{|c||}{} & \multicolumn{4}{c|}{(pb)}\\
\hline
\multicolumn{7}{|c|}{$25\le M_{4j}<50$~GeV}\\
\hline
-1.371 & - & -1.057 & 0.61 & $\pm$ 0.09 & $_{-0.36}^{+0.91}$ & $_{-0.08}^{+0.08}$\\
-1.057 & - & -0.743 & 1.76 & $\pm$ 0.23 & $_{-0.67}^{+0.85}$ & $_{-0.30}^{+0.30}$\\
-0.743 & - & -0.429 & 6.1 & $\pm$ 0.4 & $_{-1.5}^{+1.0}$ & $_{-1.0}^{+1.0}$\\
-0.429 & - & -0.114 & 13.2 & $\pm$ 0.6 & $_{-3.0}^{+2.2}$ & $_{-2.1}^{+2.1}$\\
-0.114 & - & 0.200 & 21.9 & $\pm$ 0.8 & $_{-3.6}^{+4.4}$ & $_{-4.0}^{+4.0}$\\
0.200 & - & 0.514 & 36.7 & $\pm$ 1.1 & $_{-6.7}^{+5.7}$ & $_{-6.4}^{+6.4}$\\
0.514 & - & 0.829 & 43.3 & $\pm$ 1.3 & $_{-4.6}^{+6.0}$ & $_{-7.5}^{+7.5}$\\
0.829 & - & 1.143 & 55 & $\pm$ 1 & $_{-12}^{+12}$ & $_{-9}^{+9}$\\
1.143 & - & 1.457 & 43 & $\pm$ 1 & $_{-8}^{+10}$ & $_{-8}^{+8}$\\
1.457 & - & 1.771 & 44.5 & $\pm$ 1.1 & $_{-4.1}^{+8.1}$ & $_{-9.5}^{+9.5}$\\
1.771 & - & 2.086 & 54 & $\pm$ 2 & $_{-9}^{+11}$ & $_{-12}^{+12}$\\
2.086 & - & 2.400 & 68 & $\pm$ 3 & $_{-24}^{+25}$ & $_{-13}^{+13}$\\
\hline
\multicolumn{7}{|c|}{$M_{4j}\ge50$~GeV}\\
\hline
-1.371 & - & -1.057 & 1.10 & $\pm$ 0.29 & $_{-0.70}^{+0.63}$ & $_{-0.14}^{+0.14}$\\
-1.057 & - & -0.743 & 4.9 & $\pm$ 0.6 & $_{-1.4}^{+2.3}$ & $_{-0.6}^{+0.6}$\\
-0.743 & - & -0.429 & 10.9 & $\pm$ 0.8 & $_{-2.3}^{+3.7}$ & $_{-1.2}^{+1.2}$\\
-0.429 & - & -0.114 & 20.6 & $\pm$ 0.9 & $_{-7.1}^{+5.1}$ & $_{-2.5}^{+2.5}$\\
-0.114 & - & 0.200 & 20.9 & $\pm$ 1.0 & $_{-2.8}^{+3.7}$ & $_{-2.7}^{+2.7}$\\
0.200 & - & 0.514 & 24.0 & $\pm$ 1.1 & $_{-2.5}^{+4.0}$ & $_{-3.3}^{+3.3}$\\
0.514 & - & 0.829 & 24.4 & $\pm$ 1.1 & $_{-2.5}^{+4.8}$ & $_{-3.4}^{+3.4}$\\
0.829 & - & 1.143 & 27.8 & $\pm$ 1.2 & $_{-3.4}^{+3.5}$ & $_{-3.6}^{+3.6}$\\
1.143 & - & 1.457 & 28.0 & $\pm$ 1.0 & $_{-4.7}^{+4.2}$ & $_{-4.3}^{+4.3}$\\
1.457 & - & 1.771 & 37.0 & $\pm$ 1.3 & $_{-3.7}^{+5.9}$ & $_{-6.4}^{+6.4}$\\
1.771 & - & 2.086 & 47.0 & $\pm$ 1.7 & $_{-8.9}^{+7.6}$ & $_{-8.6}^{+8.6}$\\
2.086 & - & 2.400 & 63 & $\pm$ 2 & $_{-17}^{+16}$ & $_{-11}^{+11}$\\
\hline
\end{tabular}
\caption{The measured four-jet differential cross-section $d\sigma/d\eta^{{\rm jet1}}$ in both the low- and high-mass regions. Other details as described in the caption to Table~\ref{tab:Mnj_3j_}.
\label{tab:Eta3_4j_}}
\end{center}
\end{table}
\clearpage

\begin{table}
\begin{center}
\begin{tabular}{|l c l||l l l l|}
\hline
\multicolumn{3}{|c||}{$\eta^{{\rm jet2}}$ range} & $d\sigma/d\eta^{{\rm jet2}}$ & $\pm$stat. & $\pm$syst. & $\pm$E-scale\\
\multicolumn{3}{|c||}{} & \multicolumn{4}{c|}{(pb)}\\
\hline
\multicolumn{7}{|c|}{$25\le M_{4j}<50$~GeV}\\
\hline
-1.371 & - & -1.057 & 0.67 & $\pm$ 0.18 & $_{-0.50}^{+0.43}$ & $_{-0.11}^{+0.11}$\\
-1.057 & - & -0.743 & 3.15 & $\pm$ 0.37 & $_{-0.73}^{+0.87}$ & $_{-0.35}^{+0.35}$\\
-0.743 & - & -0.429 & 11.3 & $\pm$ 0.5 & $_{-6.9}^{+6.5}$ & $_{-1.0}^{+1.0}$\\
-0.429 & - & -0.114 & 17.0 & $\pm$ 0.8 & $_{-2.8}^{+2.4}$ & $_{-1.9}^{+1.9}$\\
-0.114 & - & 0.200 & 23.5 & $\pm$ 0.9 & $_{-3.6}^{+4.1}$ & $_{-3.5}^{+3.5}$\\
0.200 & - & 0.514 & 37.9 & $\pm$ 1.2 & $_{-5.8}^{+4.9}$ & $_{-6.3}^{+6.3}$\\
0.514 & - & 0.829 & 42.4 & $\pm$ 1.3 & $_{-5.7}^{+4.9}$ & $_{-7.5}^{+7.5}$\\
0.829 & - & 1.143 & 46.2 & $\pm$ 1.3 & $_{-6.4}^{+4.2}$ & $_{-9.0}^{+9.0}$\\
1.143 & - & 1.457 & 40.3 & $\pm$ 1.1 & $_{-6.9}^{+8.7}$ & $_{-8.3}^{+8.3}$\\
1.457 & - & 1.771 & 43.3 & $\pm$ 1.1 & $_{-5.5}^{+8.7}$ & $_{-9.0}^{+9.0}$\\
1.771 & - & 2.086 & 52 & $\pm$ 1 & $_{-7}^{+10}$ & $_{-11}^{+11}$\\
2.086 & - & 2.400 & 57 & $\pm$ 2 & $_{-15}^{+19}$ & $_{-11}^{+11}$\\
\hline
\multicolumn{7}{|c|}{$M_{4j}\ge50$~GeV}\\
\hline
-1.686 & - & -1.371 & 0.08 & $\pm$ 0.04 & $_{-0.08}^{+0.21}$ & $_{-0.03}^{+0.03}$\\
-1.371 & - & -1.057 & 3.1 & $\pm$ 0.3 & $_{-1.3}^{+5.2}$ & $_{-0.6}^{+0.6}$\\
-1.057 & - & -0.743 & 7.6 & $\pm$ 0.7 & $_{-1.4}^{+1.9}$ & $_{-1.0}^{+1.0}$\\
-0.743 & - & -0.429 & 15.2 & $\pm$ 0.9 & $_{-1.7}^{+2.3}$ & $_{-2.5}^{+2.5}$\\
-0.429 & - & -0.114 & 22.9 & $\pm$ 1.2 & $_{-6.0}^{+4.2}$ & $_{-3.5}^{+3.5}$\\
-0.114 & - & 0.200 & 20.9 & $\pm$ 1.1 & $_{-2.5}^{+3.8}$ & $_{-2.5}^{+2.5}$\\
0.200 & - & 0.514 & 26.4 & $\pm$ 1.2 & $_{-5.2}^{+2.7}$ & $_{-3.2}^{+3.2}$\\
0.514 & - & 0.829 & 33.8 & $\pm$ 1.3 & $_{-9.4}^{+5.6}$ & $_{-3.9}^{+3.9}$\\
0.829 & - & 1.143 & 28.1 & $\pm$ 1.3 & $_{-3.0}^{+5.0}$ & $_{-3.7}^{+3.7}$\\
1.143 & - & 1.457 & 25.4 & $\pm$ 1.1 & $_{-4.7}^{+5.5}$ & $_{-3.6}^{+3.6}$\\
1.457 & - & 1.771 & 29.4 & $\pm$ 1.1 & $_{-3.6}^{+4.1}$ & $_{-4.4}^{+4.4}$\\
1.771 & - & 2.086 & 43.3 & $\pm$ 1.6 & $_{-7.0}^{+8.4}$ & $_{-6.7}^{+6.7}$\\
2.086 & - & 2.400 & 51 & $\pm$ 2 & $_{-12}^{+15}$ & $_{-10}^{+10}$\\
\hline
\end{tabular}
\caption{The measured four-jet differential cross-section $d\sigma/d\eta^{{\rm jet2}}$ in both the low- and high-mass regions. Other details as described in the caption to Table~\ref{tab:Mnj_3j_}.
\label{tab:Eta4_4j_}}
\end{center}
\end{table}
\clearpage

\begin{table}
\begin{center}
\begin{tabular}{|l c l||l l l l|}
\hline
\multicolumn{3}{|c||}{$\eta^{{\rm jet3}}$ range} & $d\sigma/d\eta^{{\rm jet3}}$ & $\pm$stat. & $\pm$syst. & $\pm$E-scale\\
\multicolumn{3}{|c||}{} & \multicolumn{4}{c|}{(pb)}\\
\hline
\multicolumn{7}{|c|}{$25\le M_{4j}<50$~GeV}\\
\hline
-1.371 & - & -1.057 & 0.5 & $\pm$ 0.1 & $_{-0.4}^{+1.1}$ & $_{-0.3}^{+0.3}$\\
-1.057 & - & -0.743 & 3.7 & $\pm$ 0.4 & $_{-0.9}^{+1.1}$ & $_{-0.9}^{+0.9}$\\
-0.743 & - & -0.429 & 9.4 & $\pm$ 0.6 & $_{-1.5}^{+1.9}$ & $_{-1.2}^{+1.2}$\\
-0.429 & - & -0.114 & 20.0 & $\pm$ 0.8 & $_{-7.8}^{+7.2}$ & $_{-3.2}^{+3.2}$\\
-0.114 & - & 0.200 & 26.2 & $\pm$ 0.9 & $_{-3.3}^{+2.6}$ & $_{-4.6}^{+4.6}$\\
0.200 & - & 0.514 & 37.2 & $\pm$ 1.2 & $_{-4.4}^{+6.4}$ & $_{-6.5}^{+6.5}$\\
0.514 & - & 0.829 & 45.7 & $\pm$ 1.4 & $_{-5.8}^{+4.3}$ & $_{-7.8}^{+7.8}$\\
0.829 & - & 1.143 & 47.8 & $\pm$ 1.4 & $_{-5.6}^{+5.0}$ & $_{-8.5}^{+8.5}$\\
1.143 & - & 1.457 & 37.8 & $\pm$ 1.0 & $_{-4.1}^{+8.2}$ & $_{-7.3}^{+7.3}$\\
1.457 & - & 1.771 & 40.0 & $\pm$ 1.0 & $_{-5.5}^{+7.2}$ & $_{-8.6}^{+8.6}$\\
1.771 & - & 2.086 & 51 & $\pm$ 1 & $_{-6}^{+8}$ & $_{-11}^{+11}$\\
2.086 & - & 2.400 & 61 & $\pm$ 2 & $_{-19}^{+16}$ & $_{-13}^{+13}$\\
\hline
\multicolumn{7}{|c|}{$M_{4j}\ge50$~GeV}\\
\hline
-1.686 & - & -1.371 & 0.59 & $\pm$ 0.20 & $_{-0.59}^{+0.97}$ & $_{-0.28}^{+0.28}$\\
-1.371 & - & -1.057 & 6.1 & $\pm$ 0.6 & $_{-2.0}^{+2.0}$ & $_{-1.4}^{+1.4}$\\
-1.057 & - & -0.743 & 19 & $\pm$ 1 & $_{-10}^{+5}$ & $_{-3}^{+3}$\\
-0.743 & - & -0.429 & 20.4 & $\pm$ 1.1 & $_{-4.3}^{+4.6}$ & $_{-4.6}^{+4.6}$\\
-0.429 & - & -0.114 & 21.0 & $\pm$ 1.1 & $_{-3.4}^{+4.0}$ & $_{-3.9}^{+3.9}$\\
-0.114 & - & 0.200 & 22.7 & $\pm$ 1.2 & $_{-4.3}^{+4.8}$ & $_{-3.6}^{+3.6}$\\
0.200 & - & 0.514 & 24.8 & $\pm$ 1.2 & $_{-4.1}^{+6.0}$ & $_{-3.7}^{+3.7}$\\
0.514 & - & 0.829 & 27.6 & $\pm$ 1.3 & $_{-3.4}^{+3.8}$ & $_{-4.1}^{+4.1}$\\
0.829 & - & 1.143 & 27.3 & $\pm$ 1.3 & $_{-4.1}^{+6.3}$ & $_{-4.1}^{+4.1}$\\
1.143 & - & 1.457 & 24.0 & $\pm$ 1.1 & $_{-2.7}^{+4.1}$ & $_{-3.9}^{+3.9}$\\
1.457 & - & 1.771 & 26.3 & $\pm$ 1.0 & $_{-2.7}^{+4.3}$ & $_{-3.9}^{+3.9}$\\
1.771 & - & 2.086 & 33.5 & $\pm$ 1.4 & $_{-2.9}^{+4.3}$ & $_{-4.8}^{+4.8}$\\
2.086 & - & 2.400 & 55 & $\pm$ 2 & $_{-13}^{+11}$ & $_{-5}^{+5}$\\
\hline
\end{tabular}
\caption{The measured four-jet differential cross-section $d\sigma/d\eta^{{\rm jet3}}$ in both the low- and high-mass regions. Other details as described in the caption to Table~\ref{tab:Mnj_3j_}.
\label{tab:Eta5_4j_}}
\end{center}
\end{table}
\clearpage

\begin{table}
\begin{center}
\begin{tabular}{|l c l||l l l l|}
\hline
\multicolumn{3}{|c||}{$\eta^{{\rm jet4}}$ range} & $d\sigma/d\eta^{{\rm jet4}}$ & $\pm$stat. & $\pm$syst. & $\pm$E-scale\\
\multicolumn{3}{|c||}{} & \multicolumn{4}{c|}{(pb)}\\
\hline
\multicolumn{7}{|c|}{$25\le M_{4j}<50$~GeV}\\
\hline
-1.371 & - & -1.057 & 1.3 & $\pm$ 0.2 & $_{-1.1}^{+1.0}$ & $_{-0.3}^{+0.3}$\\
-1.057 & - & -0.743 & 4.4 & $\pm$ 0.4 & $_{-1.6}^{+1.1}$ & $_{-0.4}^{+0.4}$\\
-0.743 & - & -0.429 & 18 & $\pm$ 1 & $_{-13}^{+10}$ & $_{-3}^{+3}$\\
-0.429 & - & -0.114 & 18.8 & $\pm$ 0.8 & $_{-1.9}^{+2.1}$ & $_{-3.6}^{+3.6}$\\
-0.114 & - & 0.200 & 26.3 & $\pm$ 0.9 & $_{-2.4}^{+4.0}$ & $_{-4.5}^{+4.5}$\\
0.200 & - & 0.514 & 36.6 & $\pm$ 1.2 & $_{-4.3}^{+4.5}$ & $_{-6.4}^{+6.4}$\\
0.514 & - & 0.829 & 44.1 & $\pm$ 1.3 & $_{-6.3}^{+7.1}$ & $_{-7.7}^{+7.7}$\\
0.829 & - & 1.143 & 45.0 & $\pm$ 1.4 & $_{-5.0}^{+7.3}$ & $_{-7.9}^{+7.9}$\\
1.143 & - & 1.457 & 37.8 & $\pm$ 1.0 & $_{-4.1}^{+7.9}$ & $_{-7.0}^{+7.0}$\\
1.457 & - & 1.771 & 39.2 & $\pm$ 1.0 & $_{-3.5}^{+7.5}$ & $_{-8.2}^{+8.2}$\\
1.771 & - & 2.086 & 48.3 & $\pm$ 1.3 & $_{-6.8}^{+7.6}$ & $_{-9.8}^{+9.8}$\\
2.086 & - & 2.400 & 60 & $\pm$ 2 & $_{-15}^{+14}$ & $_{-12}^{+12}$\\
\hline
\multicolumn{7}{|c|}{$M_{4j}\ge50$~GeV}\\
\hline
-1.686 & - & -1.371 & 2.8 & $\pm$ 0.4 & $_{-1.8}^{+2.0}$ & $_{-2.2}^{+5.6}$\\
-1.371 & - & -1.057 & 7.3 & $\pm$ 0.7 & $_{-2.2}^{+2.8}$ & $_{-1.0}^{+1.0}$\\
-1.057 & - & -0.743 & 15.2 & $\pm$ 0.9 & $_{-2.2}^{+3.8}$ & $_{-2.0}^{+2.0}$\\
-0.743 & - & -0.429 & 21.6 & $\pm$ 1.2 & $_{-4.0}^{+4.1}$ & $_{-3.5}^{+3.5}$\\
-0.429 & - & -0.114 & 22.7 & $\pm$ 1.2 & $_{-5.7}^{+5.5}$ & $_{-3.5}^{+3.5}$\\
-0.114 & - & 0.200 & 24.3 & $\pm$ 1.1 & $_{-4.1}^{+4.1}$ & $_{-3.9}^{+3.9}$\\
0.200 & - & 0.514 & 23.3 & $\pm$ 1.2 & $_{-3.1}^{+4.9}$ & $_{-3.7}^{+3.7}$\\
0.514 & - & 0.829 & 27.3 & $\pm$ 1.2 & $_{-4.0}^{+4.1}$ & $_{-4.0}^{+4.0}$\\
0.829 & - & 1.143 & 34.6 & $\pm$ 1.3 & $_{-9.3}^{+6.3}$ & $_{-4.7}^{+4.7}$\\
1.143 & - & 1.457 & 25.8 & $\pm$ 1.0 & $_{-3.5}^{+2.8}$ & $_{-3.7}^{+3.7}$\\
1.457 & - & 1.771 & 25.6 & $\pm$ 1.0 & $_{-4.2}^{+5.3}$ & $_{-4.2}^{+4.2}$\\
1.771 & - & 2.086 & 34.0 & $\pm$ 1.4 & $_{-5.8}^{+7.4}$ & $_{-5.4}^{+5.4}$\\
2.086 & - & 2.400 & 44.3 & $\pm$ 2.0 & $_{-9.7}^{+9.9}$ & $_{-6.1}^{+6.1}$\\
\hline
\end{tabular}
\caption{The measured four-jet differential cross-section $d\sigma/d\eta^{{\rm jet4}}$ in both the low- and high-mass regions. Other details as described in the caption to Table~\ref{tab:Mnj_3j_}.
\label{tab:Eta6_4j_}}
\end{center}
\end{table}
\clearpage

\begin{table}
\begin{center}
\begin{tabular}{|l c l||l l l l||l l l l|}
\hline
\multicolumn{3}{|c||}{$cos(\psi_{3})$ range} & $d\sigma/dcos(\psi_{3})$ & $\pm$stat. & $\pm$syst. & $\pm$E-scale & \multicolumn{2}{c}{had. corr.}&\multicolumn{2}{c|}{MPI corr.}\\
\multicolumn{3}{|c||}{} & \multicolumn{4}{c||}{(pb)} & & & &\\
\hline
\multicolumn{11}{|c|}{$25\le M_{3j}<50$~GeV}\\
\hline
-1.00 & - & -0.88 & 1494 & $\pm$ 13 & $_{-175}^{+217}$ & $_{-188}^{+188}$ &0.864& $\pm$0.012&1.76& $\pm$0.43\\
-0.88 & - & -0.72 & 856 & $\pm$ 9 & $_{-105}^{+117}$ & $_{-107}^{+107}$ &0.821& $\pm$0.017&1.51& $\pm$0.28\\
-0.72 & - & -0.56 & 804 & $\pm$ 8 & $_{-77}^{+77}$ & $_{-99}^{+99}$ &0.793& $\pm$0.019&1.43& $\pm$0.20\\
-0.560 & - & -0.336 & 766 & $\pm$ 7 & $_{-76}^{+75}$ & $_{-94}^{+94}$ &0.776& $\pm$0.008&1.43& $\pm$0.19\\
-0.336 & - & -0.112 & 759 & $\pm$ 7 & $_{-64}^{+70}$ & $_{-94}^{+94}$ &0.763& $\pm$0.005&1.42& $\pm$0.20\\
-0.112 & - & 0.112 & 746 & $\pm$ 6 & $_{-57}^{+70}$ & $_{-92}^{+92}$ &0.761& $\pm$0.006&1.41& $\pm$0.19\\
0.112 & - & 0.336 & 812 & $\pm$ 7 & $_{-80}^{+80}$ & $_{-98}^{+98}$ &0.767& $\pm$0.012&1.42& $\pm$0.20\\
0.336 & - & 0.560 & 851 & $\pm$ 7 & $_{-98}^{+101}$ & $_{-99}^{+99}$ &0.780& $\pm$0.015&1.48& $\pm$0.22\\
0.56 & - & 0.72 & 886 & $\pm$ 9 & $_{-103}^{+99}$ & $_{-100}^{+100}$ &0.809& $\pm$0.017&1.48& $\pm$0.22\\
0.72 & - & 0.88 & 898 & $\pm$ 9 & $_{-107}^{+114}$ & $_{-102}^{+102}$ &0.839& $\pm$0.013&1.51& $\pm$0.27\\
0.88 & - & 1.00 & 1475 & $\pm$ 13 & $_{-196}^{+212}$ & $_{-173}^{+173}$ &0.894& $\pm$0.006&1.64& $\pm$0.38\\
\hline
\multicolumn{11}{|c|}{$M_{3j}\ge50$~GeV}\\
\hline
-1.00 & - & -0.88 & 542 & $\pm$ 9 & $_{-94}^{+99}$ & $_{-55}^{+55}$ &0.824& $\pm$0.018&1.15& $\pm$0.15\\
-0.88 & - & -0.72 & 263 & $\pm$ 5 & $_{-45}^{+44}$ & $_{-29}^{+29}$ &0.812& $\pm$0.031&1.11& $\pm$0.11\\
-0.72 & - & -0.56 & 215 & $\pm$ 5 & $_{-40}^{+38}$ & $_{-25}^{+25}$ &0.807& $\pm$0.021&1.10& $\pm$0.10\\
-0.560 & - & -0.336 & 164 & $\pm$ 3 & $_{-24}^{+26}$ & $_{-19}^{+19}$ &0.790& $\pm$0.026&1.10& $\pm$0.09\\
-0.336 & - & -0.112 & 156 & $\pm$ 3 & $_{-30}^{+31}$ & $_{-18}^{+18}$ &0.793& $\pm$0.012&1.10& $\pm$0.10\\
-0.112 & - & 0.112 & 149 & $\pm$ 3 & $_{-27}^{+33}$ & $_{-17}^{+17}$ &0.785& $\pm$0.013&1.11& $\pm$0.11\\
0.112 & - & 0.336 & 143 & $\pm$ 3 & $_{-22}^{+25}$ & $_{-15}^{+15}$ &0.784& $\pm$0.009&1.12& $\pm$0.11\\
0.336 & - & 0.560 & 151 & $\pm$ 3 & $_{-28}^{+25}$ & $_{-16}^{+16}$ &0.802& $\pm$0.017&1.12& $\pm$0.11\\
0.56 & - & 0.72 & 169 & $\pm$ 4 & $_{-34}^{+32}$ & $_{-17}^{+17}$ &0.823& $\pm$0.015&1.10& $\pm$0.11\\
0.72 & - & 0.88 & 217 & $\pm$ 5 & $_{-45}^{+39}$ & $_{-22}^{+22}$ &0.847& $\pm$0.006&1.10& $\pm$0.11\\
0.88 & - & 1.00 & 370 & $\pm$ 7 & $_{-61}^{+67}$ & $_{-38}^{+38}$ &0.865& $\pm$0.009&1.12& $\pm$0.12\\
\hline
\end{tabular}
\caption{The measured three-jet differential cross-section $d\sigma/dcos(\psi_{3})$ in both the low- and high-mass regions. Other details as described in the caption to Table~\ref{tab:Mnj_3j_}.
\label{tab:cos_psi3_3j_}}
\end{center}
\end{table}
\clearpage

\newpage 

%%%%%%%%%%%%%%%%%%%%%%%%%%%%%%%%%%%%%%%%%%%%%%%%%%%%%%%%%%%%%%%%%%%%%%%%%

%%%%%%%%%%%%%%%%%%%%%%%%%%%%%%%%%%%%%%%%%%%%%%%%%%%%%%%%%%%%%%%%%%%%%%%%%

%%%%%%%%%%%%%%%%%%%%%%%%%%%%%%%%%%%%%%%%%%%%%%%%%%%%%%%%%%%%%%%%%%%%%%%%%

%%%%%%%%%%%%%%%%%%%%%%%%%%%%%%%%%%%%%%%%%%%%%%%%%%%%%%%%%%%%%%%%%%%%%%%%%

%%%%%%%%%%%%%%%%%%%%%%%%%%%%%%%%%%%%%%%%%%%%%%%%%%%%%%%%%%%%%%%%%%%%%%%%%

%%%%%%%%%%%%%%%%%%%%%%%%%%%%%%%%%%%%%%%%%%%%%%%%%%%%%%%%%%%%%%%%%%%%%%%%%

%%%%%%%%%%%%%%%%%%%%%%%%%%%%%%%%%%%%%%%%%%%%%%%%%%%%%%%%%%%%%%%%%%%%%%%%

%%%%%%%%%%%%%%%%%%%%%%%%%%%%%%%%%%%%%%%%%%%%%%%%%%%%%%%%%%%%%%%%%%%%%%%%

%%%%%%%%%%%%%%%%%%%%%%%%%%%%%%%%%%%%%%%%%%%%%%%%%%%%%%%%%%%%%%%%%%%%%%%%

%%%%%%%%%%%%%%%%%%%%%%%%%%%%%%%%%%%%%%%%%%%%%%%%%%%%%%%%%%%%%%%%%%%%%%%%%

%%%%%%%%%%%%%%%%%%%%%%%%%%%%%%%%%%%%%%%%%%%%%%%%%%%%%%%%%%%%%%%%%%%%%%%%%

%%%%%%%%%%%%%%%%%%%%%%%%%%%%%%%%%%%%%%%%%%%%%%%%%%%%%%%%%%%%%%%%%%%%%%%%%

%%%%%%%%%%%%%%%%%%%%%%%%%%%%%%%%%%%%%%%%%%%%%%%%%%%%%%%%%%%%%%%%%%%%%%%%

%%%%%%%%%%%%%%%%%%%%%%%%%%%%%%%%%%%%%%%%%%%%%%%%%%%%%%%%%%%%%%%%%%%%%%%%%

%%%%%%%%%%%%%%%%%%%%%%%%%%%%%%%%%%%%%%%%%%%%%%%%%%%%%%%%%%%%%%%%%%%%%%%%%

\end{document}